 \documentclass[iop,apj,numberedappendix,twocolappendix]{emulateapj}

 \usepackage{apjfonts}

\slugcomment{to appear in the Astrophysical Journal}

\shorttitle{Light Curve Analysis of Slow Novae}
\shortauthors{Hachisu \& Kato}

\begin{document}

\title{A Light Curve Analysis of Classical Novae: \\
Free-free Emission vs. Photospheric Emission}


\author{Izumi Hachisu}
\affil{Department of Earth Science and Astronomy, 
College of Arts and Sciences, The University of Tokyo,
3-8-1 Komaba, Meguro-ku, Tokyo 153-8902, Japan} 
\email{hachisu@ea.c.u-tokyo.ac.jp}

\and

\author{Mariko Kato}
\affil{Department of Astronomy, Keio University, 
Hiyoshi, Kouhoku-ku, Yokohama 223-8521, Japan} 
\email{mariko@educ.cc.keio.ac.jp}

%




\begin{abstract}
    We analyzed light curves of seven relatively slower novae, 
PW~Vul, V705~Cas, GQ~Mus,  RR~Pic, V5558~Sgr, HR~Del, and V723~Cas,
based on an optically thick wind theory of nova outbursts.  For fast novae,
free-free emission dominates the spectrum in optical bands rather
than photospheric emission and nova optical light curves follow the
universal decline law.  Faster novae blow stronger winds with larger
mass loss rates.  Since the brightness of free-free emission depends
directly on the wind mass loss rate, faster novae show brighter
optical maxima.   In slower novae, however, we must take into account
photospheric emission because of their lower wind mass loss rates.
We calculated three model light curves of free-free emission,
photospheric emission, and the sum of them for various WD masses with
various chemical compositions of their envelopes, and fitted reasonably
with observational data of optical, near-IR (NIR), and UV bands.
From light curve fittings of the seven novae,
we estimated their absolute magnitudes, distances, and WD masses.
In PW~Vul and V705~Cas, free-free emission still dominates the spectrum
in the optical and NIR bands.  In the very slow novae, RR~Pic, V5558~Sgr,
HR~Del, and V723~Cas, photospheric emission dominates the spectrum rather
than free-free emission, which makes a deviation from the universal
decline law.  We have confirmed that the absolute brightnesses of
our model light curves are consistent with the distance moduli of four
classical novae with known distances (GK~Per, V603~Aql, RR~Pic, and DQ~Her).
We also discussed the reason why the very slow novae are about $\sim 1$ mag
brighter than the proposed maximum magnitude
vs. rate of decline relation.
\end{abstract}




\section{Introduction}
A classical nova is a thermonuclear runaway event on a mass-accreting
white dwarf (WD) in a binary.  When the mass of the hydrogen-rich
envelope on the WD reaches a critical value,
hydrogen ignites to trigger a nova outburst.
Optical light curves of novae have a wide variety of timescales and shapes 
\citep[e.g.,][]{pay57,due81,str10,hac14k}.  \citet{hac06kb} found that,
in terms of free-free emission, optical and near-infrared (NIR) light
curves of several novae follow a universal decline law.
Their time-normalized light curves are almost independent of the WD mass,
chemical composition of ejecta, and wavelength.  \citet{hac06kb} also
found that their UV 1455 \AA\  model light curves \citep{cas02},
interpreted as photospheric blackbody emission, are also time-normalized
by the same factor as in the optical and NIR light curves.  Using the fact
that the time-scaling factor is closely related to the WD mass, the authors
determined the WD mass and other parameters for a number of well-observed
novae \citep[e.g.,][]{hac07k, hac10k, hac14k, hac08kc, kat09hc}.

On the basis of the universal decline law, \citet{hac10k} further obtained
absolute magnitudes of their model light curves and derived
their maximum magnitude vs. rate of decline (MMRD) relation.  Such
MMRD relations were empirically proposed, e.g., by \citet{sch57}, 
\citet{del95}, and \citet{dow00}.  For individual novae, however, 
there is large scatter around the proposed trends \citep[e.g.,][]{dow00}.
Hachisu \& Kato's (2010) theoretical MMRD relation is governed
by two parameters, one is the WD mass and the other is the initial
envelope mass at the nova outburst, i.e., the ignition mass.  
The ignition mass depends on the mass-accretion rate to the WD.  
The higher the mass-accretion rate is, the smaller the ignition mass is 
\citep[e.g.,][]{nom82, pri95, kat14shn}.  In other words, the smaller 
the mass accretion rate is, the brighter the maximum magnitude is.
So, they concluded that this second parameter (the ignition mass) explains 
scatter of the MMRD distribution of individual novae from the averaged
trend that was determined mainly by the WD mass.  Thus, the main 
trends of nova speed class were theoretically clarified. 

In this way, the main properties of fast novae have been theoretically
explained, in which free-free emission dominates the continuum spectra
in optical and NIR bands.  As far as free-free emission is the dominant
source of nova optical light curves, there should be the universal
decline law and we expect that novae follow Hachisu \& Kato's (2010)
theoretical MMRD relation with the intrinsic scatter mentioned above.
For slow novae, however, the universal decline law could not be applied
because photospheric emission contributes substantially to the
continuum spectra rather than free-free emission \citep{hac14k}.
Our aim of this paper is to analyze light curves of seven relatively
slower novae, PW~Vul, V705~Cas, GQ~Mus, RR~Pic, V5558~Sgr, HR~Del,
and V723~Cas, and to clarify how deviate their light curves from
the universal decline law.

We organize the present paper as follows.
Section \ref{slow_vs_fast} describes our strategy of light curve analysis.
In Section \ref{result_pw_vul}, we start with a study of well-observed
multi-wavelength light-curves of the slow nova PW~Vul and, through our
method,  we determine its relevant physical parameters such as the WD mass.  
Our method for nova light curves is also applied to the moderately fast 
nova V705~Cas in Section \ref{v705_cas}, to the fast nova
GQ~Mus in Section \ref{gq_mus}, and to the very slow novae, RR~Pic,
V5558~Sgr, HR~Del, and V723~Cas in Section \ref{very_slow_novae}.  
Discussion and conclusions follow in Sections \ref{discussion_section}
and \ref{conclusions}.  Appendix \ref{absolute_magnitudes} is devoted
to a calibration of the absolute magnitude of our free-free model
light curves and a theoretical MMRD relation.  Appendix 
\ref{time-stretching_method_appendix} presents
our time-stretching method for PW~Vul.

\begin{figure}
\epsscale{1.0}
\plotone{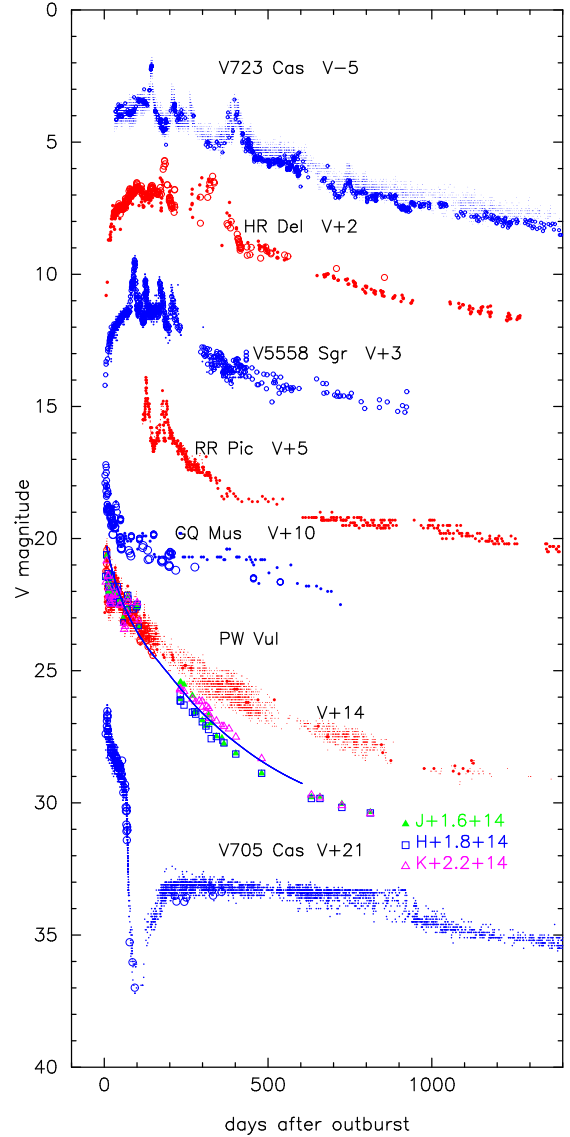}
\caption{
Visual and $V$ light curves for seven novae, from top to bottom,
V723~Cas, HR~Del, V5558~Sgr, RR~Pic, GQ~Mus, PW~Vul, and V705~Cas
in the order of global decline rates.
Red or blue open circles denote $V$ magnitudes for each nova.
Green filled triangles are $J$, blue open squares are $H$,
and magenta open triangles are $K$ magnitudes of PW~Vul.
Their reference sources are found in the section of each object.
Red or blue small dots represent visual magnitudes, all taken from
the American Association of Variable
Star Observers (AAVSO) archive except for GQ~Mus.  
Blue small dots for GQ~Mus are visual magnitude data collected
by the Royal Astronomical Society of New Zealand.
A blue solid line is our theoretical free-free emission
light curve for a $0.83~M_\sun$ WD (see Section \ref{result_pw_vul}),
which nicely fits with the near-infrared (NIR) light curves of PW~Vul but not
with the optical data that is contaminated by strong emission lines
in the nebular phase.
\label{light_curve_rr_pic_v723_cas_hr_del_v5558_sgr_v705_cas_pw_vul}}
\end{figure}


\begin{figure}
\epsscale{1.15}
\plotone{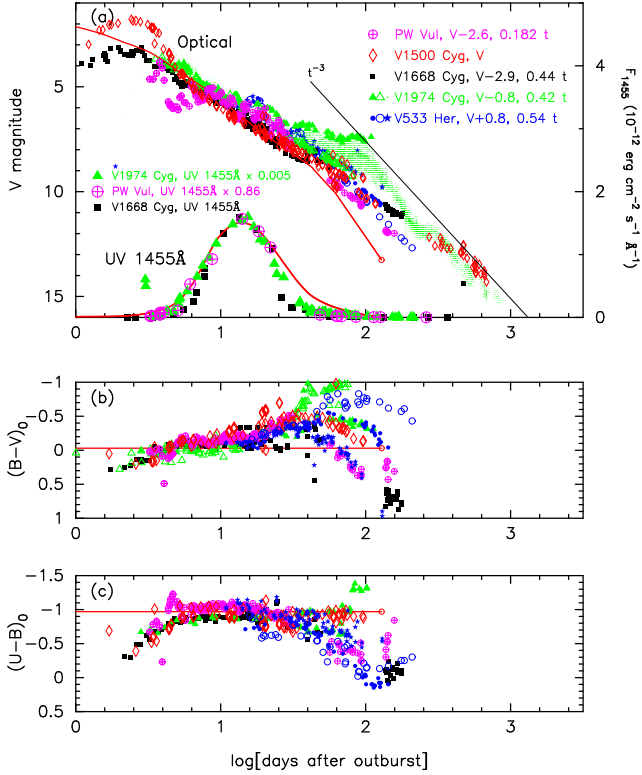}
\caption{
(a) $V$ band and UV 1455\AA\ narrow band light curves,
(b) $(B-V)_0$, and (c) $(U-B)_0$ color curves,
for PW~Vul (magenta open circles with plus sign inside),
V1500~Cyg (red open diamonds), V1668~Cyg (black filled squares),
and V1974~Cyg (green open and filled triangles).  The reference
sources are the same as those in \citet{hac06kb, hac10k, hac14k}.
Here, $(B-V)_0$ and $(U-B)_0$ denote dereddened colors.
Each color is dereddened with
$(B-V)_0= B-V - E(B-V)$ and $(U-B)_0= U-B - 0.64 E(B-V)$ \citep{rie85}.
Those of V533~Her are also added:
blue filled circles are taken from \citet{gen63},
blue open circles are from \citet{chi64}, and blue star symbols
are from \citet{she64}.  
To make them overlap in the early decline phase,
we shift horizontally their logarithmic times of PW~Vul, V1668~Cyg,
V1974~Cyg, and V533~Her by $-0.74=\log 0.182$, 
$-0.36=\log 0.44$, $-0.38=\log 0.42$, and $-0.28=\log 0.54$,
and vertically their magnitudes by $-2.6$, $-2.9$, $-0.8$,
and $+0.8$ mag, respectively, as indicated in the figure.
UV~1455\AA\  fluxes of each nova are also rescaled against that of
V1668~Cyg as indicated
in the figure.  Here, we assume the start of the day ($t=0$)
as JD 2445910.0 for PW~Vul and JD 2438052.0 for V533~Her
to correctly overlap them with other light curves.  
Red solid lines are our theoretical free-free emission and UV~1455\AA\ 
light curves for a $0.83~M_\sun$ WD (see Section \ref{result_pw_vul}).
We also draw two red horizontal lines of (b) $B-V=-0.03$ and 
(c) $U-B=-0.97$, both of which are the colors of optically
thick free-free emission \citep[see, e.g.,][]{hac14k}.
\label{pw_vul_v1668_cyg_v1500_cyg_v1974_cyg_v_bv_ub_color_curve_logscale_no3}}
\end{figure}


\begin{figure}
\epsscale{1.15}
\plotone{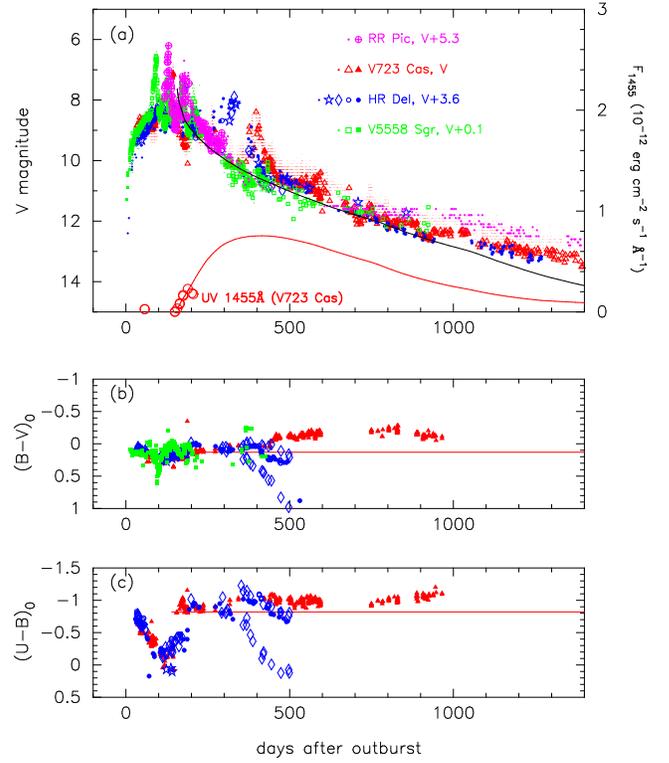}
\caption{
(a) Visual, $V$ band, and UV 1455\AA\ narrow band light curves,
(b) $(B-V)_0$, and (c) $(U-B)_0$ color curves,
for RR~Pic (magenta open circles with plus sign inside, magenta dots),
V723~Cas (red open and filled triangles, red dots),
HR~Del (blue open stars, open circles, filled circles, blue dots),
and V5558~Sgr (green open and filled squares, green dots).  Here,
$(B-V)_0$ and $(U-B)_0$ denote dereddened colors.
To make them overlap in the early decline phase,
we shift horizontally their times 
and vertically their magnitudes by $+5.3$, $0.0$, $+3.6$,
and $+0.1$ mag, respectively, as indicated in the figure.
UV~1455\AA\  fluxes of V723~Cas are plotted by large open circles.
Black/Red solid lines are our theoretical $V$/UV~1455\AA\ 
light curves for a $0.51~M_\sun$ WD (see Section \ref{v723_cas}).
We add two red horizontal lines of (b) $B-V=+0.13$ and 
(c) $U-B=-0.82$, both of which are the colors of optically
thin free-free emission \citep[see, e.g.,][]{hac14k}.
\label{v723_cas_hr_del_v5558_sgr_rr_pic_ub_bv_color_light_curve_revised_no2}}
\end{figure}


\begin{figure}
\epsscale{1.15}
\plotone{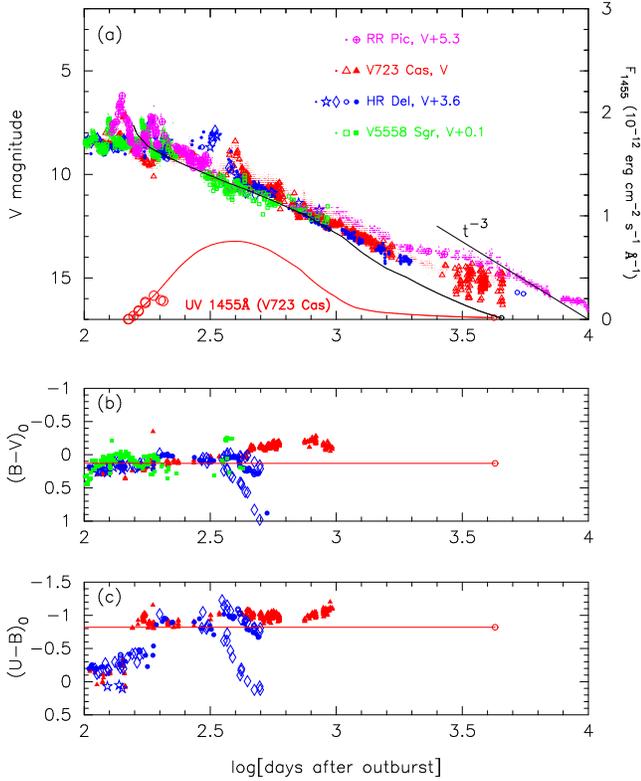}
\caption{
Same as Figure
\ref{v723_cas_hr_del_v5558_sgr_rr_pic_ub_bv_color_light_curve_revised_no2},
but on a logarithmic timescale.
\label{v723_cas_hr_del_v5558_sgr_rr_pic_ub_bv_color_light_curve_revised_logscale_no2}}
\end{figure}


\begin{figure}
\epsscale{1.15}
\plotone{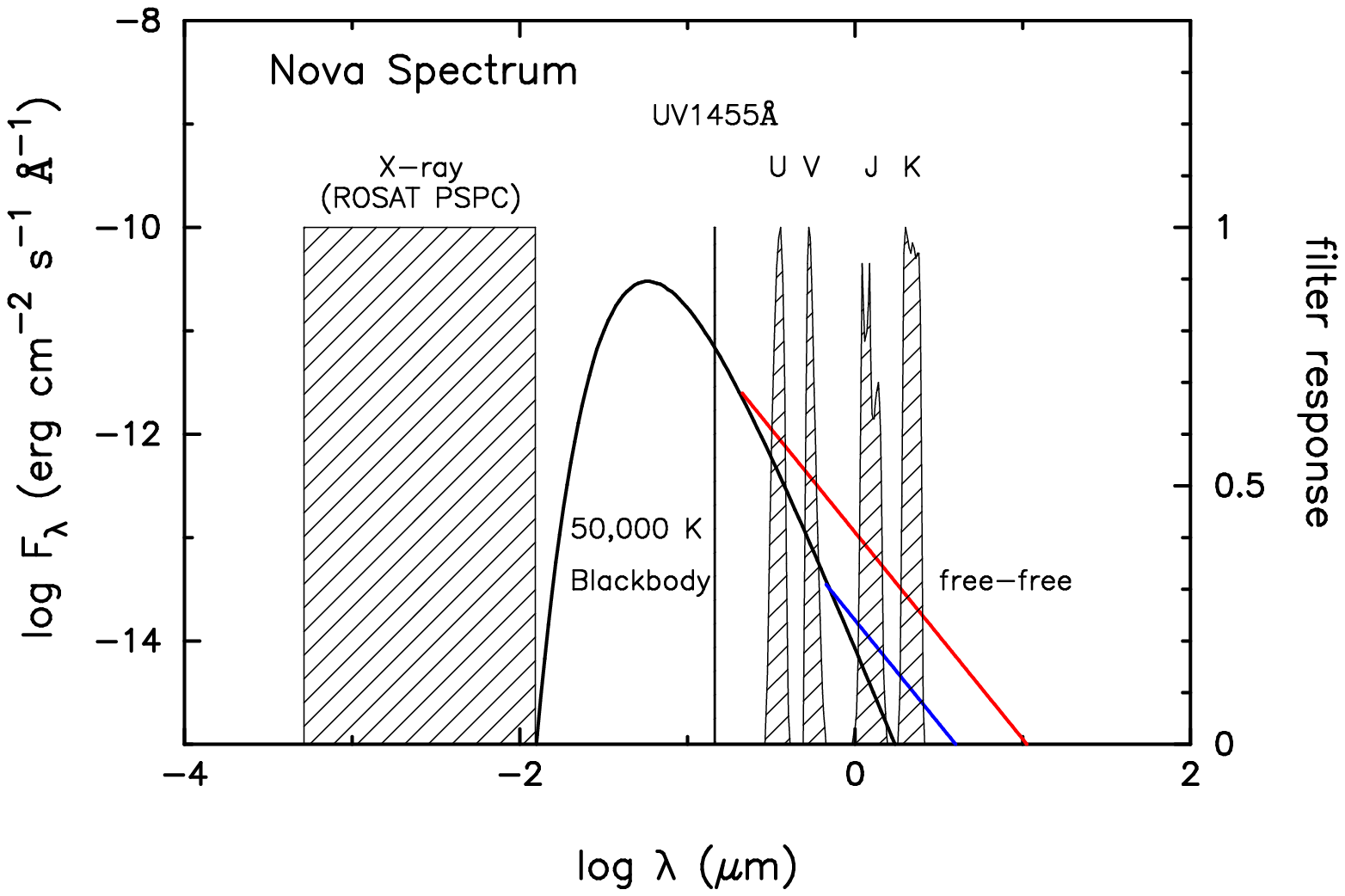}
\caption{
An illustrative example of spectral energy distribution of a classical nova
with a photospheric temperature of $T_{\rm ph}=50,000$~K as well as
passbands of the photometric filters used in this work.
The UV 1455\AA\  (1445--1465\AA) and supersoft X-ray (0.1--2.4~keV) 
fluxes are calculated from blackbody spectrum while the $U$, $V$, $J$, 
and $K$ magnitudes are calculated from free-free emission spectrum.
Note that the flux of free-free emission depends on the wind mass-loss
rate, $\dot M_{\rm wind}$.  If the mass-loss rate is large, 
the flux of free-free emission dominates the spectrum 
in the optical and IR region as indicated by a red solid line.
If it is small, the flux of free-free
emission may dominate only in the IR region as indicated by a blue line.
Supersoft X-ray flux is negligibly small for
the adopted photospheric temperature.
\label{sed_uv_opt_ir_ijhk_filter_no2}}
\end{figure}

\section{Analysis on Various Light Curve Shapes of Classical Novae}
\label{slow_vs_fast}

Figure \ref{light_curve_rr_pic_v723_cas_hr_del_v5558_sgr_v705_cas_pw_vul}
shows optical light curves of our target novae, V723~Cas, HR~Del, V5558~Sgr, 
RR~Pic, GQ~Mus, PW~Vul, and V705~Cas on a linear timescale.
These seven novae are plotted in the order of global decline rate.
The light curves show a rich variety of shapes, so one might think that
there are no common physical properties like the universal decline law.
However, we can find common properties hidden in the complicated 
light curve shapes.  For example, PW~Vul shows an oscillatory behavior
in the light curve but the overall decline trend and color evolution
are very similar to other smoothly-declining classical novae,
as shown in Figure 
\ref{pw_vul_v1668_cyg_v1500_cyg_v1974_cyg_v_bv_ub_color_curve_logscale_no3}
(see Section \ref{result_pw_vul} for details).
Figure
\ref{pw_vul_v1668_cyg_v1500_cyg_v1974_cyg_v_bv_ub_color_curve_logscale_no3}
depicts the time-normalized light curves of PW~Vul and
well-observed fast novae.   Figures 
\ref{v723_cas_hr_del_v5558_sgr_rr_pic_ub_bv_color_light_curve_revised_no2}
and
\ref{v723_cas_hr_del_v5558_sgr_rr_pic_ub_bv_color_light_curve_revised_logscale_no2} 
also show the light curves and colors of V723~Cas, HR~Del, V5558~Sgr, and
RR~Pic.  Closely looking at the light curves in Figures
\ref{pw_vul_v1668_cyg_v1500_cyg_v1974_cyg_v_bv_ub_color_curve_logscale_no3},
\ref{v723_cas_hr_del_v5558_sgr_rr_pic_ub_bv_color_light_curve_revised_no2},
and
\ref{v723_cas_hr_del_v5558_sgr_rr_pic_ub_bv_color_light_curve_revised_logscale_no2},
we can see common properties as follows.
\begin{itemize}
\item[(1)] There is a linear decline phase in the middle part of
optical light curves in Figure
\ref{pw_vul_v1668_cyg_v1500_cyg_v1974_cyg_v_bv_ub_color_curve_logscale_no3}.
This phase is explained by free-free emission based on the optically thick
winds (red solid line) and the light curves follow the universal decline law
\citep{hac06kb, hac10k}.
\item[(2)] In the linear decline phase, the colors are constant at
$B-V=-0.03$ and $U-B=-0.97$, as indicated by a horizontal solid line
in Figure
\ref{pw_vul_v1668_cyg_v1500_cyg_v1974_cyg_v_bv_ub_color_curve_logscale_no3}(b)
and 
\ref{pw_vul_v1668_cyg_v1500_cyg_v1974_cyg_v_bv_ub_color_curve_logscale_no3}(c),
respectively.  They are the colors of optically thick free-free emission
spectra \citep{hac14k}. 
\item[(3)] There is another linear decline of a slope $t^{-3}$
in the very late phase in Figure
\ref{pw_vul_v1668_cyg_v1500_cyg_v1974_cyg_v_bv_ub_color_curve_logscale_no3}(a).
This phase corresponds to free expansion of a nebula with no additional
wind mass loss \citep{hac06kb}.
\item[(4)] The UV~1455\AA\  narrow band flux basically follows a 
time-normalized universal shape unless the flux is absorbed by dust.
The theoretical flux (red solid line) represents a blackbody flux of
the pseudophotosphere \citep{kat94h}. 
\end{itemize}

Novae blow optically thick winds, which are the origin of free-free
emission.  As far as the free-free emission dominates the continuum
spectra, we can expect that novae follow the universal decline law.
Even if there are wavy-structured or dust-blackout shapes,
the overall light curves follow the universal decline law
\citep[e.g.,][]{hac06kb, hac07k, hac10k, hac14k, hac08kc, kat09hc}.
On the other hand, if photospheric emission dominates the
nova continuum spectra, light curves do not follow the universal 
decline law.  

Figure \ref{sed_uv_opt_ir_ijhk_filter_no2} shows 
a schematic illustration of nova continuum spectrum superposed
on the various wavelength bands.  Free-free spectra are 
plotted for a high wind mass loss rate (red solid line) and
a low wind mass loss rate (blue solid line).  
In general, slower novae are related to less massive WDs which blow 
optically thick winds with relatively smaller wind mass loss rates
\citep{kat94h}.
Thus, in such cases, we apply the universal decline law only to
light curves in the NIR region but not to light curves
in the optical regions.

In this paper, we analyze the light curves of relatively
slower novae in the following way.
\begin{itemize}
\item[(a)] First, we determine the WD mass by applying properties
(1)--(4) above.
\item[(b)] For this specified WD mass, we calculate the composite
light curve of free-free plus photospheric emissions.
From the fitting with the $V$ data, we determine the $V$ band
distance modulus of $(m-M)_V$.
We further estimate the distance to the nova
if the color excess $E(B-V)$ is known.
\item[(c)] We compare our results with various properties in literature. 
\end{itemize}

We first analyze PW~Vul, V705~Cas, and GQ~Mus based on the method 
mentioned above.  They are the three novae in the lower part of Figure
\ref{light_curve_rr_pic_v723_cas_hr_del_v5558_sgr_v705_cas_pw_vul}.
Then, we go to the other four novae, RR~Pic, V5558~Sgr, HR~Del,
and V723~Cas.  These four novae show more or less similar light curves
in their optical maximum and decline phases as shown in Figures 
\ref{light_curve_rr_pic_v723_cas_hr_del_v5558_sgr_v705_cas_pw_vul},
\ref{v723_cas_hr_del_v5558_sgr_rr_pic_ub_bv_color_light_curve_revised_no2},
and
\ref{v723_cas_hr_del_v5558_sgr_rr_pic_ub_bv_color_light_curve_revised_logscale_no2}.


\begin{deluxetable*}{llllll}
\tabletypesize{\scriptsize}
\tablecaption{Chemical composition of selected novae 
\label{pw_vul_chemical_abundance}}
\tablewidth{0pt}
\tablehead{
\colhead{object} & 
\colhead{H} & 
\colhead{CNO} & 
\colhead{Ne} &
\colhead{Na-Fe} &
\colhead{reference} \\ 
\colhead{} & 
\colhead{} & 
\colhead{} & 
\colhead{} & 
\colhead{} & 
\colhead{}  
} 
\startdata
HR Del 1967 & 0.45 & 0.074 & 0.0030  & \nodata & \citet{tyl78} \\
DQ Her 1934 & 0.27 & 0.57 & \nodata  & \nodata & \citet{pet90} \\
DQ Her 1934 & 0.34 & 0.56 & \nodata  & \nodata & \citet{wil78} \\
V705 Cas 1993 \#2 & 0.57 & 0.25 & \nodata  & 0.0009 & \citet{ark00} \\
V723 Cas 1995 & 0.52 & 0.064 & 0.052  & 0.042 & \citet{iij06} \\
GQ Mus 1983 & 0.37 & 0.24 & 0.0023  & 0.0039 & \citet{mor96} \\
GQ Mus 1983 & 0.27 & 0.40 & 0.0034  & 0.023 & \citet{has90} \\
GQ Mus 1983 & 0.43 & 0.19 & \nodata  & \nodata & \citet{and90} \\
RR Pic 1925 & 0.53 & 0.032 & 0.011 & \nodata & \citet{wil79} \\
PW Vul 1984 \#1 & 0.69 & 0.066 & 0.00066  & \nodata & \citet{sai91} \\
PW Vul 1984 \#1 & 0.47 & 0.30 & 0.0040  & 0.0048 & \citet{and94} \\
PW Vul 1984 \#1 & 0.62 & 0.13 & 0.001  & 0.0027 & \citet{sch97} \\
PW Vul 1984 \#1 & 0.49 & 0.28 & 0.0019  & \nodata & \citet{and90} \\
\enddata
\end{deluxetable*}

\section{PW~Vul 1984\#1}
\label{result_pw_vul}
     PW~Vul (Nova Vulpeculae 1984\#1) was discovered by Wakuda
on UT 1984 July 27.7 \citep{kos84} about a week before
its optical maximum of $m_{V, {\rm max}} = 6.3$ on UT 1984 August 4.1.
The light curve is plotted in Figure
\ref{light_curve_rr_pic_v723_cas_hr_del_v5558_sgr_v705_cas_pw_vul}
on a linear timescale and in Figure 
\ref{pw_vul_v1668_cyg_v1500_cyg_v1974_cyg_v_bv_ub_color_curve_logscale_no3}
on a logarithmic timescale.
The X-ray flux increased during the first year \citep{oge87} and
faded before the {\it ROSAT} observation (1990 -- 1999).  
No X-ray data are available in the supersoft X-ray phase.

\subsection{Reddening and distance}
\label{distance_reddening_estimate_pw_vul}
     \citet{and91} obtained $E(B-V)=0.58 \pm 0.06$ from \ion{He}{2} 
$\lambda1640/\lambda4686$ ratio and $E(B-V)=0.55 \pm 0.1$ from
the interstellar absorption feature at 2200\AA\  for the reddening toward 
PW~Vul.  \citet{sai91} reported $E(B-V)=0.60 \pm 0.06$ from \ion{He}{2} 
$\lambda1640/\lambda4686$ ratio.   \citet{due84} estimated the extinction
to be $E(B-V)=0.45 \pm 0.1$ from galactic extinction in the direction
toward the nova, whose galactic coordinates are 
$(l, b)=(61\fdg0983,+5\fdg1967)$. 
For the galactic extinction, we examined the galactic dust absorption
map in the NASA/IPAC Infrared Science
Archive\footnote{http://irsa.ipac.caltech.edu/applications/DUST/},
which is calculated on the basis of recent data from \citet{schl11}. 
It gives $E(B-V)=0.43 \pm 0.02$ in the direction of PW~Vul.
The arithmetic mean of these four values is $E(B-V)=0.55 \pm 0.05$.

Recently, \citet{hac14k} proposed a new method for determining 
reddening of classical novae.  They identified a general course of
$UBV$ color-color evolution and determined reddenings of novae
by matching the track of a target nova with their general course.
They obtained $E(B-V)=0.55\pm0.05$ for PW~Vul, which agrees
well with the above mean value. 

     As for the distance to PW~Vul, a reliable estimate, 
$d=1.8 \pm 0.05$~kpc, was obtained by \citet{dow00}
through the nebular expansion parallax method.
Adopting this value, together
with $E(B-V)=0.55\pm0.05$, the distance modulus of PW~Vul is
\begin{eqnarray}
(m-M)_V &=& 5 \log \left({d \over{\rm 10~pc}}\right) + 3.1 E(B-V)\cr\cr
 &=& 5 \log (180\pm5) + 3.1 (0.55\pm0.05)\cr\cr
 &=& 13.0\pm0.2.
\label{distance_modulus_eq_pw_vul_no1}
\end{eqnarray}
We also obtained $(m-M)_{V,{\rm PW~Vul}}=13.0\pm0.1$
(see Appendix \ref{time-stretching_method_appendix}) from ``the 
time-stretching method'' \citep{hac10k} of nova light curves.  This value
is consistent with Equation (\ref{distance_modulus_eq_pw_vul_no1}).


\begin{figure}
\epsscale{1.0}
\plotone{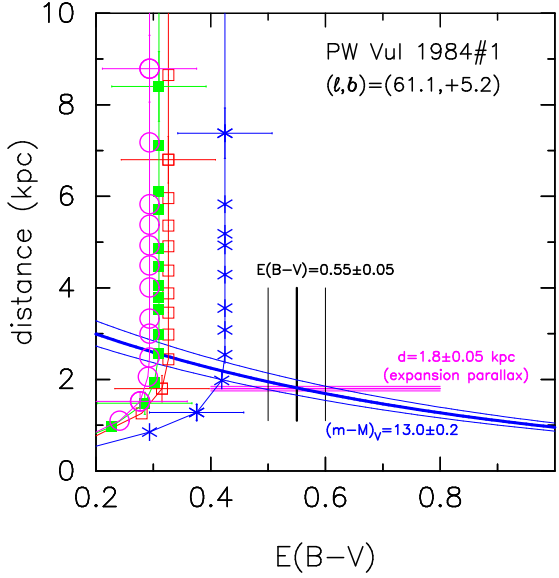}
\caption{
Distance--reddening relation toward PW~Vul.
Blue thick solid line with flanking thin solid lines
denotes the distance--reddening relation
calculated from the distance modulus of Equation 
(\ref{distance_modulus_eq_pw_vul_no1}), i.e., $(m-M)_V
= 5 \log (d /{\rm 10~pc}) + 3.1 E(B-V)= 13.0\pm0.2$.
Vertical thick black solid line with flanking thin lines represents
the color excess of $E(B-V)=0.55\pm0.05$.
Horizontal magenta thick solid line with flanking thin lines represents
the distance estimate of $d=1.8\pm0.05$~kpc
from an expansion parallax method \citep{dow00}.
Distance--reddening relations in four directions close to PW~Vul
are shown by four different sets of data with error bars,
taken from \citet{mar06}, i.e.,
$(l, b)=(61\fdg00,5\fdg00)$ (red open squares),
$(61\fdg25,5\fdg00)$ (green filled squares),
$(61\fdg00,5\fdg25)$ (blue asterisks), and
$(61\fdg25,5\fdg25)$ (magenta open circles).
Typical error bars are shown only two points for each set to avoid
complexity of lines.  
\label{distance_reddening_pw_vul_only_no2}}
\end{figure}

Figure \ref{distance_reddening_pw_vul_only_no2} shows
various distance-reddening relations for comparison.
A horizontal magenta thick solid line with flanking thin lines represents
the distance estimate of $d=1.8\pm0.05$~kpc \citep{dow00} mentioned above.
A vertical black solid line with flanking thin lines represents
the color excess of $E(B-V)=0.55\pm0.05$.
The reddening-distance relation of 
Equation (\ref{distance_modulus_eq_pw_vul_no1})
is plotted by a blue thick solid line with flanking thin solid lines.

\citet{mar06} published a three-dimensional dust extinction map
of our galaxy in the direction of $-100\fdg0 \le l \le 100\fdg0$
and $-10\fdg0 \le b \le +10\fdg0$ with grids of $\Delta l=0\fdg25$
and $\Delta b=0\fdg25$, where $(l,b)$ are the galactic coordinates.
Four sets of data with error bars in Figure 
\ref{distance_reddening_pw_vul_only_no2} show distance--reddening relations
in four directions close to PW~Vul:
$(l, b)=(61\fdg00,5\fdg00)$ (red open squares),
$(61\fdg25,5\fdg00)$ (green filled squares),
$(61\fdg00,5\fdg25)$ (blue asterisks), and
$(61\fdg25,5\fdg25)$ (magenta open circles).
The closest one is the blue asterisk relation that gives
$E(B-V)=0.42\pm0.08$ at $d\approx1.8\pm0.05$~kpc.
This value is consistent with $E(B-V)=0.43\pm0.02$ calculated 
from the NASA/IPAC dust map in the direction of PW~Vul.
Our value of $E(B-V)=0.55$ obtained above is larger than these values.
However, the reddening trend of blue asterisks suggests a large
deviation from the other three trends by $\Delta E(B-V)\approx0.1$, i.e., 
reddening has patchy structure in this direction and further
variation of $\Delta E(B-V)\sim0.1$ is possible.
Thus, we adopt $E(B-V)=0.55$ and $d\approx1.8$~kpc in this paper.


\begin{deluxetable*}{llllllll}
\tabletypesize{\scriptsize}
\tablecaption{Chemical composition of the present models
\label{chemical_composition}}
\tablewidth{0pt}
\tablehead{
\colhead{novae case} & 
\colhead{$X$} & 
\colhead{$Y$} & 
\colhead{$X_{\rm CNO}$} & 
\colhead{$X_{\rm Ne}$} & 
\colhead{$Z$\tablenotemark{a}}  & 
\colhead{mixing\tablenotemark{b}}  & 
\colhead{comments}
} 
\startdata
CO nova 1
 & 0.35 & 0.13 & 0.50 & 0.0 & 0.02 & 100\% & DQ Her \\
CO nova 2\tablenotemark{c} & 0.35 & 0.33 & 0.30 & 0.0 & 0.02 & 100\% & GQ Mus \\
CO nova 3 & 0.45 & 0.18 & 0.35 & 0.0 & 0.02 & 55\% & V1668 Cyg \\ 
CO nova 4 & 0.55 & 0.23 & 0.20 & 0.0 & 0.02 & 25\% & PW Vul \\
Ne nova 1 & 0.35 & 0.33 & 0.20 & 0.10 & 0.02 & 100\% & V351 Pup \\
Ne nova 2\tablenotemark{d}
 & 0.55 & 0.30 & 0.10 & 0.03 & 0.02 & 25\% & V1500 Cyg \\
Ne nova 3 & 0.65 & 0.27 & 0.03 & 0.03 & 0.02 & 8\% & QU Vul \\
Solar & 0.70 & 0.28 & 0.0 & 0.0 & 0.02 & 0\% &  
\enddata
\tablenotetext{a}{Carbon, nitrogen, oxygen, and neon are also
included in $Z=0.02$ with the same ratio as the solar composition
\citep{gre89}.}
\tablenotetext{b}{Mixing between the helium layer + core material
and the accreted matter with solar composition, which is calculated from
$\eta_{\rm mix}=(0.7/X)-1$.}
\tablenotetext{c}{Free-free light curves for this chemical composition
are tabulated in Table 2 of \citet{hac10k}.}
\tablenotetext{d}{Free-free light curves for this chemical composition
are tabulated in Table 3 of \citet{hac10k}.}
\end{deluxetable*}


\begin{figure*}
\epsscale{0.75}
\plotone{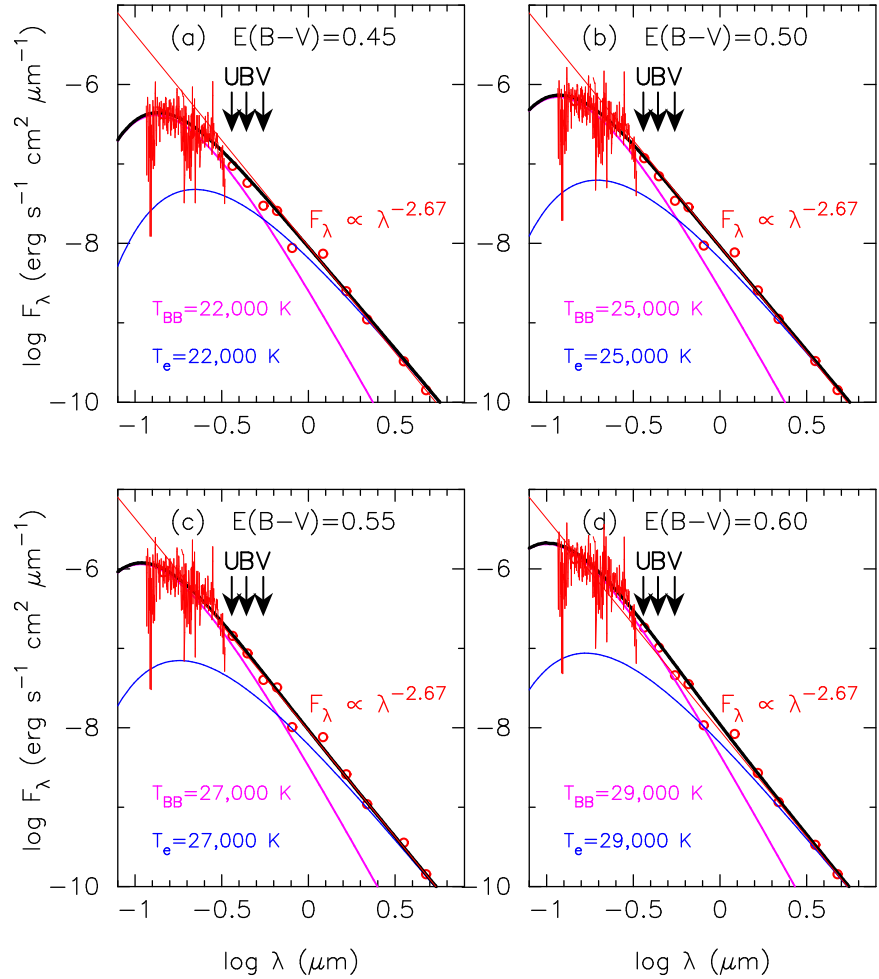}
\caption{
Dereddened spectrum of PW~Vul, 64 days after the outburst
(UT 1984 Sep 30 $=$ JD~2\,445\,973.5) for different
extinctions: (a) $E(B-V)=0.45$, (b) $E(B-V)=0.50$, 
(c) $E(B-V)=0.55$, and (d) $E(B-V)=0.60$.  Red solid line:
{\it IUE} spectra, SWP24088 and LWP04458, are taken from the INES data
archive server.  Open red
circles: $UBVRI$ data from \citet{rob95}
and $JHKLM$ data from \citet{geh88}.
Global features of spectrum can be fitted with a combination
(thick black solid line) of the blackbody with a temperature
of (a) $T_{\rm BB}=22,000$~K, (b) $T_{\rm BB}=25,000$~K, 
(c) $T_{\rm BB}=27,000$~K, and (d) $T_{\rm BB}=29,000$~K
(thick magenta line) and the optically-thick,
free-free emission with the same electron
temperature of $T_{\rm e}$ (thick blue line).
We also add a straight red thin solid line
of $F_\lambda \propto \lambda^{-2.67}$, which corresponds to 
Equations (\ref{free_free_thick_eq3_infity})
and (\ref{free_free_thick_eq2_simple}),
as the limiting case of $T_{\rm e}=\infty$ for
optically-thick free-free emission.
\label{sed_pw_vul_iue_ubvrijhk_free_free_bb_4fig}}
\end{figure*}

\subsection{Chemical composition of ejecta}
\label{chemical_composition_pw_vul}

     One of the most intriguing properties of classical novae is the
metal-enrichment of ejecta  \citep[e.g.,][]{geh98}, which is ascribed 
to  mixing with WD core material during outburst \citep[e.g.,][]{pri95}.
PW~Vul is not an exception of this general trend, as summarized
in Table \ref{pw_vul_chemical_abundance}.  There is a noticeable scatter
in the abundance estimates, from $X=0.47$ to $X=0.69$.
The arithmetic mean is $X=0.57$, $Y=0.22$, and $X_{\rm CNO}=0.19$
for $Z=0.02$.  Here, $X$, $Y$, $Z$, and $X_{\rm CNO}$ are
hydrogen, helium, heavy elements with solar abundance, and
carbon-nitrogen-oxygen fractions in weight, respectively.
 
     We took a simple parameterization for the degree of mixing
between core material and accreted matter as $\eta_{\rm mix}=(0.7/X)-1$.
Here we assume the solar composition for the accreted matter.
Table \ref{chemical_composition} shows seven representative cases
of degree of mixing, that is,
100\% (denoted by ``CO nova 1'', ``CO nova 2'', and ``Ne nova 1''), 
55\% (``CO nova 3''), 25\% (``CO nova 4'' and ``Ne nova 2''),
and 8\% (``Ne nova 3'').  We first adopt ``CO nova 4,'' because
it is closest to the above averaged values of PW~Vul.
Then, we discuss the dependence of light curves on the chemical composition.

\subsection{Contribution from photospheric emission}
\label{contribution_of_blackbody}

Now we analyze the spectra of PW~Vul assuming that the continuum flux
$F_\nu$ is simply the sum of a blackbody spectrum
of the temperature $T_{\rm ph}$ and an optically thick free-free emission
with the electron temperature of $T_{\rm e}$, i.e., 
\begin{equation}
F_{\nu} = f_1 B_{\nu}(T_{\rm ph}) + f_2 S_{\nu}(T_{\rm e}),
\label{composite_free_free_blackbody}
\end{equation}
where $\nu$ is the frequency,
$B_{\nu}(T_{\rm ph})$ is the Planckian of the photospheric
temperature $T_{\rm ph} = T_{\rm BB}$, and $S_{\nu}(T_{\rm e})$ is
the free-free spectrum of the electron temperature $T_{\rm e}$,
$f_1$ and $f_2$ are numeric constants \citep[e.g.,][]{nis08}.
Following \citet{wri75}, the free-free spectrum can be expressed as
\begin{equation}
S_{\nu}(T_{\rm e}) = B_{\nu}(T_{\rm e}) K^{2/3}_{\nu}(T_{\rm e}),
\label{free_free_thick_eq1}
\end{equation}
where the linear free-free absorption coefficient $K_{\nu}(T_{\rm e})$
is given by
\begin{equation}
K_{\nu}(T_{\rm e}) = 3.7 \times 10^8 \left[ 1- 
\exp\left(-{{h \nu} \over {k T_{\rm e}}}\right)
\right] Z^2 g_{\nu}(T_{\rm e}) T^{-1/2}_{\rm e} \nu^{-3}
\end{equation}
in cgs units; $g_\nu(T_{\rm e})$ is the Gaunt factor.
In general, the Gaunt factor depends weakly on the frequency
and temperature, but we assume it to be unity following \citet{hac14k}.  
So there are four fitting parameters,
i.e., $f_1$, $f_2$, $T_{\rm ph}= T_{\rm BB}$, and $T_{\rm e}$. 
When $h\nu \ll k T_e$,
Equation (\ref{free_free_thick_eq1}) can be expressed as \citep{wri75}
\begin{equation}
S_\nu = 23.2 \left( {{\dot M_{\rm wind}} \over {\mu v_{\infty}}} \right)^{4/3}
{{\nu^{2/3}} \over {D^2}} \gamma^{2/3} g^{2/3} Z^{4/3} {~\rm Jy},
\label{free_free_thick_eq3_infity}
\end{equation}
where 1 Jy $=10^{-26}$~W~m$^{-2}$~Hz$^{-1}$;
the ion number density is assumed equal to 
the total gas number density $n$ and the electron number density 
is equal to $\gamma$ times the ion number density;
$\dot M_{\rm wind}$ is the wind mass-loss rate in units of
$M_\sun$~yr$^{-1}$; $D$ is the distance in units of kpc; $v_\infty$ is
the terminal wind velocity in units of km~s$^{-1}$; $\mu$ is the mean
molecular weight; $Z$ is the charged-degree of ion (only in this formula);
$\nu$ is the frequency in units of Hz; and $g$ is the Gaunt factor.
Equation (\ref{free_free_thick_eq3_infity}) can be further simplified as
\begin{equation}
S_\nu \propto \nu^{2/3}, {\rm ~or~} S_\lambda \propto \lambda^{-8/3},
\label{free_free_thick_eq2_simple}
\end{equation}
where $\lambda$ is the wavelength.

Figure \ref{sed_uv_opt_ir_ijhk_filter_no2} schematically shows
the contributions of $B_\lambda$ and $S_\lambda$.
When the wind mass loss rate is small, contribution of free-free emission
is relatively small in Equation (\ref{composite_free_free_blackbody}).
\citet{hac14k} decomposited the spectrum of PW~Vul about 64 days
after the outburst using Equation (\ref{composite_free_free_blackbody})
and concluded that the free-free flux is comparable to the pseudo-photospheric
flux in $V$ band.  Here, we reanalyzed the same data and show them
in Figure \ref{sed_pw_vul_iue_ubvrijhk_free_free_bb_4fig},
assuming four different extinctions, i.e., (a) $E(B-V)=0.45$,
(b) $E(B-V)=0.50$, (c) $E(B-V)=0.55$, and (d) $E(B-V)=0.60$.
In our decomposition process, we simply assume
that $T_{\rm BB}=T_e$ and change the temperature in steps of 1000~K.
We see that the blackbody emission gives a good approximation
to the UV region and the free-free emission is a good fit to the IR region.
In the region between them, we have a comparable contribution
from the blackbody and free-free components.

These four decompositions of different sets of $T_{\rm e}=T_{\rm BB}$
and $E(B-V)$ more or less show similar good agreement.
Among these four cases, $T_{\rm BB}=27,000$~K in Figure 
\ref{sed_pw_vul_iue_ubvrijhk_free_free_bb_4fig}(c)
is in best agreement with the temperature deduced from the light curve
analysis in Section \ref{calibration_co4_pw_vul}.
It should be noted that the dereddened spectrum with $E(B-V)=0.55$
is closest to the straight line (red thin solid line) of 
$F_\lambda\propto\lambda^{-2.67}$ of Equation 
(\ref{free_free_thick_eq2_simple}). 
\citet{hau97} calculated synthetic NLTE nova spectra (see their
Figure 10), in which the continuum flux has a slope of
$F_\lambda\propto\lambda^{-2.7}$ in the range of $\lambda=0.2$--2~$\mu$m
for $T_{\rm eff}=25,000$~K and $T_{\rm eff}=30,000$~K.
If we apply this slope directly to PW~Vul, the spectrum is in best
agreement with the reddening of $E(B-V)=0.55$ in Figure
\ref{sed_pw_vul_iue_ubvrijhk_free_free_bb_4fig}(c).

The decomposition in Figure \ref{sed_pw_vul_iue_ubvrijhk_free_free_bb_4fig}(c)
indicates that the photospheric emission may substantially 
contribute in the optical light curve.  In the next subsection, we
calculate theoretical light curves, from the sum of free-free plus
photospheric emission for the optical bands, and from only
photospheric emission for the UV~1455\AA\  band.


\begin{figure}
\epsscale{1.15}
\plotone{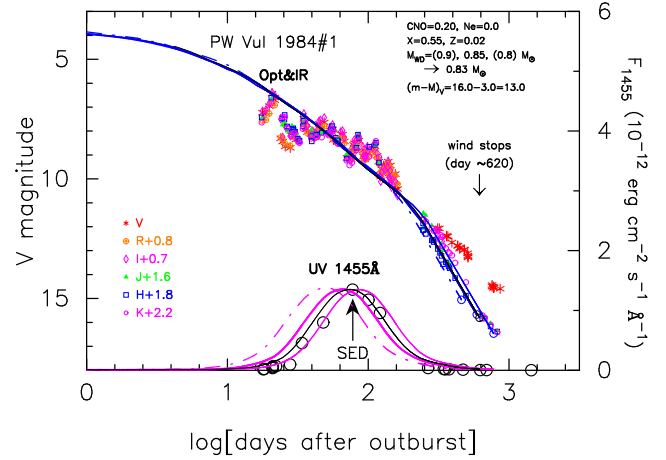}
\caption{
Optical, NIR, and UV 1455 \AA\  light curves of PW~Vul.
Here, we assume the start of the day ($t=0$) as JD 2445897.0.
Red asterisks represent $V$ magnitudes, which were taken
from IAU Circular Nos. 3971 and 4091, and from \citet{rob95} 
and the AAVSO archive.  Orange open circles with plus sign inside
denote $R$ magnitudes shifted down
by 0.8 mag, taken from \citet{rob95}.  Magenta open diamonds are 
$I$ magnitudes shifted down by 0.7 mag, taken from \citet{eva90}, 
\citet{geh88}, \citet{rob95}, and \citet{wil91}.
Green filled triangles are $J$ magnitudes shifted down by 1.6 mag,
taken from \citet{eva90}, \citet{geh88}, and \citet{wil96}.
Blue open squares are $H$ magnitudes shifted down by 1.8 mag,
taken from \citet{eva90}, \citet{geh88}, \citet{rob95}, and \citet{wil96}.
Magenta open circles are $K$ magnitudes shifted down by 2.2 mag, taken
from \citet{eva90}, \citet{geh88}, and \citet{wil96}.
Black large open circles are the UV 1455\AA\  fluxes observed with {\it IUE}, 
taken from \citet{cas02}.  We plot three different WD mass models, 
$0.80~M_\sun$ (blue/magenta solid lines), $0.85~M_\sun$ (blue/magenta
thick solid lines), and  $0.90~M_\sun$ (blue/magenta dash-dotted lines),
for the envelope chemical composition of ``CO nova 4.''
We also added a fine mass model of $0.83~M_\sun$ (black thin solid lines).
Optically thick winds stopped about 620 days after the outburst
for the case of $0.83~M_\sun$ WD.  The spectrum in Figure
\ref{sed_pw_vul_iue_ubvrijhk_free_free_bb_4fig} was secured at
the day denoted by an arrow labeled ``SED.''  
\label{all_mass_pw_vul_x55z02c10o10_vrijhk_calib_no2}}
\end{figure}


\begin{figure}
\epsscale{0.95}
\plotone{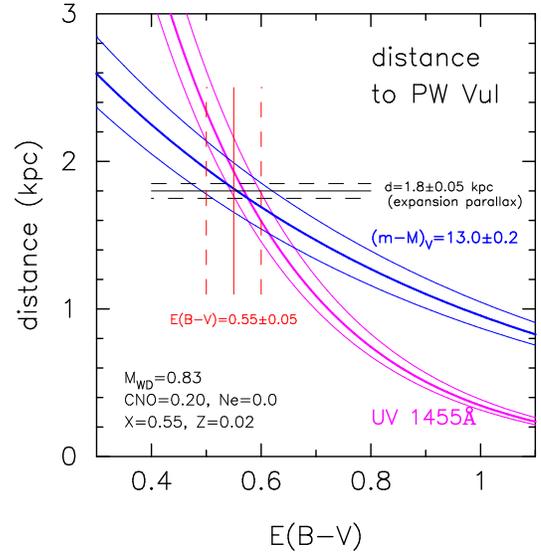}
\caption{
Same as Figure \ref{distance_reddening_pw_vul_only_no2}, but for
the $0.83~M_\sun$ WD model with the chemical abundance of ``CO nova 4.''
Blue solid line with flanking thin solid lines: distance-reddening
relation calculated from Equation (\ref{distance_modulus_eq_pw_vul_no1}),
labeled ``$(m-M)_V=13.0\pm0.2$.''  Magenta solid line with flanking
thin solid lines: distance-reddening relation
calculated from Equation (\ref{pw_vul_uv1455_fit_co4}), i.e.,
the UV 1455\AA\   flux fitting (labeled ``UV 1455 \AA\  '') of the
the $0.83~M_\sun$ WD model.
\label{pw_vul_distance_reddening_x55z02c10o10}}
\end{figure}


\begin{figure}
\epsscale{1.15}
\plotone{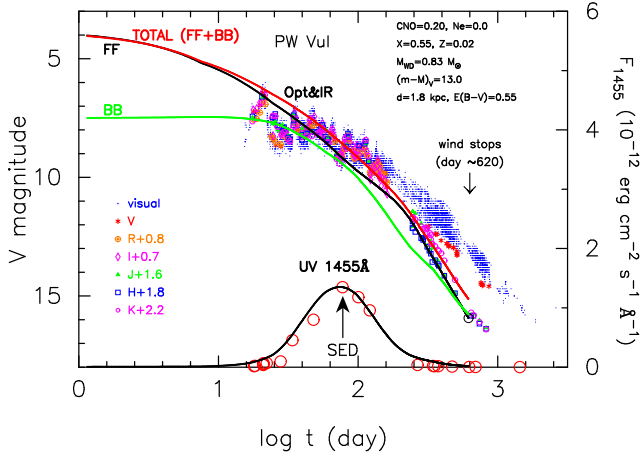}
\caption{
Same as Figure \ref{all_mass_pw_vul_x55z02c10o10_vrijhk_calib_no2}, but
we added blackbody $V$ magnitude (green solid line labeled ``BB'')
and total $V$ magnitude (red solid line labeled ``TOTAL'') for
the distance modulus of $(m-M)_V=13.0$ in $V$ band.
We determine the absolute magnitude of free-free
emission light curve (black solid line labeled ``FF'')
by $M_{\rm w}=3.0$, i.e., $m_{\rm w}= 3.0 + 13.0=16.0$,
at the end of wind phase
(denoted by a black open circle) from fitting of our model light curve
of total $V$ (red solid line) with the $V$ observation (red asterisks).
We also added visual magnitudes (blue dots), which are taken from
the AAVSO archive.  Our model light curve reasonably fits with the early
$V$ light curve but deviates from the visual observation in the later 
phase, that is, in the nebular phase.  
This deviation is due to strong emission lines
such as [\ion{O}{3}], which are not included in our model light curves.
The spectrum in Figure
\ref{sed_pw_vul_iue_ubvrijhk_free_free_bb_4fig} was secured at
the day denoted by an arrow labeled ``SED.''
Our $0.83~M_\sun$ WD model shows the photospheric temperature of
about $T_{\rm ph}=28,000$~K on this day, being
consistent with the fitting temperature of $T_{\rm BB}=27,000$~K
in Figure \ref{sed_pw_vul_iue_ubvrijhk_free_free_bb_4fig}.
See text for more details.
\label{all_mass_pw_vul_x55z02c10o10_vrijhk_composite}}
\end{figure}

\subsection{Model Light Curves of ``CO Nova 4''}
\label{calibration_co4_pw_vul}

Now we make light curve models for PW~Vul assuming
the chemical composition of ``CO Nova 4.''
We calculated nova light curves for various WD masses
and fitted them to the observational data. 
The flux in the UV~1455\AA\  band \citep[a narrow band of
1445--1465\AA, see][]{cas02} was calculated as blackbody 
emission  from the nova pseudophotosphere ($R_{\rm ph}$ and $T_{\rm ph}$)
using the optically-thick wind solutions in \citet{kat94h}.
In our fitting process, we changed the WD mass from
$0.80~M_\sun$ to $0.90~M_\sun$ in steps of $0.01~M_\sun$.
In Figure \ref{all_mass_pw_vul_x55z02c10o10_vrijhk_calib_no2}, 
the $0.83~M_\sun$ WD model (black thin solid lines) shows reasonable
agreement with the optical, NIR, and UV observations,
in particular with the UV~1455\AA\  observations.  An arrow labeled 
``SED'' indicates the date (Day 64) at which the spectrum in
Figure \ref{sed_pw_vul_iue_ubvrijhk_free_free_bb_4fig} was secured.
Our $0.83~M_\sun$ WD model has $T_{\rm ph}=28,000$~K on this day, which
is consistent with the blackbody temperature of $T_{\rm BB}=27,000$~K
determined in Section \ref{contribution_of_blackbody}.
 
From the UV 1455\AA\   light curve fitting of the $0.83~M_\sun$ WD in Figure
\ref{all_mass_pw_vul_x55z02c10o10_vrijhk_calib_no2},
we obtained the following distance-reddening relation, i.e.,
\begin{eqnarray}
& & 2.5 \log F_{\lambda 1455}^{\rm mod} 
- 2.5 \log F_{\lambda 1455}^{\rm obs} \cr\cr
&=& 2.5 \log(3.38 \times 10^{-12})  
- 2.5 \log((1.35\pm0.27) \times 10^{-12}) \cr\cr
&=&  5 \log \left({d \over {10\mbox{~kpc}}} \right)  + 8.3 \times E(B-V),
\label{pw_vul_uv1455_fit_co4}
\end{eqnarray}
where $F_{\lambda 1455}^{\rm mod}= 3.38 \times 
10^{-12}$~erg~cm$^{-2}$~s$^{-1}$~\AA$^{-1}$ is the calculated 
UV~1455\AA\  band flux at maximum of the $0.83~M_\sun$ WD model for an
assumed distance of 10~kpc and $F_{\lambda 1455}^{\rm obs}=
(1.35 \pm0.27) \times 10^{-12}$~erg~cm$^{-2}$~s$^{-1}$~\AA$^{-1}$ is
the maximum observed flux \citep{cas02}.
Here we assume an absorption of
$A_\lambda= 8.3 \times E(B-V)$ at $\lambda= 1455$\AA\   \citep{sea79}.
The distance-reddening relation of Equation
(\ref{pw_vul_uv1455_fit_co4}) is plotted by magenta solid lines
in Figure \ref{pw_vul_distance_reddening_x55z02c10o10}
(labeled ``UV~1455\AA'').  Two relations of Equations 
(\ref{distance_modulus_eq_pw_vul_no1}) and (\ref{pw_vul_uv1455_fit_co4})
cross each other at the point of
$\left(E(B-V), d\right)= (0.57\pm0.05~{\rm mag}, ~1.75\pm0.3~{\rm kpc})$,
being consistent with the observations summarized in Section
\ref{distance_reddening_estimate_pw_vul}.
Therefore, we safely conclude that the WD mass of PW~Vul is 
as massive as $\sim0.83~M_\sun$ if the chemical composition is
close to $X=0.55$, $Y=0.23$, $Z=0.02$, and $X_{\rm CNO}=0.20$.

Contrary to the UV~1455\AA\  blackbody flux, our free-free 
model light curves are not yet calibrated.
Optical light curves in Figure
\ref{all_mass_pw_vul_x55z02c10o10_vrijhk_calib_no2}
are freely shifted in vertical direction to fit the observation
because the proportionality constant in Equation (9) of
\citet{hac06kb} is unknown for ``CO nova 4.''
To fix the absolute magnitude of each light curve, we use
the absolute magnitude of PW~Vul as follows. 

First, we calculate the light curve model of the $0.83~M_\sun$ WD
and obtain the blackbody light curve in $V$ band
(green solid line labeled ``BB'')
as shown in Figure \ref{all_mass_pw_vul_x55z02c10o10_vrijhk_composite}.
Here, we adopt the distance modulus of $(m-M)_V=13.0$.
Assuming a trial value for the proportionality constant $C$ in Equation 
(\ref{free-free_calculation_original}) of Appendix 
\ref{free_free_model_light_curve}, which is the same as
Equation (9) of \citet{hac06kb}, we obtain the absolute 
magnitude of the free-free model light curve (black solid line
labeled ``FF'').  The total flux (red solid line labeled ``TOTAL'')
is the sum of these two fluxes.  However, in general, this total 
$V$ magnitude light curve does not fit well with the observed data.
Then, we change the proportionality constant until the total $V$ flux
fits well with the observed $V$ light curve.
Figure \ref{all_mass_pw_vul_x55z02c10o10_vrijhk_composite}
shows our final best fit model.  We directly read $m_{\rm w}=16.0$
from Figure \ref{all_mass_pw_vul_x55z02c10o10_vrijhk_composite},
where $m_{\rm w}$ is the apparent magnitude at the end of wind phase
(open circle at the end of black solid line labeled ``FF'').
Then, we obtain $M_{\rm w}= m_{\rm w} - (m-M)_V= 16.0 - 13.0=3.0$,
where $M_{\rm w}$ is the absolute magnitude of free-free model
light curve at the end of wind phase.
Thus, the proportionality constant can be specified by $M_{\rm w}=3.0$
of the $0.83~M_\sun$ WD for PW~Vul.

Based on the result of the $0.83~M_\sun$ WD and applying 
the time-scaling law of free-free light curves to other WD mass models,
we obtain the absolute magnitudes of free-free light curves for other
WD masses with the chemical composition of ``CO nova 4'' 
(see Appendix \ref{absolute_magnitudes}).  The absolute magnitudes
are specified by the value of $M_{\rm w}$ and listed
in Table \ref{light_curves_of_novae_co4} 
for $0.55$--$1.2~M_\sun$ WDs in steps of $0.05~M_\sun$.

It should be noted that our model light curve reasonably fits with
the early $V$ light curve but deviates from the visual observation
(small blue dots in Figure 
\ref{all_mass_pw_vul_x55z02c10o10_vrijhk_composite})
in the nebular phase.
On the other hand, our free-free model light curve
almost perfectly fits with the NIR light curves even in the later phase.
This deviation in visual magnitudes is owing to strong emission lines
such as [\ion{O}{3}], which are not included in our model
\citep[see][for details]{hac06kb}.  In the NIR region,
free-free emission dominates the spectrum and our free-free light curve 
works well.

Thus, we may conclude that the effect of photospheric emission in $V$ band
can be neglected in novae much faster than PW~Vul because
the mass-loss rate is large enough for free-free emission to dominate
the spectrum in the optical and NIR region.  We will discuss
such examples of fast novae in Sections \ref{discussion_gk_per} and 
\ref{discussion_v603_aql}.
However, in less massive WDs or in novae much slower than PW~Vul,
we must take into account the contribution of photospheric emission.
In this sense, PW~Vul is a critical one between them.


\begin{figure}
\epsscale{1.15}
\plotone{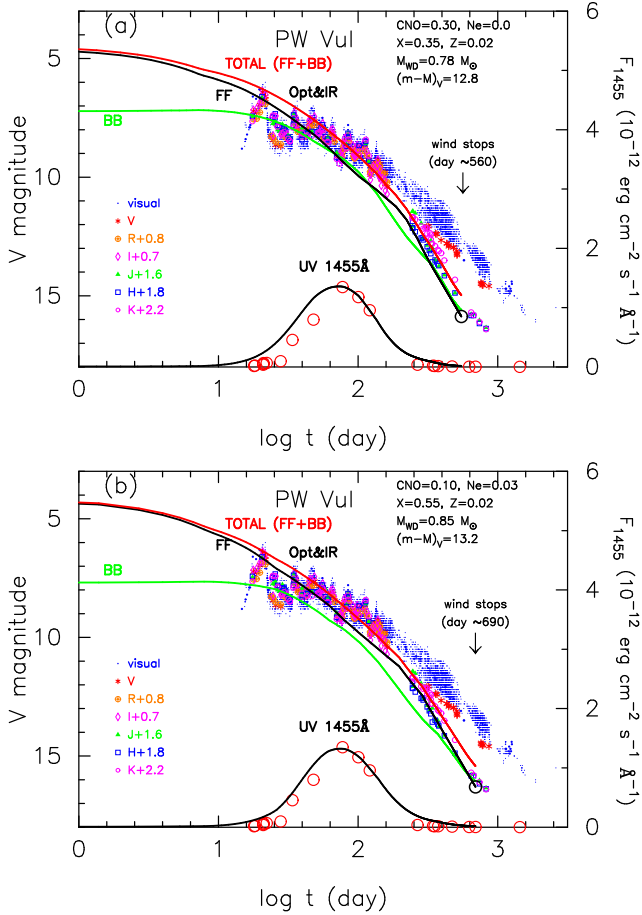}
\caption{
Same as Figure \ref{all_mass_pw_vul_x55z02c10o10_vrijhk_composite},
but for (a) the $0.78~M_\sun$ WD model with the ``CO nova 2'' chemical 
composition and (b) the $0.85~M_\sun$ WD model with the ``Ne nova 2''
chemical composition.  Optically thick winds stopped (a) about 560 days
and (b) about 690 days after the outburst.
We obtained (a) $(m-M)_V=12.8$ and (b) $(m-M)_V=13.2$ 
from the total $V$ flux fitting with the $V$ observation.
\label{all_mass_pw_vul_x35z02c10o20_x55z02o10ne03_vrijhk_composite}}
\end{figure}


\begin{figure}
\epsscale{0.95}
\plotone{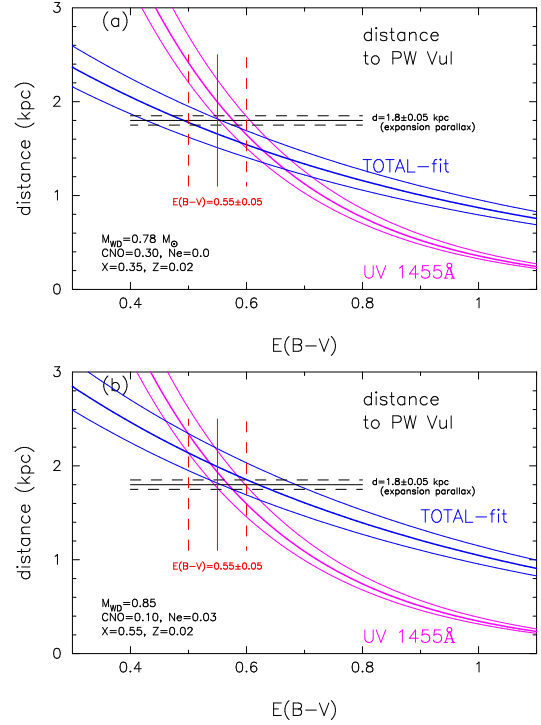}
\caption{
Distance-reddening relation toward PW~Vul for the model of
(a) $0.78~M_\sun$ WD with the chemical composition of 
``CO nova 2'' and (b) $0.85~M_\sun$ WD with the chemical
composition of ``Ne nova 2.''
Blue solid line with flanking thin solid lines:
distance-reddening relation calculated from the fitting
of total (free-free emission plus photospheric emission) $V$ 
model light curve (labeled ``TOTAL-fit'') obtained from Figure 
\ref{all_mass_pw_vul_x35z02c10o20_x55z02o10ne03_vrijhk_composite}.
Magenta solid line with flanking thin solid lines:
distance-reddening relation
calculated from the UV 1455\AA\   flux fitting
(labeled ``UV 1455 \AA\  '') obtained from Figure 
\ref{all_mass_pw_vul_x35z02c10o20_x55z02o10ne03_vrijhk_composite}.
Two other constraints are also plotted; one is the distance of 
$d=1.8\pm0.05$~kpc determined by \citet{dow00} 
with nebular expansion parallaxes and
the other is the reddening of $E(B-V)=0.55\pm0.05$
determined from various constrains described in Section 
\ref{distance_reddening_estimate_pw_vul}.
\label{pw_vul_distance_reddening_x35z02c10o20_x55z02o10ne03}}
\end{figure}

\subsection{Effect of chemical composition}
\label{model_light_curve_pw_vul}

The chemical composition of ejecta is usually not so accurately constrained
as described in Section \ref{chemical_composition_pw_vul} and
as tabulated in Table \ref{pw_vul_chemical_abundance}. 
If we adopt a chemical composition different from the true one,
we could miss the WD mass and distance modulus of a nova.
Therefore, we examine the dependence of our model light curve
on the chemical composition, i.e., the degree of mixing.
We adopted two other chemical compositions of ``CO nova 2,'' 
a high degree of mixing, $\eta_{\rm mix}=1.0$ (100\%) and
``Ne nova 2,'' a low degree of mixing, $\eta_{\rm mix}=0.25$ (25\%),
mainly because their absolute magnitudes of free-free emission
model light curves were already calibrated by \citet{hac10k}
independently of the light curve of PW~Vul.

In a similar way to that in Section \ref{calibration_co4_pw_vul},
we have obtained best fit models for the two chemical compositions
mentioned above.  
Figure \ref{all_mass_pw_vul_x35z02c10o20_x55z02o10ne03_vrijhk_composite}
shows our model light curves for (a) ``CO Nova 2''
and (b) ``Ne Nova 2.''  In our fitting process, we changed the WD mass from
$0.70~M_\sun$ to $0.90~M_\sun$ in steps of $0.01~M_\sun$.  In Figure 
\ref{all_mass_pw_vul_x35z02c10o20_x55z02o10ne03_vrijhk_composite}(a),
the $0.78~M_\sun$ WD model shows reasonable
agreement with the $V$ and UV observations.
Also a good agreement is found for the $0.85~M_\sun$ WD in Figure 
\ref{all_mass_pw_vul_x35z02c10o20_x55z02o10ne03_vrijhk_composite}(b).
It should be noted again that our model light curve fits well 
with the early $V$ light curve but deviates from the visual observation
in the later phase, that is, in the nebular phase.  This deviation is
due to strong emission lines such as [\ion{O}{3}], 
which are not included in our model light curves.

The flux in UV~1455\AA\  band was calculated as blackbody 
emission  from the nova pseudophotosphere using the optically-thick
wind solutions in \citet{kat94h} and \citet{hac06kb, hac10k}.
From the UV 1455\AA\   flux fitting, we obtained the distance-reddening
relation, which is plotted by magenta solid lines (labeled ``UV 1455\AA'')
in Figure \ref{pw_vul_distance_reddening_x35z02c10o20_x55z02o10ne03}.
Both the distance-reddening relations of UV 1455\AA\  
in Figure \ref{pw_vul_distance_reddening_x35z02c10o20_x55z02o10ne03}
are similar to that for ``CO nova 4'' in Figure 
\ref{pw_vul_distance_reddening_x55z02c10o10}.

The total $V$ fluxes are calculated from the sum of free-free and
blackbody fluxes.   From the fitting, we also obtained
the distance-reddening relation of (a) $(m-M)_V=12.8$ for the 
$0.78~M_\sun$ WD and (b) $(m-M)_V=13.2$  for the $0.85~M_\sun$ WD.
The both are barely consistent with our adopted value of
$(m-M)_V=13.0\pm0.2$ for PW~Vul.  The two distance-reddening relations
of total $V$ magnitude fitting are plotted by blue solid lines 
(labeled ``TOTAL'') in Figure 
\ref{pw_vul_distance_reddening_x35z02c10o20_x55z02o10ne03}.
It is remarkable that our fitting misses the distance modulus
only by 0.2 mag even if we assume a different chemical composition
of $\Delta X=0.55 - 0.35=0.2$.

For a lower value of the hydrogen content $X$,
the evolution timescale becomes shorter
even if the WD mass is the same \citep[see][]{kat94h, hac01kb}.
Therefore, a less massive WD of $0.78~M_\sun$ is fitted with
the observation for a lower value of $X=0.35$ as shown in Figure 
\ref{all_mass_pw_vul_x35z02c10o20_x55z02o10ne03_vrijhk_composite}(a).
The wind mass loss rate is smaller for a less
massive WD of $0.78~M_\sun$.  The lower wind mass loss rate leads to
fainter free-free emission and, as a result, a fainter
total $V$ light curve.  This is the reason for $(m-M)_V=12.8$,
which is a bit smaller than the original value of $(m-M)_V=13.0$.

For the chemical composition of ``Ne nova 2,'' however,
the hydrogen content of $X$ is the same as that of ``CO nova 4.''
The difference is between $X_{\rm CNO}=0.20$ and
$X_{\rm CNO}=0.10$.  The CNO abundance is relevant to the nuclear
burning rate and a lower value of $X_{\rm CNO}=0.10$ makes 
the evolution timescale longer.  This requires a more massive
WD of $0.85~M_\sun$ than the $0.83~M_\sun$ WD of $X_{\rm CNO}=0.20$
 as shown in Figure 
\ref{all_mass_pw_vul_x35z02c10o20_x55z02o10ne03_vrijhk_composite}(b).
The more massive WD blows stronger winds.  This results in a brighter
light curve of free-free emission and, as a result, a brighter
total $V$ light curve.  This is the reason for $(m-M)_V=13.2$,
which is a bit larger than the original value of $(m-M)_V=13.0$.

In Figure 
\ref{pw_vul_distance_reddening_x35z02c10o20_x55z02o10ne03}(a),
the two distance-reddening relations, i.e.,
``UV~1455\AA'' and ``TOTAL-fit,''  cross each other at the point of
$\left(E(B-V), d\right)= (0.63~{\rm mag}, ~1.5~{\rm kpc})$,
being not consistent with $E(B-V)=0.55\pm0.05$ and $d=1.8\pm0.05$~kpc.
The degree of mixing may not be as high as 100\% ($X\approx0.35$)
in PW~Vul.

Figure \ref{pw_vul_distance_reddening_x35z02c10o20_x55z02o10ne03}(b)
shows that the two relations cross at the point of
$\left(E(B-V), d\right)= (0.54~{\rm mag}$, $2.0~{\rm kpc})$.
This value is consistent with the reddening estimate
of $E(B-V) = 0.55 \pm 0.05$ although the distance estimate is 
a bit larger than the distance estimate of $d=1.8 \pm 0.05$~kpc.
We may conclude that the lower degree of mixing (25\% mixing)
is more reasonable in PW~Vul.

To summarize, we reached a reasonable distance-reddening 
result for a 25\% mixing of ``Ne Nova 2'' but not for a 100\% mixing
of ``CO Nova 2.'' 
Note that enrichment of neon with hydrogen mass fraction being unchanged
hardly influences the nova light curves
because neon is not relevant either to nuclear burning (CNO-cycle)
or to opacity \citep[e.g.,][]{kat94h,hac06kb,hac10k}.
Therefore, the agreement in the lower 25\% mixing model suggests
that a lower degree of mixing ($\sim 25$\%) is reasonable rather
than a higher degree of mixing ($\sim 100$\%).
Unfortunately, there is significant scatter in the abundance 
determinations (see Table \ref{pw_vul_chemical_abundance}), but
their averaged values of chemical composition,  which show a 23\% mixing,
are close enough to those of ``CO nova 4.''
Therefore, we may conclude that, through our
method of model light curve fitting, one might discriminate between
different degrees of mixing, at least, in terms of the hydrogen mass 
fraction $X$.
We summarize our fitting result for PW~Pul in Table \ref{physical_parameters}.


\begin{figure}
\epsscale{1.15}
\plotone{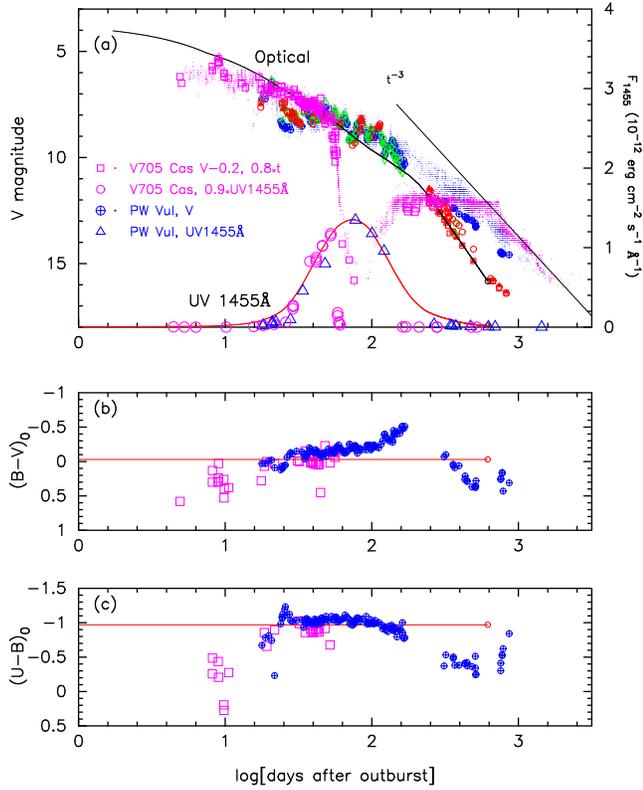}
\caption{
Same as Figure
\ref{pw_vul_v1668_cyg_v1500_cyg_v1974_cyg_v_bv_ub_color_curve_logscale_no3},
but for V705~Cas (magenta open squares) and
PW~Vul (blue open circles with plus sign inside).
The $UBV$ data of V705~Cas are taken from 
\citet{mun94b}, \citet{hri98}, and IAU Circular Nos. 5920 and 5929.  
Visual magnitudes (magenta/blue small dots) are taken from the AAVSO archive.
We also added $IJHK$ light curves of PW~Vul (small green and red symbols),
the data of which are the same as those in Figure
\ref{all_mass_pw_vul_x55z02c10o10_vrijhk_calib_no2}.  Panel (a) 
also shows the UV1455\AA\  light curves of V705~Cas (magenta open 
circles) and PW~Vul (blue open triangles).
In order to overlap the optical light curve of V705~Cas with that of
PW~Vul, we squeeze the light curve of V705~Cas by a factor of 0.8 in the
direction of time and shifted the $V$ magnitudes of V705~Cas by $-0.2$
mag up as indicated in the figure.  Model light curves of a 
$0.83~M_\sun$ WD are added: black solid line is the free-free $V$
and red solid line is the blackbody UV~1455\AA\ 
(same as those in Figure
\ref{all_mass_pw_vul_x55z02c10o10_vrijhk_composite}).
\label{v705_cas_pw_vul_v_bv_ub_color_curve_logscale}}
\end{figure}


\begin{figure}
\epsscale{1.15}
\plotone{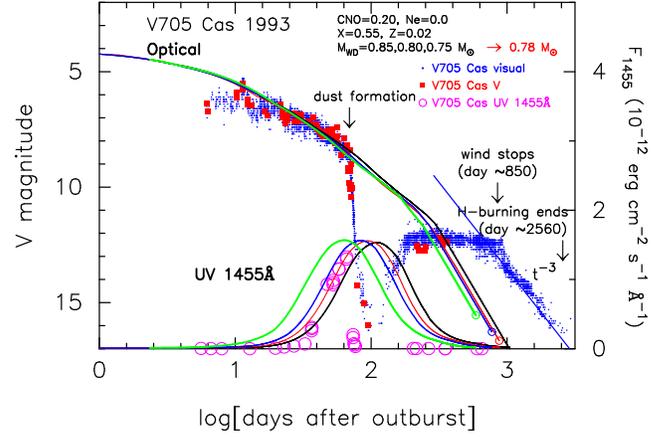}
\caption{
Optical and UV 1455 \AA\  light curves of V705~Cas 1993.
The observed $V$ magnitudes (filled red squares) are taken from
\citet{mun94b}, \citet{hri98}, and IAU Circular Nos. 5920 and 5929. 
The visual magnitudes (small blue dots) are taken from the AAVSO archive.
The UV~1455\AA\  flux (magenta open circles) are taken from \citet{cas02}.
We plot three different WD mass models: $0.75~M_\sun$ (black solid lines), 
$0.80~M_\sun$ (blue solid lines), and  $0.85 ~M_\sun$ (green solid lines)
for the envelope chemical composition of ``CO nova 4.''
We add a $0.78~M_\sun$ WD model (red thin solid lines), which shows
better fit to the UV~1455\AA\  observation.
\label{all_mass_v705_cas_x55z02c10o10}}
\end{figure}


\begin{figure}
\epsscale{0.95}
\plotone{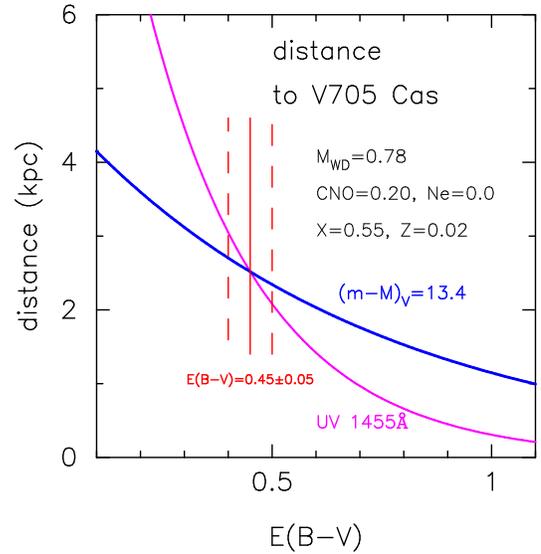}
\caption{
Distance-reddening relation for V705~Cas.  We plot three
relations, i.e., the color excess estimate of $E(B-V)=0.45\pm0.05$,
distance modulus in $V$ band, and distance modulus in UV~1455\AA\  band,
where we adopt the WD mass of $M_{\rm WD}=0.78~M_\sun$ with the
elemental abundance of ``CO nova 4.''
Three trends almost cross at the point of $E(B-V)\approx0.45$ and
$d\approx2.5$~kpc. 
\label{v705_cas_distance_reddening_x55z02c10o10}}
\end{figure}


\begin{figure}
\epsscale{1.15}
\plotone{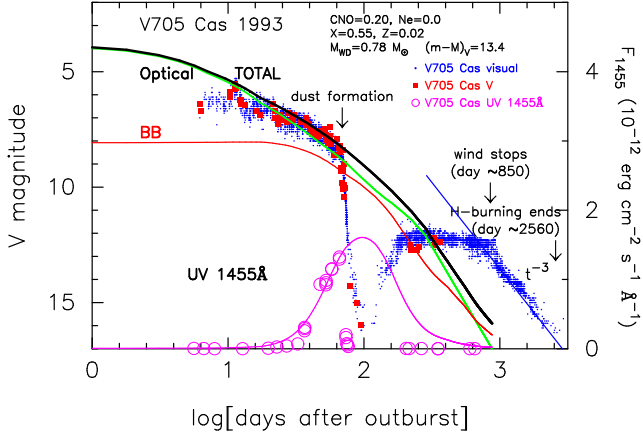}
\caption{
Same as Figure \ref{all_mass_v705_cas_x55z02c10o10}, but
we added the blackbody flux in $V$ band (red solid line labeled ``BB'') 
and the total flux (thick black solid line labeled ``TOTAL'') 
of free-free (thick green solid line) plus blackbody in $V$ band
for the $0.78~M_\sun$ WD model.  Here, we assumed
the distance modulus of $(m-M)_V=13.4$.  
\label{all_mass_v705_cas_x55z02c10o10_composite}}
\end{figure}

\section{V705~Cas 1993}
\label{v705_cas}
Next, we analyze V705~Cas that shows a similar optical decline rate 
to PW~Vul until a deep dust blackout started (see Figures
\ref{light_curve_rr_pic_v723_cas_hr_del_v5558_sgr_v705_cas_pw_vul}
and \ref{v705_cas_pw_vul_v_bv_ub_color_curve_logscale}). 
The chemical composition is also similar to PW~Vul,
as listed in Table \ref{pw_vul_chemical_abundance}.
V705~Cas was discovered by Kanatsu on UT 1993 December 7 at about
6.5 mag \citep{nak93}.   It rose up to $m_V=5.5$ on UT December 17.  
\citet{hri98} estimated the decline rate of $t_{2,V}=33$~days,
so V705~Cas is a moderately-fast nova. 

Figure \ref{v705_cas_pw_vul_v_bv_ub_color_curve_logscale} compares
the light curve and color evolutions of V705~Cas with those of PW~Vul. 
Here, we squeeze the light curves of V705~Cas by a factor of 0.8.
We see that these two novae show similar evolution.
Using the time-stretching method, \citet{hac14k}
estimated the absolute magnitude of V705~Cas as
$(m-M)_{V,\rm V705~Cas} = 13.4$ \citep[see Table 2 of][]{hac14k}.
We reanalyzed the data in Figure 
\ref{v705_cas_pw_vul_v_bv_ub_color_curve_logscale} in the same way as
adopted for PW~Vul in Appendix \ref{time-stretching_method_appendix},
and obtained
\begin{eqnarray}
(m-M)_{V,\rm V705~Cas} &=& (m-M)_{V,\rm PW~Vul} - \Delta V - 2.5 \log 0.8
 \cr\cr
&\approx& 13.0 + 0.2 + 0.24 \approx 13.4\cr\cr
&=& 5 \log \left( {d \over {\rm 10~pc}} \right)
+ 3.1 \times E(B-V),
\label{v705_cas_pw_vul_distance_modulus}
\end{eqnarray}
where we use $(m-M)_{V,\rm PW~Vul}= 13.0$ 
determined in Section \ref{result_pw_vul}.
This value of $(m-M)_{V,\rm V705~Cas} = 13.4$ is consistent with
that obtained by \citet{hac14k}.

The distance modulus in UV 1455\AA\ 
is estimated from our model light curve fitting.
Figure \ref{all_mass_v705_cas_x55z02c10o10} shows
three model light curves of free-free emission for
$M_{\rm WD}=0.75$, 0.80, and $0.85~M_\sun$ WDs in steps of $0.05~M_\sun$
as well as the fine grid model of $0.78~M_\sun$ WD (red thin solid line)
in steps of $0.01~M_\sun$.
The distance-reddening relation of the $0.78~M_\sun$ WD model
is derived from the UV 1455\AA\ flux fitting, i.e.,
\begin{eqnarray}
& & 2.5 \log F_{\lambda 1455}^{\rm mod} 
- 2.5 \log F_{\lambda 1455}^{\rm obs} \cr\cr
&=& 2.5 \log(2.61 \times 10^{-12}) - 2.5 \log(1.32 \times 10^{-12}) \cr\cr
&=& 5 \log \left({d \over {10\mbox{~kpc}}} \right)  + 8.3 \times E(B-V),
\label{v705_cas_uv1455a}
\end{eqnarray}
where $F_{\lambda 1455}^{\rm obs}= 1.32 \times 
10^{-12}$~erg~cm$^{-2}$~s$^{-1}$~\AA$^{-1}$ is the
observed peak flux in Figure \ref{all_mass_v705_cas_x55z02c10o10} and
$F_{\lambda 1455}^{\rm mod}= 2.61 \times 
10^{-12}$~erg~cm$^{-2}$~s$^{-1}$~\AA$^{-1}$
is the calculated flux of the $0.78~M_\sun$ model corresponding
to the observed maximum at the distance of 10~kpc.
Figure \ref{v705_cas_distance_reddening_x55z02c10o10} shows
these two distance-reddening relations, i.e., Equation 
(\ref{v705_cas_pw_vul_distance_modulus}), labeled ``$(m-M)_V=13.4$,''
and Equation (\ref{v705_cas_uv1455a}), labeled ``UV 1455 \AA.''  
These two lines cross at $E(B-V)\approx0.45$ and $d\approx2.5$~kpc.

The reddening toward V705~Cas was estimated by \citet{hri98}
to be $E(B-V)=0.38$ from the intercomparison of color indexes
of the stars surrounding the nova selected from SAO catalog.  They
also obtained $E(B-V)= (B-V)_{\rm ss}-(B-V)_{0, \rm ss}=0.32-(-0.11)=0.43$
from the intrinsic color at the stabilization stage \citep{mir88}.
\citet{hau95} obtained $E(B-V)=0.5$ from an assumption that
the total (optical + UV) luminosity in an early phase is constant
\citep[see also][]{sho94}.  A simple arithmetic mean of these values
is $E(B-V)=0.44\pm 0.05$.  The galactic dust absorption map of NASA/IPAC
gives $E(B-V)=0.48 \pm 0.02$ in the direction toward V705~Cas, whose
galactic coordinates are $(l,b)=(113\fdg6595,-4\fdg0959)$.
\citet{hac14k} obtained $E(B-V)=0.45\pm0.05$ from the general course
of novae in the color-color diagram.
These values are all consistent with $E(B-V)=0.45\pm0.05$.
Therefore, we use $E(B-V)=0.45\pm0.05$ in the present paper.
Combining the distance modulus of $(m-M)_V=13.4$ in $V$ band
and $E(B-V)=0.45$, we obtain the distance of $d=2.5$~kpc.
This reddening estimate is very consistent with our $E(B-V)=0.45$ as
shown in Figure \ref{v705_cas_distance_reddening_x55z02c10o10}.
This consistency strongly support the validity of our UV~1455\AA\  
light curve and time-stretching method of $V$ light curve.

Finally, we check the contribution of photospheric emission.
Using this $0.78~M_\sun$ WD model, we calculated the brightness of
photospheric emission in $V$ band (red solid line labeled
``BB'') and the total flux of free-free plus blackbody in $V$ band
(thick black solid line labeled ``TOTAL''), as shown in Figure 
\ref{all_mass_v705_cas_x55z02c10o10_composite}.
Here we use $M_{\rm w}=3.5$ for the $0.78~M_\sun$ WD 
from a linear interpolation between $M_{\rm w} = 3.3$ ($0.80~M_\sun$)
and $M_{\rm w} = 3.8$ ($0.75~M_\sun$) in Table 
\ref{light_curves_of_novae_co4}.
For the distance modulus of $(m-M)_V=13.4$, 
the total $V$ light curve (thick black solid line)
nicely fits with the $V$ observation.
The contribution of photospheric emission is relatively smaller for
V705~Cas.  The obtained physical parameters are summarized in 
Table \ref{physical_parameters}.


\begin{figure}
\epsscale{1.15}
\plotone{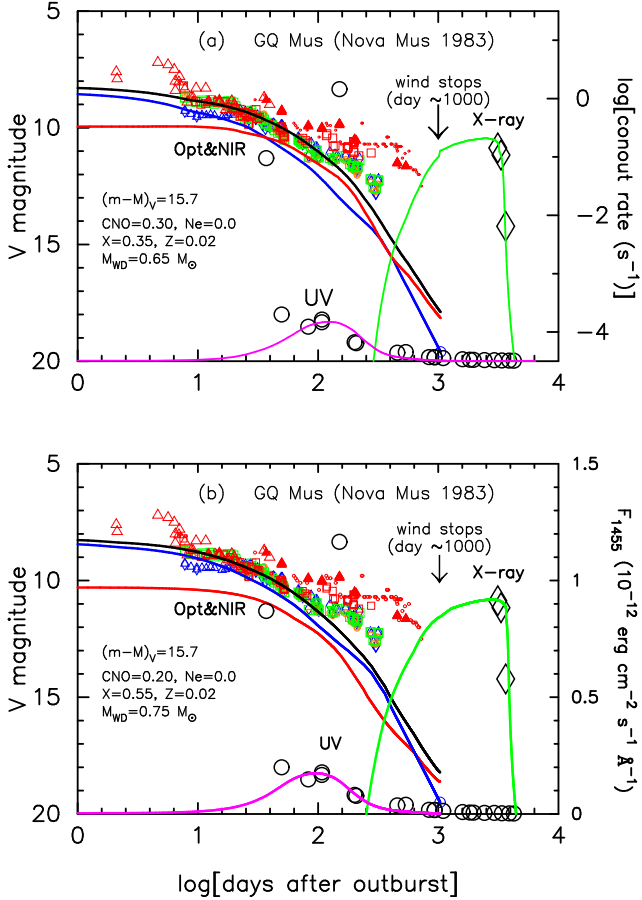}
\caption{
Multi-wavelength light curves for GQ~Mus.
The blackbody $V$ (red solid line), free-free (blue solid line),
total $V$ magnitude (black solid line), UV~1455\AA\   (magenta
solid line), and X-ray (0.1--2.4 keV, green solid line) light curves
for (a) $0.65~M_\sun$ WD with the chemical composition of ``CO Nova 2.''
and (b) $0.75~M_\sun$ WD with the chemical composition of ``CO Nova 4.''
The distance modulus of $(m-M)_V=15.7$ is obtained for the both models, 
(a) and (b), from light curve fitting of the total $V$ flux 
(photospheric emission plus free-free emission).  The observational
data are the same as those cited in \citet{hac08kc}.  We shifted the 
observed $J$ (blue symbols), $H$ (orange symbols), and $K$ (green symbols)
magnitudes down by 3.0, 2.4, and 3.0 mag, respectively.  
\label{all_mass_gq_mus_x35z02c10o20_x55z02c10o10_combine_bb}}
\end{figure}


\begin{figure}
\epsscale{0.95}
\plotone{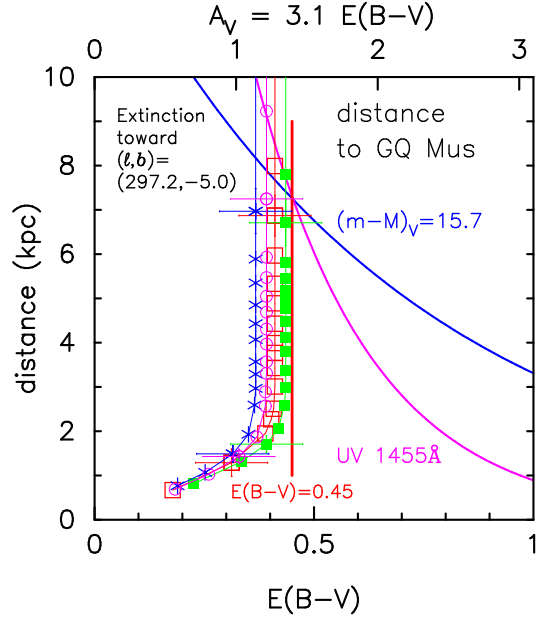}
\caption{
Distance-reddening relations toward GQ~Mus.  A blue thick solid line
denotes Equation (\ref{gq_mus_distance_modulus_vmag}), i.e.,
$(m-M)_V= 15.7$.  A magenta thick solid line represents
Equation (\ref{gq_mus_distance_modulus_uv1455a}), i.e.,
the UV~1455\AA\   fit.  Vertical red thick solid line
indicates the color excess of $E(B-V)=0.45$.
Four sets of data with error bars show distance--reddening relations
in four directions close to GQ~Mus of $(l, b)=(297\fdg2118,  -4\fdg9959)$:
$(l, b)=(297\fdg00,  -4\fdg75)$ (red open squares),
$(297\fdg25,  -4\fdg75)$ (green filled squares),
$(297\fdg00,  -5\fdg00)$ (blue asterisks), and
$(297\fdg25,  -5\fdg00)$ (magenta open circles),
taken from \citet{mar06}.  
\label{gq_mus_distance_reddening_no3}}
\end{figure}

\section{GQ~Mus 1983}
\label{gq_mus}
GQ~Mus is a fast nova with $t_2\sim18$~days \citep{whi84}.
Its peak was missed, so we assume
$m_{V,\rm max}\approx7.2$ after \citet{war95}.
We plot the $V$, visual, $J$, $H$, $K$, UV~1455\AA, and X-ray light curves 
in Figure \ref{all_mass_gq_mus_x35z02c10o20_x55z02c10o10_combine_bb}.
The $V$ data of GQ~Mus are taken from \citet{bud83} (red open triangles), 
\citet{whi84} (red open squares),
and the Fine Error Sensor monitor on board {\it IUE} (red filled triangles)
whereas the visual data are from the Royal
Astronomical Society of New Zealand (small red open circles)
and AAVSO (small red open circles) 
\citep[see][for more details]{hac08kc}.
The $J$ (blue symbols), $H$ (orange symbols), $K$ (green symbols)
light curves are taken from \citet{whi84} and \citet{kra84}.  
The UV~1455\AA\  data are the same as those in \citet{hac08kc}.
The supersoft X-ray fluxes are taken from \citet{sha95} and \citet{ori01}.
\citet{kra84} suggested that the outburst took place 3--4 days
before the discovery.  In absence of precise estimates,
we assumed that the outburst took place at $t_{\rm OB} =$
~JD~ 2,445,348.0 (1983 January 13.5 UT), i.e., 4.6 days
before the discovery by Liller on January 18.14, and adopted
$t_{\rm OB}=$ JD 2,445,348.0 as day zero in the following analysis.

\citet{pac85} determined the color excess of GQ~Mus to be $E(B-V)=0.43$
and \citet{peq93} obtained $E(B-V)=0.50\pm0.05$  both from
the hydrogen Balmer lines.  Similar values were reported by
\citet{kra84} and \citet{has90}, who found $E(B-V) = 0.45$ and 0.50,
respectively, on the basis of the 2175 \AA\  feature
in the early {\it IUE} spectra.  \citet{hac08kc} obtained
$E(B-V) = 0.55\pm 0.05$ on the basis of the 2175\AA\  feature
and various line ratios.  \citet{hac14k} redetermined the color excess
to be $E(B-V)=0.45\pm0.05$ by fitting the general tracks 
with that of GQ~Mus in the $UBV$ color-color diagram.  
We adopt $E(B-V)=0.45\pm0.05$ in this paper, because 
the above estimates are all consistent with $E(B-V)=0.45\pm0.05$.

The chemical composition of GQ~Mus were estimated by a few groups
but scattered from $X=0.27$ to $X=0.43$ as listed in Table
\ref{pw_vul_chemical_abundance}.  So, we adopt two sets of
chemical composition, i.e., ``CO nova 2'' and ``CO nova 4.'' 
Figure \ref{all_mass_gq_mus_x35z02c10o20_x55z02c10o10_combine_bb}
shows theoretical light curves for the chemical composition of
(a) ``CO nova 2'' and (b) ``CO nova 4.''  These light curves are
the best fit ones obtained by \citet{hac08kc} based on 
the free-free emission, UV~1455\AA, and supersoft X-ray model light curves.  
We calculated the photospheric emission and total emission model
$V$ light curves and added them to the figure.

We calculated the total $V$ magnitudes for the $0.65~M_\sun$ WD
with ``CO nova 2.''  Here we used the absolute magnitudes of
free-free emission model light curves in Table 2 of \citet{hac10k}.
Figure 
\ref{all_mass_gq_mus_x35z02c10o20_x55z02c10o10_combine_bb}(a)
shows that the photospheric emission significantly
contributes to the total flux in $V$ band and its effect
improved the fitting. 
We obtain $(m-M)_V=15.7$ from fitting, i.e.,
\begin{eqnarray}
(m-M)_V &=& 15.7 \cr
&=& 5 \log \left( {d \over {\rm 10~pc}} \right)
+ 3.1 \times E(B-V).
\label{gq_mus_distance_modulus_vmag}
\end{eqnarray}
The UV~1455\AA\  flux fitting gives
\begin{eqnarray}
& & 2.5 \log F_{\lambda 1455}^{\rm mod} 
- 2.5 \log F_{\lambda 1455}^{\rm obs} \cr\cr
&=& 2.5 \log(2.97 \times 10^{-12}) - 2.5 \log(1.78 \times 10^{-13}) \cr\cr
&=& 5 \log \left({d \over {10\mbox{~kpc}}} \right)  + 8.3 \times E(B-V),
\label{gq_mus_distance_modulus_uv1455a}
\end{eqnarray}
where $F_{\lambda 1455}^{\rm mod}= 2.97 \times 
10^{-12}$~erg~cm$^{-2}$~s$^{-1}$~\AA$^{-1}$
is the calculated peak flux of the $0.65~M_\sun$ model at the distance 
of 10~kpc corresponding to the magenta solid line in Figure 
\ref{all_mass_gq_mus_x35z02c10o20_x55z02c10o10_combine_bb}(a)
and $F_{\lambda 1455}^{\rm obs}= 1.78 \times 
10^{-13}$~erg~cm$^{-2}$~s$^{-1}$~\AA$^{-1}$ is the 
corresponding observed flux at the same epoch.
We plot these two distance-reddening relations of
Equations (\ref{gq_mus_distance_modulus_vmag}) and
(\ref{gq_mus_distance_modulus_uv1455a}) in Figure
\ref{gq_mus_distance_reddening_no3} together with Marshal et al.'s
(2006) relation and $E(B-V)=0.45$ toward GQ~Mus. 
All trends cross consistently at $d\approx7.3$~kpc and $E(B-V)\approx0.45$.
The galactic dust absorption map of NASA/IPAC
gives $E(B-V)=0.42 \pm 0.01$ in the direction toward GQ~Mus,
whose galactic coordinates are $(l, b)=(297\fdg2118, -4\fdg9959)$,
being consistent with our obtained value of $E(B-V)=0.45\pm0.05$.

For the $0.75~M_\sun$ WD of ``CO Nova 4'' in Figure
\ref{all_mass_gq_mus_x35z02c10o20_x55z02c10o10_combine_bb}(b),
we also obtain $(m-M)_V=15.7$ for the total $V$ light curve fitting.
The UV~1455\AA\  fitting also shows a similar relation as Equation 
(\ref{gq_mus_distance_modulus_uv1455a}).
Therefore, the distance-reddening relations
are almost the same as those for ``CO Nova 2.''
These fitting results are summarized in Table \ref{physical_parameters}.

\citet{hac08kc} obtained $(m-M)_V=14.7$ mainly from various MMRD relations.
This old value is 1.0 mag smaller than our new value, suggesting that 
the MMRD relations are not reliable for individual novae (see 
Section \ref{discussion_mmrd} and Figure 
\ref{max_t3_point_B_scale_pw_vul_no3} for the MMRD values of GQ~Mus).

Figure \ref{all_mass_gq_mus_x35z02c10o20_x55z02c10o10_combine_bb}
shows two UV flashes around Day 37 and Day 151, the 
latter of which was a secondary outburst noticed by \citet{has90}.
The secondary outburst around Day 151 had actually
the appearance of a ``UV flash'' because of its especially large amplitude
at short wavelengths. Indeed, compared with the {\it IUE} low resolution
observations obtained just before and after this event (Days 108 and 202),
the UV flux increased by a factor of 9 at 1455 \AA\ and by a factor 2.2 at
2885 \AA, while the visual flux increased only by a factor of 1.5.  
\citet{hac08kc} discussed this UV flash in more detail.
Therefore, we excluded these points on Days 37, 49, and 151
from our UV light curve fittings because our model light curves
follow only gradual increase and decrease in the UV flux.

 As discussed in Sections \ref{result_pw_vul} and \ref{v705_cas},
$V$ and visual magnitudes are contaminated by strong emission
lines, causing an upward deviation from our free-free models.
In GQ Mus, forbidden [\ion{O}{3}] $\lambda\lambda$4959, 5007 emission
lines already appeared on Day 39 \citep{kra84}.
At about this date the observed visual light curve did actually
start to show an upward deviation from the total flux model.
On the other hand, $JHK$ bands are not so
heavily contaminated by emission lines, as shown in Figure
\ref{all_mass_gq_mus_x35z02c10o20_x55z02c10o10_combine_bb}.

GQ Mus is considered to be a super-bright nova.
The observed $V$ magnitudes in 
Figure \ref{all_mass_gq_mus_x35z02c10o20_x55z02c10o10_combine_bb}
shows about 1.5 mag brighter than our model light curve 
in the earliest phase (until Day 8).
A similar excess is present in the super-bright nova V1500~Cyg
as shown in Figure
\ref{pw_vul_v1668_cyg_v1500_cyg_v1974_cyg_v_bv_ub_color_curve_logscale_no3}(a)
\citep[see also][]{del91, hac06kb}.
We regard the early excess in $V$ magnitude ($<$ Day 8) of GQ~Mus
as the super-bright phase and exclude this phase from fitting,
because the spectra in these super-bright phase are similar to blackbody
rather than free-free emission \citep[e.g.,][for V1500~Cyg]{gal76}.
V1500~Cyg is a polar system \citep[see, e.g.,][]{sch87a, sch87b}.
GQ~Mus is also suggested to be a polar system \citep{dia89, dia94}.
This suggests a possibility that some of
polar systems become a super-bright nova.


\begin{deluxetable*}{lllllllll}
\tabletypesize{\scriptsize}
\tablecaption{Physical parameters of the present models
\label{physical_parameters}}
\tablewidth{0pt}
\tablehead{
\colhead{object} & 
\colhead{WD mass} & 
\colhead{$E(B-V)$} & 
\colhead{$(m-M)_V$} & 
\colhead{distance} & 
\colhead{chem.comp.\tablenotemark{a}}  & 
\colhead{$m_{V,\rm max}$}  & 
\colhead{$t_2$}  &
\colhead{$t_3$} \cr  
\colhead{}  & 
\colhead{$(M_\sun)$} & 
\colhead{} & 
\colhead{} & 
\colhead{(kpc)} & 
\colhead{}  & 
\colhead{}  & 
\colhead{(day)}  &  
\colhead{(day)}
} 
\startdata
PW Vul & 0.83 & 0.55 & 13.0 & 1.8 & CO Nova 4 & 6.3\tablenotemark{b} & 82\tablenotemark{b} & 126\tablenotemark{b} \\
V705 Cas & 0.78 & 0.45 & 13.4 & 2.6 & CO Nova 4 &  5.5\tablenotemark{c} & 33\tablenotemark{c} & 61\tablenotemark{c}\\
GQ Mus & 0.65 & 0.45 & 15.7 & 7.3 & CO Nova 2 & 7.2\tablenotemark{d} & 18\tablenotemark{e} & 40\tablenotemark{e}\\ 
GQ Mus & 0.75 & 0.45 & 15.7 & 7.3 & CO Nova 4 & 7.2 & 18 & 40 \\ 
RR Pic & 0.5--0.60 & 0.04\tablenotemark{f} & 8.7 & 0.52\tablenotemark{f} & CO Nova 4 & 1.1\tablenotemark{f} & 78\tablenotemark{f} & 136\tablenotemark{f} \\
V5558 Sgr & 0.5--0.55 & 0.70 & 13.9 & 2.2 & CO Nova 4 & 6.5\tablenotemark{g} & 125\tablenotemark{h} & 170\tablenotemark{g} \\
HR Del & 0.5--0.55 & 0.15 & 10.4 & 0.97 & CO Nova 4 & 3.76\tablenotemark{b} & 172\tablenotemark{b} & 230\tablenotemark{b} \\
V723 Cas & 0.5--0.55 & 0.35 & 14.0 & 3.9 & CO Nova 4 & 7.1\tablenotemark{i}& (102)\tablenotemark{i}& 173\tablenotemark{i}
\enddata
\tablenotetext{a}{Chemical composition: see Table \ref{chemical_composition}.}
\tablenotetext{b}{\citet{dow00}}
\tablenotetext{c}{\citet{hri98}}
\tablenotetext{d}{\citet{war95}}
\tablenotetext{e}{\citet{whi84}}
\tablenotetext{f}{\citet{har13}}
\tablenotetext{g}{\citet{pog10}}
\tablenotetext{h}{\citet{sch11}}
\tablenotetext{i}{\citet{cho97b}}
\end{deluxetable*}

\section{Very Slow Novae, RR~Pic, V5558~Sgr, HR~Del, and V723~Cas}
\label{very_slow_novae}

In this section we analyze the very slow novae,
RR~Pic, V5558~Sgr, HR~Del, and V723~Cas.
These novae have similar complex light curves of multiple peak
as shown in Figures
\ref{light_curve_rr_pic_v723_cas_hr_del_v5558_sgr_v705_cas_pw_vul} and 
\ref{v723_cas_hr_del_v5558_sgr_rr_pic_ub_bv_color_light_curve_revised_no2}.
Here, we adopt Kato \& Hachisu's (2009, 2011) explanation for the
multiple peak.  They showed that there are two kinds of
envelope solutions for the same envelope mass and WD mass,
one is a static and the other is a wind mass loss solution,
in a narrow range of WD mass, $0.5 ~M_\sun \lesssim M_{\rm WD} 
\lesssim 0.7 ~M_\sun$.  On these WDs, nova outbursts begin
quasi-statically and then undergo a transition from static to wind
evolution.  During the transition, the nova accompanies oscillatory
activity and begins to blow massive winds after the transition is
completed.  Thus, we apply our method of optically thick wind solutions
to the light curves only after the transition is completed.

\begin{figure*}
\epsscale{0.9}
\plotone{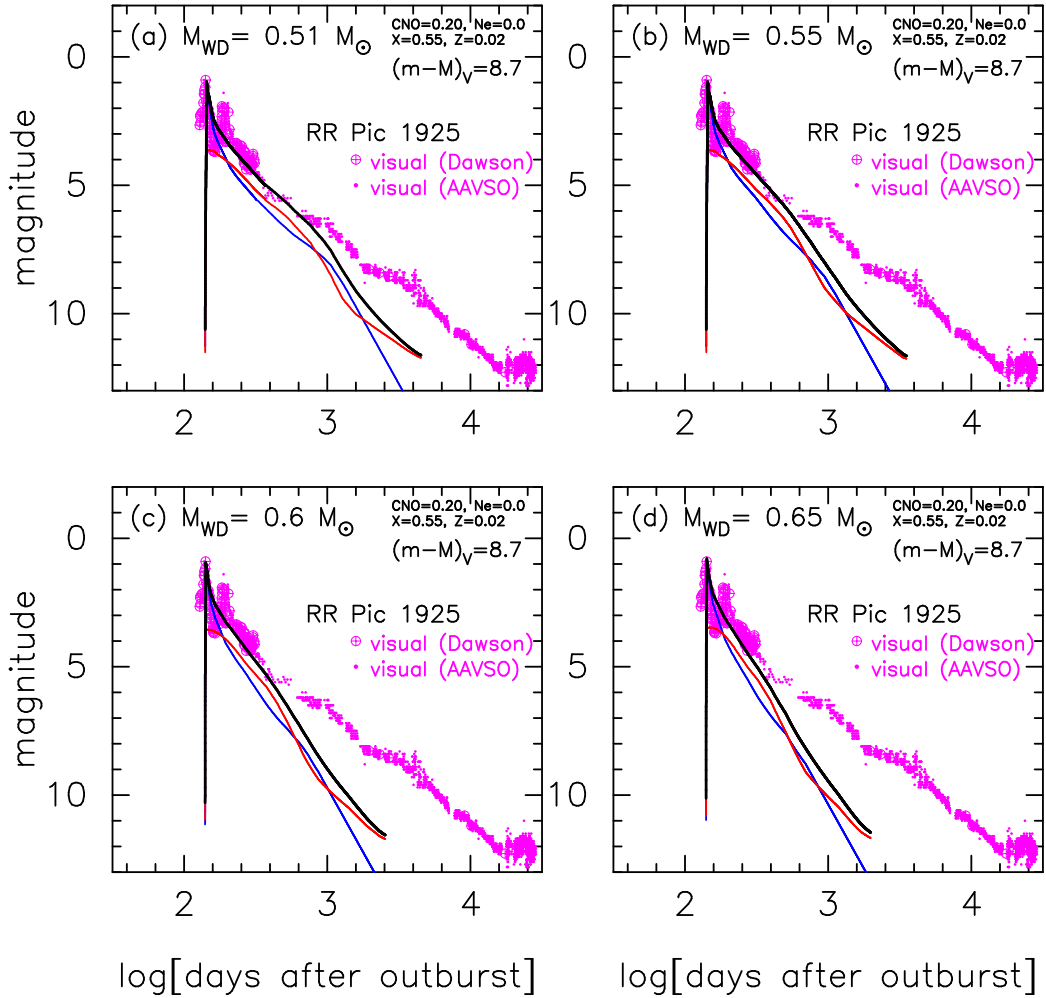}
\caption{
Visual light curves of RR~Pic.
Magenta circles with plus sign inside: visual magnitudes
taken from \citet{daw26}.
Magenta dots: visual magnitudes taken from the AAVSO archive.
We plot four different WD mass models: (a) $0.51~M_\sun$, (b) $0.55~M_\sun$, 
(c) $0.6~M_\sun$, and (d) $0.65~M_\sun$ for the envelope chemical
composition of ``CO nova 4.''
Blue solid lines denote the $V$ fluxes of free-free emission.
Red solid lines represent the $V$ fluxes of blackbody emission.
Black solid lines indicate the $V$ fluxes of total (free-free
plus blackbody) emission. 
\label{all_mass_rr_pic_x55z02c10o10_composite_4fig}}
\end{figure*}

\begin{figure}
\epsscale{1.15}
\plotone{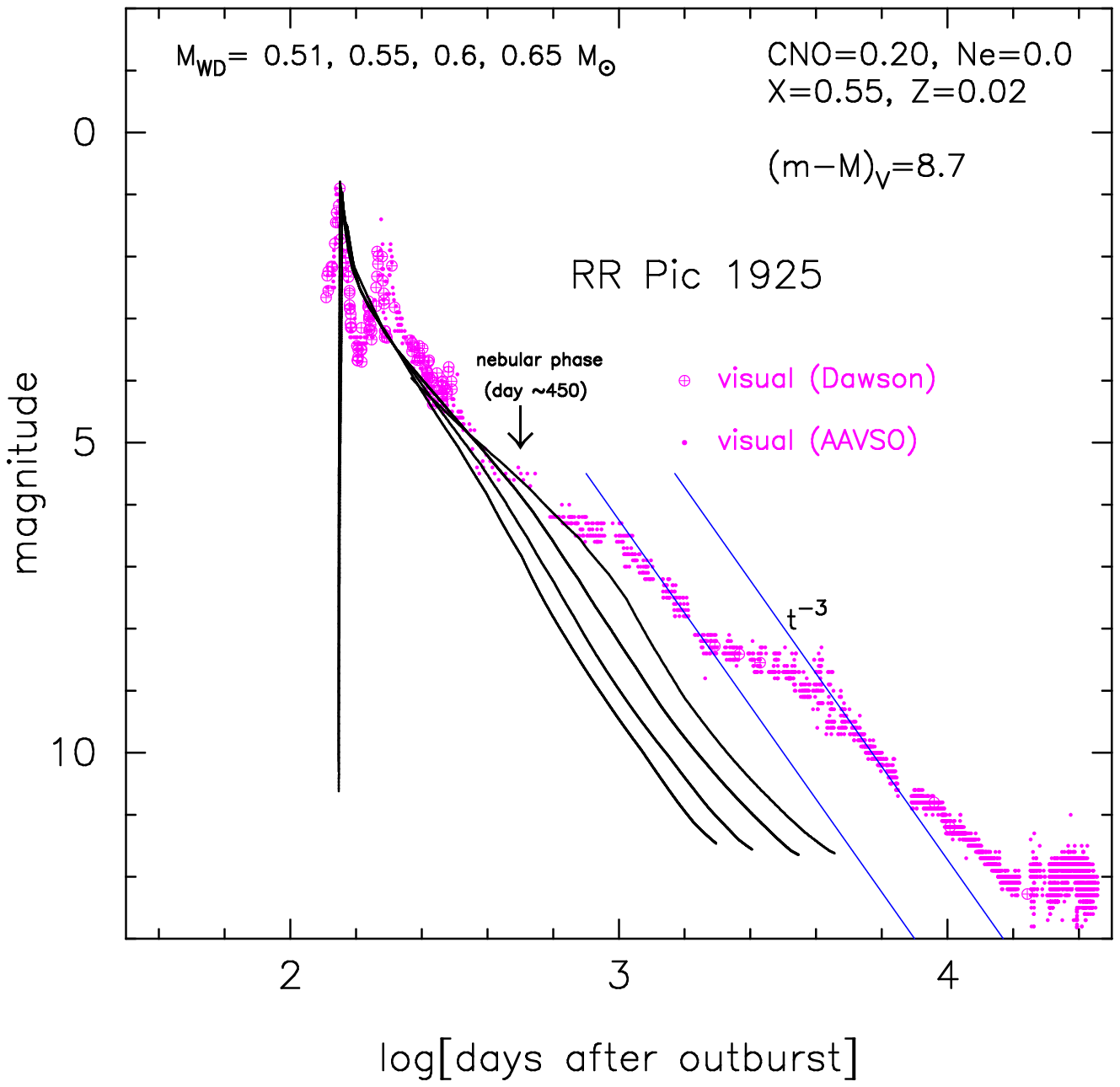}
\caption{
Same as Figure \ref{all_mass_rr_pic_x55z02c10o10_composite_4fig}, 
but only for the total (free-free plus blackbody) 
$V$ flux model light curves.  Black thick solid lines: total $V$ 
flux light curves for the 0.51, 0.55, 0.60, and $0.65~M_\sun$ WD models.
\label{all_mass_rr_pic_x55z02c10o10_composite_one}}
\end{figure}

\begin{figure}
\epsscale{0.95}
\plotone{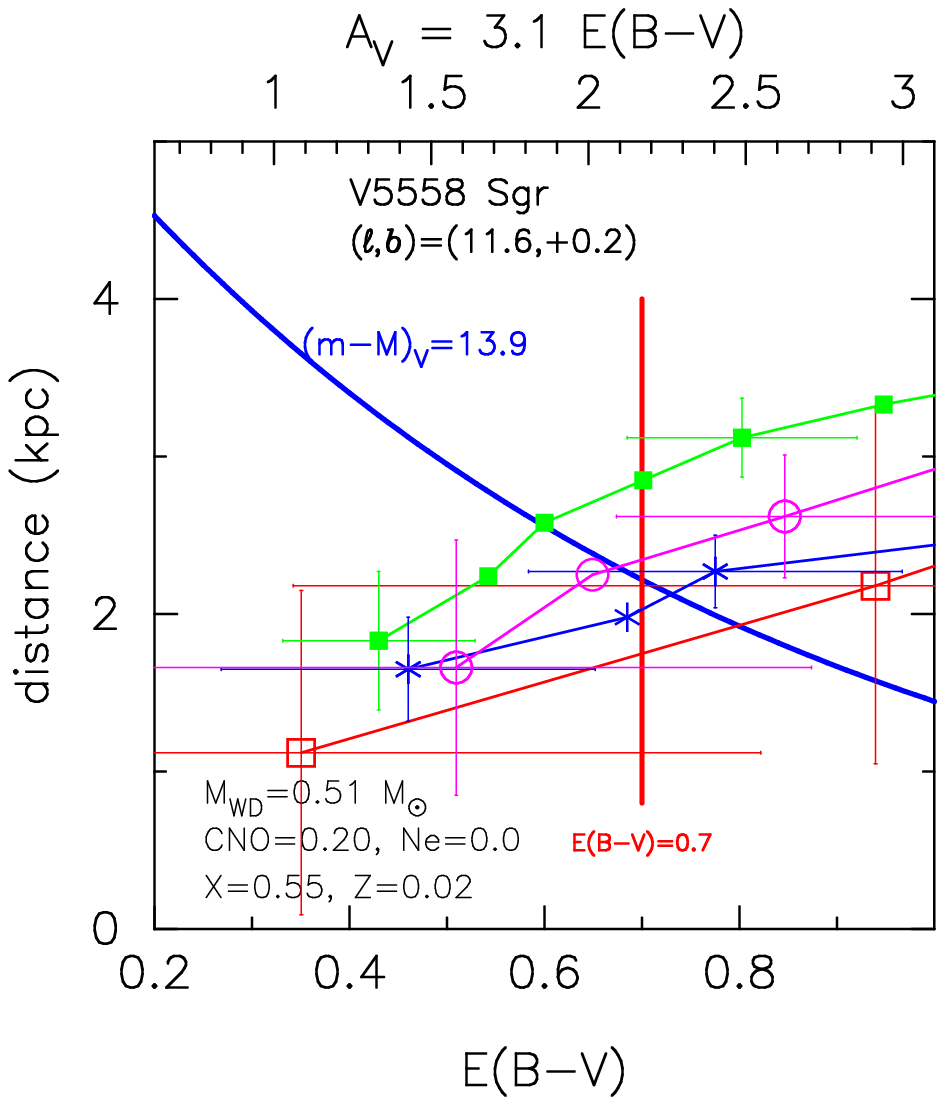}
\caption{
Various distance-reddening relations toward V5558~Sgr, whose
galactic coordinates are $(l, b)=(11\fdg6107,+0\fdg2067)$.
We plot distance--reddening relations, which are taken from 
\citet{mar06}, in four directions close to V5558~Sgr. 
Red open squares: toward $(l, b)= (11\fdg5, 0\fdg0)$,
Green filled squares: toward $(11\fdg75, 0\fdg0)$.
Blue asterisks: toward $(11\fdg5, 0\fdg25)$.
Magenta open circles: toward $(11\fdg75, 0\fdg25)$.
We also plot $E(B-V)=0.7$ (vertical red solid line) and
Equation (\ref{distance-reddening_v5558_sgr}), i.e., 
$(m-M)_V=13.9$ (blue solid line).
\label{v5558_sgr_distance_reddening_x55z02c10o10_no3}}
\end{figure}

\begin{figure*}
\epsscale{0.9}
\plotone{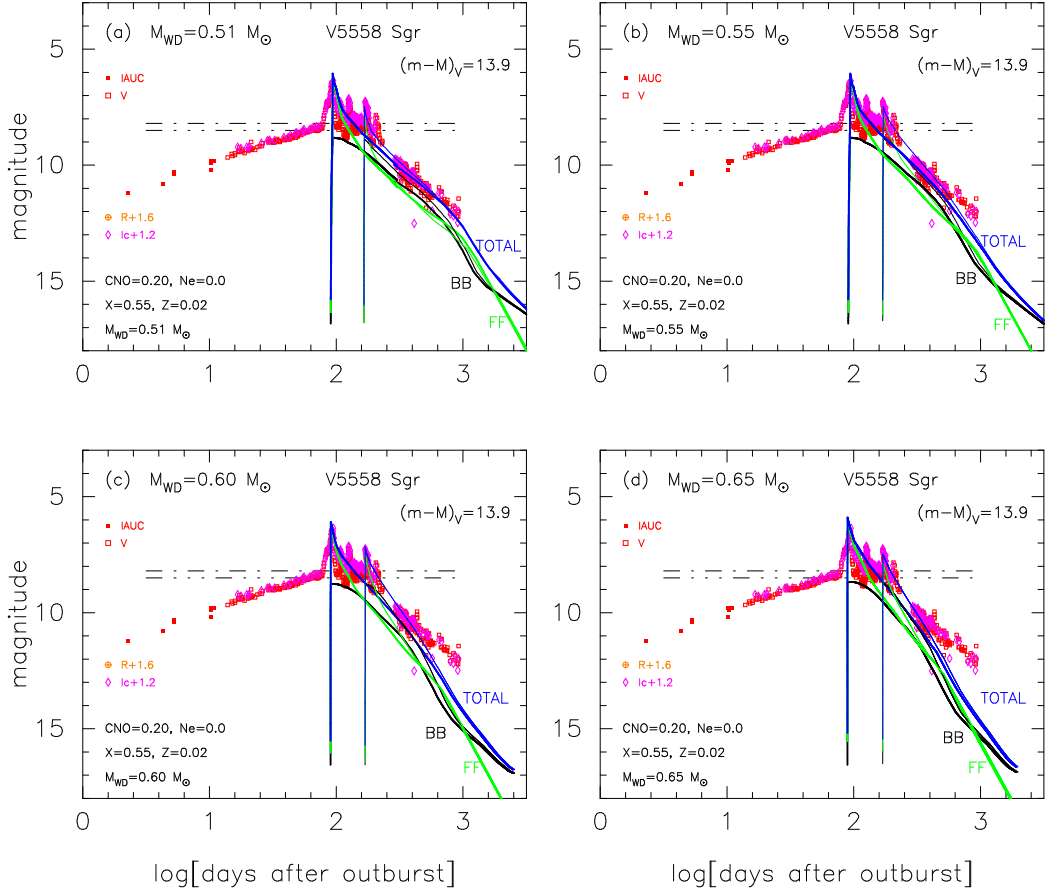}
\caption{
Optical and NIR light curves of V5558~Sgr.
We assumed the distance modulus of $(m-M)_V=13.9$ from
Equation (\ref{rr_pic_distance_mod}).
We plot four different WD mass models: (a) $0.51~M_\sun$, (b) $0.55~M_\sun$, 
(c) $0.6~M_\sun$, and (d) $0.65~M_\sun$ for the envelope chemical
composition of ``CO nova 4.''   Black solid lines labeled ``BB''
denote the $V$ fluxes of blackbody emission.
Green solid lines labeled ``FF'' represent the $V$ fluxes
of free-free emission.  Blue solid lines labeled ``TOTAL'' depict
the $V$ fluxes of total (free-free
plus blackbody) emission.  We assumed that the transition from static
to wind evolution occurred just at the optical maximum (first peak).
We also added different light curves for a different transition
time at the third peak.
Red filled squares: $V$ magnitudes taken from IAU Circular
No. 8832.  Red open squares: $V$ magnitudes taken from archives of 
AAVSO and Variable Star Observers League of Japan (VSOLJ).
Orange open circles with plus sign inside: $R_C$ magnitudes 
taken from the archives of AAVSO and VSOLJ.
Magenta open diamonds: $I_C$ magnitudes taken from the archives of 
AAVSO and VSOLJ.
Horizontal dash-dotted and dash-three-dotted lines denote
the absolute magnitudes of $M_V=-5.7$ and $-5.4$ for each panel, 
which are the absolute magnitudes of the flat peak of the symbiotic
nova PU~Vul
in 1979 and 1981-1983, respectively \citep[see Figure 15 of][]{hac14k}.
\label{all_mass_v5558_sgr_x55z02c10o10_4fig_no2}}
\end{figure*}

\begin{figure}
\epsscale{1.15}
\plotone{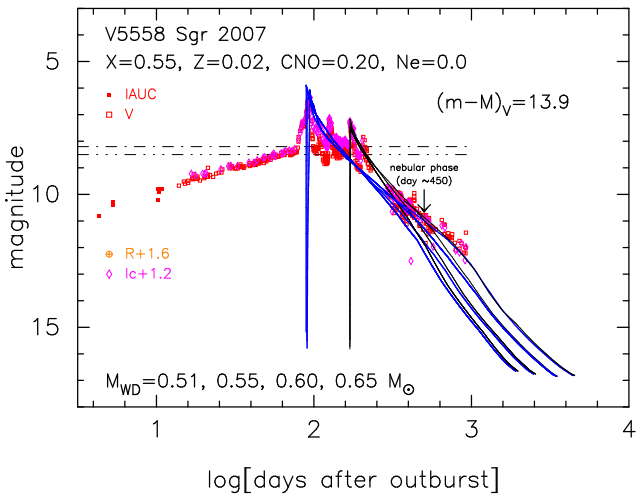}
\caption{
Same as Figure \ref{all_mass_v5558_sgr_x55z02c10o10_4fig_no2}, 
but only for the total (free-free plus blackbody) 
$V$ flux model light curves.  Blue thick solid lines: a transition
occurs from static to wind evolution at the first peak.
Black thin solid lines: a different transition time at the third peak.
\label{all_mass_v5558_sgr_x55z02c10o10_one_fig}}
\end{figure}

The light curves of these four novae are very similar to each other and
their chemical compositions were obtained to be $X=0.53$ for RR~Pic,
$X=0.45$ for HR~Del, and $X=0.52$ for V723~Cas, as listed in Table 
\ref{pw_vul_chemical_abundance}, which are close to that of ``CO nova 4.'' 
Therefore, we adopt the chemical composition of $X=0.55$,
$Y=0.23$, $X_{\rm CNO}=0.2$, and $Z=0.02$ (``CO nova 4'').
We made light curves of free-free plus blackbody emission for four
different WD masses, $M_{\rm WD}= 0.51$, $0.55$, $0.6$, and $0.65~M_\sun$,
because the transition occurs from static to wind evolution for
$0.5 ~M_\sun \lesssim M_{\rm WD} \lesssim 0.7 ~M_\sun$.
The absolute magnitudes of free-free emission light curves
are taken from Table \ref{light_curves_of_novae_co4}
for the $0.55$, $0.6$, and $0.65~M_\sun$ WD models.  The $0.51~M_\sun$
WD model is not tabulated in Table \ref{light_curves_of_novae_co4} but
is calibrated in the same way as those of $0.55$, $0.6$, and 
$0.65~M_\sun$ WD models.  We could not successfully obtain wind solutions
for $M_{\rm WD}\le0.50~M_\sun$ because of numerical difficulty
\citep{kat94h}.
Our results are shown in Figure 
\ref{all_mass_rr_pic_x55z02c10o10_composite_4fig} and
\ref{all_mass_rr_pic_x55z02c10o10_composite_one} for RR~Pic,
in Figures \ref{v5558_sgr_distance_reddening_x55z02c10o10_no3},
\ref{all_mass_v5558_sgr_x55z02c10o10_4fig_no2}, and
\ref{all_mass_v5558_sgr_x55z02c10o10_one_fig} for V5558~Sgr,
in Figures \ref{hr_del_distance_reddening_x55z02c10o10} and 
\ref{all_mass_hr_del_x55z02c10o10_composite_2fig_no2} for HR~Del,  
in Figures \ref{all_mass_v723_cas_x55z02c10o10_composite_m51} and 
\ref{v723_cas_distance_reddening_x55z02c10o10_no3} for V723~Cas.


\subsection{RR~Pic 1925}
\label{rr_pic}
RR~Pic was discovered by Watson at about 2.3 mag on UT 1925 May 25. 
The details of visual light curve of RR~Pic were found
in \citet{spe31}, in which the photographic magnitude prior to 1925 was
$m_{\rm pg}=12.75$ and the nova brightened up to 3rd magnitude between
February 18 (fainter than 11th mag) and April 13 (3rd mag), and
the first maximum was reached on UT June 7 ($m_v=1.18$).  
\citet{spe31} discussed a possibility of another peak between
April 13 (3rd mag) and May 25 (2.3 mag) before the first
peak on June 7 ($m_v=1.18$).   Such a multiple peak
has been observed also in V5558~Sgr, HR~Del, and
V723~Cas, as clearly shown in Figure 
\ref{light_curve_rr_pic_v723_cas_hr_del_v5558_sgr_v705_cas_pw_vul}.
We superpose these four novae in Figure
\ref{v723_cas_hr_del_v5558_sgr_rr_pic_ub_bv_color_light_curve_revised_no2}
to confirm that these novae have very similar decline rates.
We shift horizontally their times and vertically their magnitudes
by $+5.3$, $0.0$, $+3.6$, and $+0.1$ mag, respectively, 
as indicated in the figure.  We also superpose these four novae
in a logarithmic timescale in Figure
\ref{v723_cas_hr_del_v5558_sgr_rr_pic_ub_bv_color_light_curve_revised_logscale_no2}.
These four nova light curves almost overlap each other. 

The distance to RR~Pic was obtained using the trigonometric
parallax, i.e., $d=521^{+54}_{-45}$~pc \citep{har13}.
The distance modulus in $V$ band is calculated to
be $(m-M)_V=5\log 521^{+54}_{-45}/10 + 0.13 = 8.7\pm0.2$,
where we used $A_V=0.13$ after \citet{har13}.  Adopting this
distance modulus, we plot, in Figure
\ref{all_mass_rr_pic_x55z02c10o10_composite_4fig},
our model light curves for four WD masses, i.e.,
(a) $0.51~M_\sun$, (b) $0.55~M_\sun$, (c) $0.60~M_\sun$,
and (d) $0.65~M_\sun$ as well as the visual observation.
Here, we assumed that the transition was completed at the first peak
(UT 1925 June 7) and the outburst day was JD~2,424,170.0
about 140 days before the first peak.
Figure \ref{all_mass_rr_pic_x55z02c10o10_composite_one} shows the same
model light curves as those in 
Figure \ref{all_mass_rr_pic_x55z02c10o10_composite_4fig}, but 
only the total $V$ light curves of different WD masses. 

It is remarkable that two models of (a) $0.51~M_\sun$ and
(b) $0.55~M_\sun$ nicely fit with the
visual magnitude until the nebular phase begins about 450 days
after the outburst \citep[$\sim300$ days after the first peak,
see][]{spe31, iij06}.  Note that the nova outburst begins
quasi-statically and then undergoes a transition from static to wind
evolution at the first peak.  We think that, during the transition,
the nova accompanies oscillatory activity of relaxation.  This
corresponds to the second and third peaks of the light curve.
Thus, we confirm that the absolute
magnitudes of our model light curves (total flux of free-free
plus blackbody) are consistent with that of very slow novae
even if they do not follow the universal decline law.  Therefore,
we confidently apply our absolute magnitude estimate to very slow novae. 

It should be noted that the peak brightness of each model light curve
depends on the initial envelope mass, which is closely related to
the ignition mass of outburst.  The larger the envelope mass is,
the brighter the peak is.  Therefore, we can estimate the envelope
mass by adjusting the peak brightness to the observation
if the WD mass is fixed.  

Our model light curves of total $V$ flux have a similar brightness
in the early phase of outburst for the WD mass range of 
$0.5 ~M_\sun \lesssim M_{\rm WD} \lesssim 0.7 ~M_\sun$ (see Figure
\ref{all_mass_rr_pic_x55z02c10o10_composite_one}).
This property also can be seen in the light curve analysis of 
the slow nova DQ~Her (see Figure 
\ref{all_mass_dq_her_x35z02c10o20_m060_logscale_bb_ff_total_2fig} below).
This is a problem in our light curve analysis because we are not able to
identify the WD mass only from our model $V$ light curve fitting
in the early phase.  On the other hand, we are able to estimate
the absolute magnitudes of slow novae, independently of the WD mass,
by directly comparing them with a nova with known distance such as RR~Pic.

Using the distance modulus of RR~Pic, i.e., $(m-M)_V= 8.7$, we obtain
the distance moduli for the other three novae.  
Because these four novae have very similar decline shapes and
should have similar brightnesses in the early phase of outbursts,
we simply assumed that their brightnesses are all the same in 
the overlapping region of the light curves in Figures
\ref{v723_cas_hr_del_v5558_sgr_rr_pic_ub_bv_color_light_curve_revised_no2}
and 
\ref{v723_cas_hr_del_v5558_sgr_rr_pic_ub_bv_color_light_curve_revised_logscale_no2}.
The difference in apparent $V$ magnitude against V723~Cas is $-5.3$
for RR~Pic, $-3.6$ for HR~Del, and $-0.1$ for V5558~Sgr.
Therefore, the difference
$\Delta V$ from RR~Pic is calculated as $\Delta V= -3.6+5.3$ for HR~Del,
$\Delta V= -0.0+5.3$ for V723~Cas, and $\Delta V= -0.1+5.3$ for V5558~Sgr.
Thus we have
\begin{eqnarray}
(m-M)_{V, \rm RR~Pic} &=& 8.7\cr
&=&(m-M)_{V,\rm HR~Del} - \Delta V \cr
&=& 10.4 -(-3.6+5.3) = 8.7\cr
&=&(m-M)_{V,\rm V723~Cas} - \Delta V \cr
&=& 14.0 -(-0.0+5.3) = 8.7\cr
&=&(m-M)_{V,\rm V5558~Sgr} - \Delta V \cr
&=& 13.9 -(-0.1+5.3) = 8.7.
\label{rr_pic_distance_mod}
\end{eqnarray}
Then, the distance moduli of these three novae are 
$(m-M)_{V,\rm HR~Del}=10.4$, $(m-M)_{V,\rm V723~Cas}=14.0$,
and $(m-M)_{V,\rm V5558~Sgr}=13.9$.


\subsection{V5558~Sgr 2007}
\label{v5558_sgr}
V5558~Sgr was first detected by Sakurai
\citep{nak07a} at mag 10.3 on UT 2007 April 14.777.  Sakurai also
reported that nothing is visible on an image taken on UT April 9.8
(limiting mag 11.4).  The star was also detected by Haseda
\citep{yam07b} at mag 11.2  on UT April 11.792.  Since the outburst
day is not known, we adopted UT 2007 April 8.5 as the outburst
day, i.e., $t_{\rm OB}=$JD~2454199.0, in this paper.
The distance modulus in $V$ band was already obtained to be
\begin{eqnarray}
(m-M)_V &=& 13.9 \cr
&=& 5 \log \left( {d \over {\rm 10~pc}} \right)
+ 3.1 \times E(B-V),
\label{distance-reddening_v5558_sgr}
\end{eqnarray}
in Equation (\ref{rr_pic_distance_mod}) of Section \ref{rr_pic}.
We plot this distance-reddening relation of
Equation (\ref{distance-reddening_v5558_sgr}) in Figure
\ref{v5558_sgr_distance_reddening_x55z02c10o10_no3}.
We found in literature two different values of reddening,
one is $E(B-V)=0.36$ obtained 
by \citet{mun07c} from \ion{Na}{1}~D lines and the other is
$E(B-V)=0.8$ obtained by \citet{rud07b} from \ion{O}{1} lines.
Since these two values are largely different,
we examine other reddening estimates.
Figure \ref{v5558_sgr_distance_reddening_x55z02c10o10_no3} also shows
distance-reddening relations taken from \citet{mar06} 
in four directions close to V5558~Sgr, $(l, b)=(11\fdg6107,+0\fdg2067)$.
The closest one is that of blue asterisks, which crosses our line of 
$(m-M)_V=13.9$ at $E(B-V)\approx0.7$ and $d\approx2.2$~kpc.  This reddening
value is consistent with $E(B-V)=0.7\pm0.05$ estimated by \citet{hac14k},
who obtained the reddening by assuming that the three novae, V5558~Sgr,
HR~Del, and V723~Cas have the same intrinsic $(B-V)_0$ color in the
premaximum phase.  We adopt $(m-M)_V=13.9$, $E(B-V)=0.7$, and $d=2.2$~kpc
in this paper.

Figure \ref{all_mass_v5558_sgr_x55z02c10o10_4fig_no2} shows optical and
NIR light curves of V5558~Sgr and our model light curves for (a) $0.51$,
(b) $0.55$, (c) $0.6$, and (d) $0.65~M_\sun$ WDs on a logarithmic timescale.
Here, we assumed the distance modulus of $(m-M)_V=13.9$.
Blue solid lines show the total $V$ fluxes (labeled ``TOTAL'') of our model
light curves while green solid lines correspond to the free-free $V$ fluxes
(labeled ``FF'') and black solid lines represent the blackbody $V$ fluxes
(labeled ``BB'').  From the $V$ light curve shape, 
we assumed that the transition occurred from static to wind evolution
about 90 days after the outburst (at the first optical peak).
For comparison, we add another case of the transition at the third peak.  
Because the peak brightness of our model light curves depend on the initial
envelope mass at the outburst, we tune the initial envelope mass
to the peak brightness for each model.
We adopted a less massive envelope for the model light curve of thin
solid line that starts at the third peak of Figure 
\ref{all_mass_v5558_sgr_x55z02c10o10_4fig_no2}.
Figure \ref{all_mass_v5558_sgr_x55z02c10o10_one_fig} shows the same
model light curves as those in 
Figure \ref{all_mass_v5558_sgr_x55z02c10o10_4fig_no2}, but 
only the total $V$ light curves of different WD masses for comparison. 

Among the four WD mass models, the $0.51$ and $0.55~M_\sun$ WD models
are in good agreement with the observation while the $0.60$ and
$0.65~M_\sun$ WDs may be too steep to be compatible with the observation.
During the transition, the nova accompanies oscillatory activity of
relaxation.  This corresponds to the second, third, and fourth peaks
of the light curve.  This conclusion is unchanged even if we adopt
the transition time at the third peak (thin solid lines in Figures
\ref{all_mass_v5558_sgr_x55z02c10o10_4fig_no2} and
\ref{all_mass_v5558_sgr_x55z02c10o10_one_fig}).  
These fitting results are summarized in Table \ref{physical_parameters}.
This good agreement of the brightness supports that the absolute
magnitudes of our model light curves (total flux of free-free
plus blackbody) are reasonable even for very slow novae.
It should be noted that the nebular phase started about
450 days after the outburst \citep[a year after the fist
peak, see][]{pog12}.  However, [\ion{O}{3}] emission lines are 
too weak to contribute significantly to the $V$ flux,
so that the $V$ light curve does not deviate so much
from the model light curves as shown in Figures 
\ref{all_mass_v5558_sgr_x55z02c10o10_4fig_no2}(a)
and \ref{all_mass_v5558_sgr_x55z02c10o10_one_fig}.

As already mentioned above, \citet{kat09h, kat11h}
modeled the pre-maximum phase of these
very slow novae with a static evolution followed by
the transition from a static to a wind structure.
They predicted that this transition occurs in a narrow range of WD masses,
$0.5 ~M_\sun \lesssim M_{\rm WD} \lesssim 0.7 ~M_\sun$.
Thus, the brightness at the pre-maximum phase of these novae
should be similar to the flat peak of the symbiotic nova
PU~Vul ($M_{\rm WD}\sim0.6~M_\sun$),
whose brightness is $M_V= -5.4$ in stage 1 (in 1979) 
and $M_V=-5.7$ in stage 2 (in 1981--1983),
respectively \citep[see Figure 15 of][]{hac14k}.
We plot these two absolute magnitudes of PU~Vul in Figures
\ref{all_mass_v5558_sgr_x55z02c10o10_4fig_no2} and
\ref{all_mass_v5558_sgr_x55z02c10o10_one_fig}
(horizontal thin dash-dotted and dash-three-dotted lines).
These brightnesses are in perfect agreement with
the brightness of V5558~Sgr at the premaximum halt (flat)
phase just before the first peak, i.e.,
before the transition started.


\begin{figure}
\epsscale{0.95}
\plotone{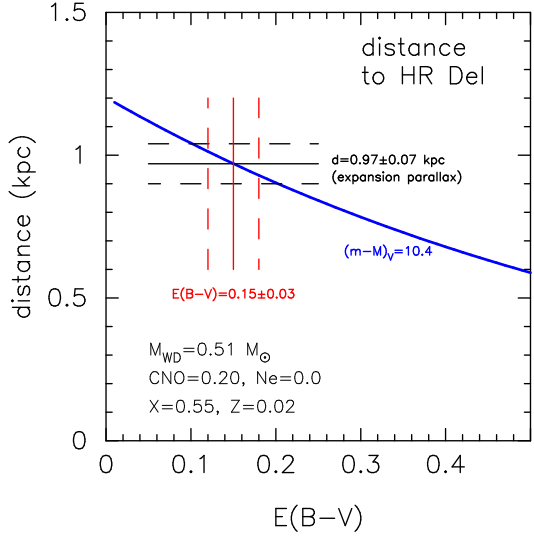}
\caption{
Distance-reddening relation toward HR~Del.
A blue solid line represents Equation (\ref{eq_ext_dis_hr_del}),
i.e., $(m-M)_V=10.4$.  The distance and reddening estimates are
taken from \citet{har03} and \citet{ver87}, respectively.
\label{hr_del_distance_reddening_x55z02c10o10}}
\end{figure}


\begin{figure}
\epsscale{1.15}
\plotone{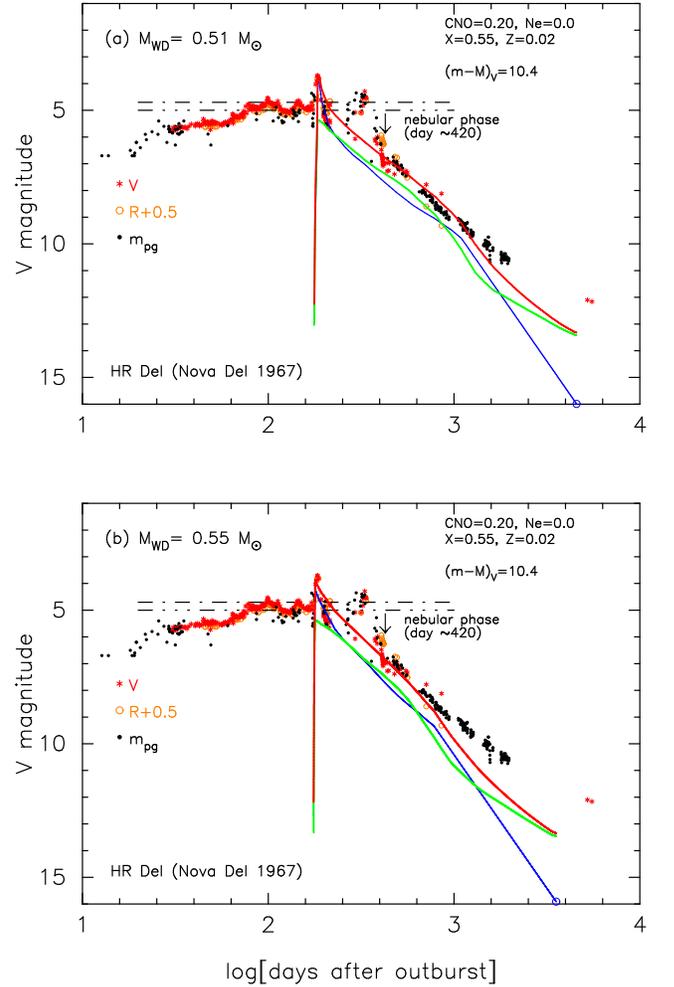}
\caption{
Optical light curves of HR~Del.
We plot two different WD mass models: (a) $0.51~M_\sun$ and 
(b) $0.55~M_\sun$ for the envelope chemical composition of ``CO nova 4.''
We assumed that the transition occurred about 170 days after the outburst
and that the distance modulus in $V$ band is $(m-M)_V=10.4$.
Red asterisks: $V$ magnitudes.  
Orange open circles:  $R$ magnitudes.
Black filled circles: photographic magnitudes, $m_{\rm pg}$.  
Observational data are taken from IAU Circular Nos. 2024, 2025, 2030, 2036, 
and from \citet{sto67}, \citet{nha67}, \citet{ond68}, \citet{oco68}, 
\citet{ter68}, \citet{gry69},
\citet{mol69}, \citet{man70}, \citet{bar70} and the AAVSO archive.  
Green solid lines indicate the blackbody $V$ flux,
blue solid lines represent the free-free $V$ flux, and
red solid lines denote the total $V$ flux of free-free plus blackbody. 
Horizontal dash-dotted and dash-three-dotted lines denote
the absolute magnitudes of $M_V=-5.7$ and $-5.4$,
which are the absolute magnitudes of the flat peak of PU~Vul.
\label{all_mass_hr_del_x55z02c10o10_composite_2fig_no2}}
\end{figure}

\subsection{HR~Del 1967}
\label{hr_del}
The slow nova HR~Del was discovered by Alcock \citep{can67}
at $m_v= 5.0$ on UT 1967 July  8.94 (JD 2439680.44).
Since the outburst day is not known, we adopt UT 1967 June 8.5
as the outburst day, i.e., $t_{\rm OB}=$JD~2439653.0,
from the light curve of \citet{rob68}.
The distance modulus of HR~Del was already obtained to be
\begin{eqnarray}
(m-M)_V &=& 10.4 \cr
&=& 5 \log \left( {d \over {\rm 10~pc}} \right)
+ 3.1 \times E(B-V),
\label{eq_ext_dis_hr_del}
\end{eqnarray}
in Equation (\ref{rr_pic_distance_mod}) of Section \ref{rr_pic}.
We plot this distance-reddening relation of
Equation (\ref{eq_ext_dis_hr_del}) in Figure
\ref{hr_del_distance_reddening_x55z02c10o10} by a blue solid line.
\citet{ver87} obtained the extinction toward HR~Del to be
$E(B-V)=0.15\pm 0.03$.  
The galactic dust absorption map of NASA/IPAC
gives $E(B-V)=0.11 \pm 0.006$ in the direction toward HR~Del,
whose galactic coordinates are $(l, b)=(63\fdg4304, -13\fdg9721)$,
being roughly consistent with Verbunt's value.
Two lines of $E(B-V)=0.15$ and Equation (\ref{eq_ext_dis_hr_del}) 
cross at a distance of $d=0.97$~kpc as shown in Figure
\ref{hr_del_distance_reddening_x55z02c10o10}.

\citet{dow00}, on the other hand, obtained
the distance to HR~Del to be $d=0.76\pm 0.13$~kpc from the nebular
expansion parallax.  More recently, \citet{har03} obtained a new
value of the distance $d=0.97\pm 0.07$~kpc also from the expansion
parallax method of {HST} imaging.  Other older estimates are
all between the above two estimates, i.e., $d=0.940\pm 0.155$~kpc for
various expansion parallax methods 
\citep{mal75, koh81, due81, sol83, coh83, sla94, sla95} 
or $d=0.835\pm 0.092$~kpc for
the other techniques \citep{dre77}.
Here we adopt the distance of $d=0.97\pm 0.07$~kpc after
\citet{har03} and the extinction 
of $E(B-V)=0.15\pm 0.03$ after \citet{ver87}.
These two values are consistent with
Equation (\ref{eq_ext_dis_hr_del}) in
Figure \ref{hr_del_distance_reddening_x55z02c10o10}.

We plot the light curve of HR~Del in Figures 
\ref{light_curve_rr_pic_v723_cas_hr_del_v5558_sgr_v705_cas_pw_vul} and 
\ref{v723_cas_hr_del_v5558_sgr_rr_pic_ub_bv_color_light_curve_revised_no2}
on a linear timescale and in Figures
\ref{v723_cas_hr_del_v5558_sgr_rr_pic_ub_bv_color_light_curve_revised_logscale_no2}
and \ref{all_mass_hr_del_x55z02c10o10_composite_2fig_no2}
on a logarithmic timescale.
Figure \ref{all_mass_hr_del_x55z02c10o10_composite_2fig_no2} shows
two model light curves for (a) $0.51~M_\sun$ and (b) $0.55~M_\sun$ WDs,
in which we assumed that the transition completed 170 days after the outburst.
Red solid lines show the total $V$ flux of free-free 
(blue solid lines) plus blackbody (green solid lines).
We calculated four model light curves of 0.51, 0.55, 0.60,
and $0.65~M_\sun$ WDs but did not plot the 0.60 and $0.65~M_\sun$ WDs
because these two are too steep to be compatible with the observation.
The $0.51~M_\sun$ WD model shows good agreement
with the observation while the $0.55~M_\sun$ WD model is marginal,
as shown in Figure \ref{all_mass_hr_del_x55z02c10o10_composite_2fig_no2}(b).
Note that the nebular phase started about
420 days after the outburst \citep[$\sim250$ days after the fist
peak, see][]{iij06}.
During the transition, the nova accompanies oscillatory activity of
relaxation.  This corresponds to the second peak of the light curve.  
The brightness of $M_V=-5.7$ in PU~Vul is in good agreement with
the brightness of HR~Del at the premaximum halt (flat) phase
just before the optical maximum, i.e., before the transition started.
This fact also confirms 
that our absolute magnitudes of optical light curves for slow novae
are reasonable.


\begin{figure*}
\epsscale{0.75}
\plotone{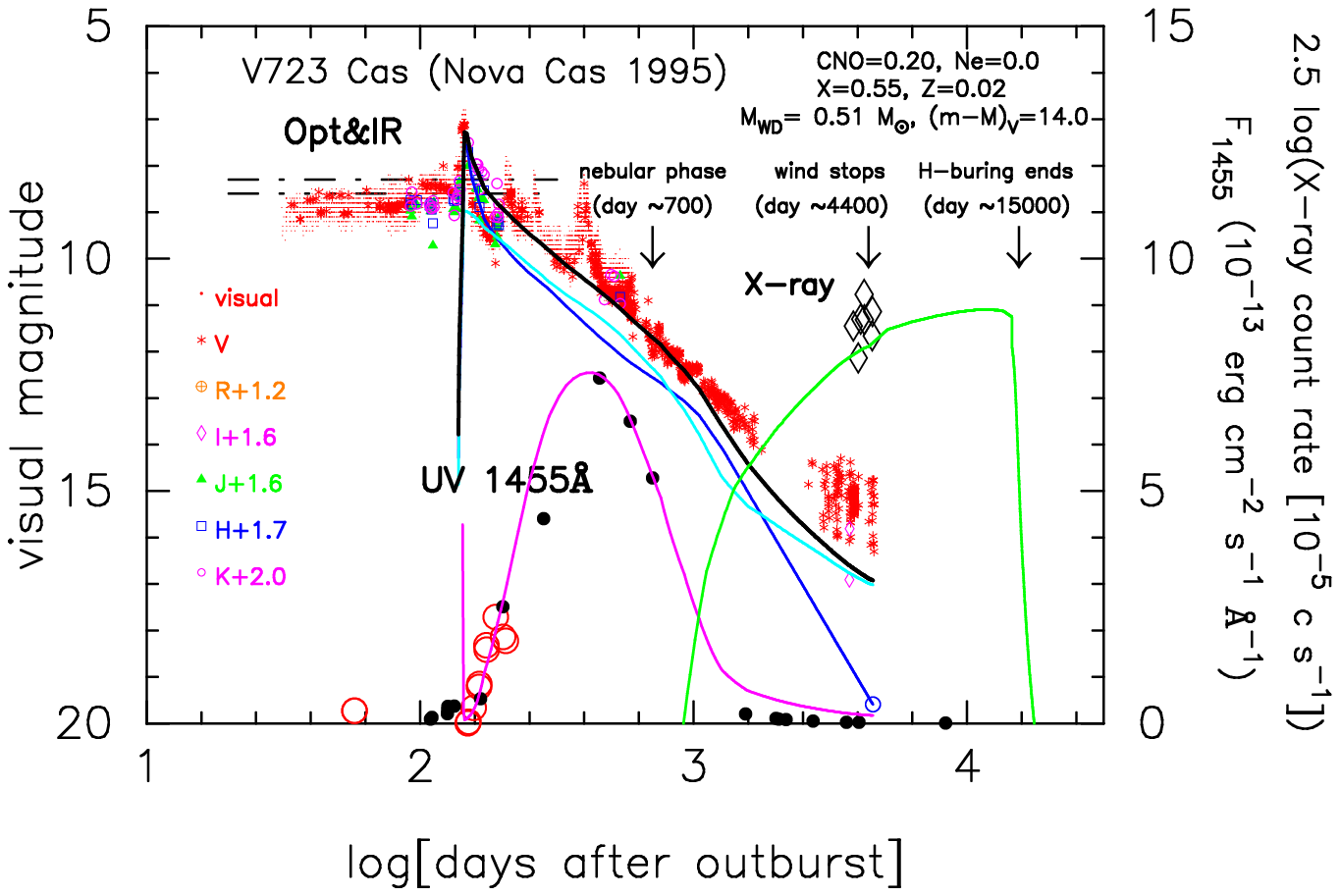}
\caption{
Optical, NIR, UV~1455\AA, and supersoft X-ray light curves of V723~Cas.
Large open red circles: {\it IUE} UV~1455\AA\  data \citep{cas02}.
Large open black diamonds: supersoft X-ray fluxes
of 0.25--0.6~keV taken from \citet{nes08}.
Other symbols show $V$, $R$, $I$, $J$, $H$, and $K$ observational
data, which are taken from \citet{cho97b, cho98}, \citet{kam99},
and the AAVSO archive.
Black filled circles: UV~1455\AA\  data of PW~Vul \citep{cas02}, but
the timescale is stretched by 5.8 times.
Horizontal dash-dotted and dash-three-dotted lines denote
the absolute magnitudes of $M_V=-5.7$ and $-5.4$
of the flat peaks of PU~Vul in 1979 and in 1981-1983, respectively. 
We plot the $0.51~M_\sun$ WD model for the envelope chemical
composition of ``CO nova 4,'' which are
the free-free emission (thick blue solid line),
blackbody emission (sky blue solid line),
total of free-free plus blackbody (black solid line),
UV~1455\AA\  (magenta solid line), and
supersoft X-ray (calculated from blackbody emission;
green solid line) light curves.
\label{all_mass_v723_cas_x55z02c10o10_composite_m51}}
\end{figure*}


\begin{figure}
\epsscale{0.95}
\plotone{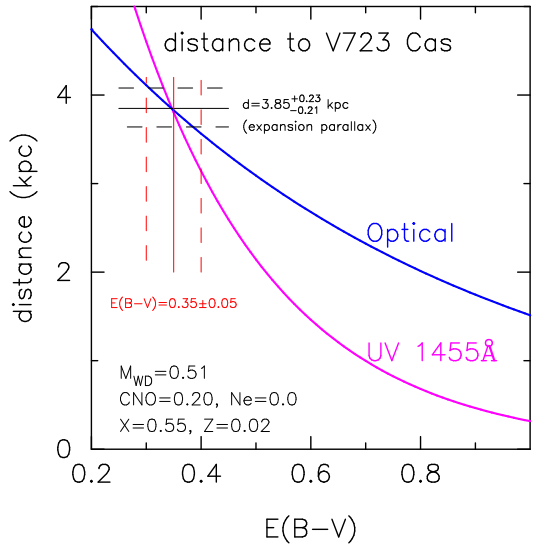}
\caption{
Distance-reddening relation toward V723~Cas.
A blue solid line represents Equation (\ref{eq_ext_dis_v723_cas}),
i.e., $(m-M)_V=14.0$. 
A magenta solid line depicts the UV~1455\AA\  distance-reddening relation
of Equation (\ref{v723_cas_distance_modulus_uv1455a}). 
The distance estimate is taken from \citet{lyk09} and 
the color excess is taken from \citet{hac14k}.
\label{v723_cas_distance_reddening_x55z02c10o10_no3}}
\end{figure}

\subsection{V723~Cas 1995}
\label{v723_cas}
V723 Cas is also a very slow nova.  It was discovered at mag 9.2
by Yamamoto on UT 1995 August 24.57 (JD~2449954.07).
\citet{mun96} proposed UT July 20.5 as the outburst day, i.e., 
$t_{\rm OB}=$JD~2449919.0, so we adopt this day in this paper.
We plot the visual, $V$, $R$, $I$, $J$, $H$, and $K$ light curves of
V723~Cas in Figure \ref{all_mass_v723_cas_x55z02c10o10_composite_m51}
together with the UV~1455\AA\  and X-ray light curves.  
The distance modulus of HR~Del was already obtained to be
\begin{eqnarray}
(m-M)_V &=& 14.0 \cr
&=& 5 \log \left( {d \over {\rm 10~pc}} \right)
+ 3.1 \times E(B-V),
\label{eq_ext_dis_v723_cas}
\end{eqnarray}
in Equation (\ref{rr_pic_distance_mod}) of Section \ref{rr_pic}.
We plot this distance-reddening relation of
Equation (\ref{eq_ext_dis_v723_cas}) in Figure
\ref{v723_cas_distance_reddening_x55z02c10o10_no3}
by a blue solid line.  The distance to V723~Cas was estimated
by \citet{lyk09} to be $d=3.85^{+0.23}_{-0.21}$~kpc 
from the expansion parallax method.  \citet{hac14k} obtained
the reddening toward V723~Cas to be 
$E(B-V) = 0.35\pm 0.05$ by fitting the general tracks with
the observed track of V723~Cas in the $UBV$ color-color diagram.
These three trends, i.e., $(m-M)_V=14.0$, $d=3.85$~kpc, and
$E(B-V) = 0.35$, cross consistently as shown in Figure
\ref{v723_cas_distance_reddening_x55z02c10o10_no3}.
Therefore, we adopt these values in this paper.

Figure \ref{all_mass_v723_cas_x55z02c10o10_composite_m51} shows
our model light curves of the $0.51~M_\sun$ WD for the chemical composition
of ``CO nova 4.''  Here, we assumed the distance modulus in $V$ band
to be $(m-M)_V=14.0$ and that the transition from static
to wind evolution occurred 155 days after the outburst,
because the UV~1455\AA\  flux started to rise on this day.
A black solid line shows the total $V$ flux of free-free (blue solid
line) plus blackbody (sky blue solid line) emission calculated from
the $0.51~M_\sun$ WD model which is the one showing the best agreement
with the observation compared with the other three WD mass models of
$0.55$, $0.60$, and $0.65~M_\sun$.  During the transition,
the nova accompanies oscillatory activity of relaxation.
This corresponds to the second, third, and fourth peaks
of the light curve.  Note that the model light curves were
fitted to the lower envelope of the $V$ light curve to avoid local
photospheric fluctuations until the nebular phase started about
700 days after the outburst \citep[$\sim550$ days after the fist
peak, see][]{iij06}.

The following additional distance-reddening relation (labeled 
``UV 1455 \AA\  '' in Figure
\ref{v723_cas_distance_reddening_x55z02c10o10_no3})
can be deduced from our UV 1455\AA\  flux fitting, i.e.,
\begin{eqnarray}
& & 2.5 \log F_{1455}^{\rm mod} 
- 2.5 \log F_{1455}^{\rm obs} \cr\cr
&=& 2.5 \log(3.2 \times 10^{-12})  - 2.5 \log(1.5 \times 10^{-12}) \cr\cr
&=&  5 \log \left({d \over {10\mbox{~kpc}}} \right)  + 8.3 \times E(B-V),
\label{v723_cas_distance_modulus_uv1455a}
\end{eqnarray}
where $F_{1455}^{\rm mod}= 3.2 \times 
10^{-12}$~erg~cm$^{-2}$~s$^{-1}$~\AA$^{-1}$ is the calculated flux
at the upper limit of the figure box at the distance of 10~kpc and
$F_{1455}^{\rm obs}= 1.5 \times 
10^{-12}$~erg~cm$^{-2}$~s$^{-1}$~\AA$^{-1}$ is the observed flux
corresponding to the upper limit of the figure box.
The two distance-reddening relations, i.e.,
Equation (\ref{eq_ext_dis_v723_cas}) and
Equation (\ref{v723_cas_distance_modulus_uv1455a}),
cross each other at the point of
$E(B-V)\approx0.34$ and $d \approx 3.9$~kpc, being consistent
with the distance of $d=3.85^{+0.23}_{-0.21}$~kpc 
\citep{lyk09} and $E(B-V)=0.35\pm0.05$ \citep{hac14k} mentioned above.

\citet{nes08} obtained $(m-M)_V= 13.7$, $E(B-V)=0.5\pm 0.1$, 
and $d=2.7^{+0.4}_{-0.3}$~kpc by assuming that the absolute magnitude
of V723~Cas is similar to that of HR~Del.  The main difference from
ours comes from the reddening estimate.
The interstellar extinction toward V723 Cas was estimated
by many authors but the values are quite scattered; that is,
in increasing order,
$E(B-V) = 0.20\pm 0.12$ in August 1999 and $0.25\pm 0.1$ in July 2000
\citep{rud02} from Paschen and Brackett lines,
$E(B-V) = 0.29$ calculated from $A_V = 0.89$ \citep*{iij98}
from reddenings of field stars near the location of V723~Cas,
$E(B-V) = 0.45$ \citep{mun96} from interstellar \ion{Na}{1}~D double lines,
$E(B-V) = 0.5\pm 0.1$ \citep{nes08} estimated
from various values in literature and their $N_{\rm H}$ value from
X-ray spectrum model fits,
$E(B-V) = 0.57$ \citep{cho97b} from intrinsic colors at maximum and
at two magnitude below maximum, $E(B-V) = 0.60$ \citep{gon96}
from the 2200\AA\  dust absorption feature,
and $E(B-V) = 0.78\pm 0.15$ \citep{eva03} from the IR \ion{H}{1} 
recombination lines.  Recently, 
Gonz\'alez-Riestra revised the value to be $E(B-V)=0.30$--$0.35$
\citep[private communication 2012, see also][]{hac14k}. 
\citet{hac14k} obtained 
$E(B-V) = 0.35\pm 0.05$ by fitting the general tracks with
the observed track of V723~Cas in the $UBV$ color-color diagram.
The recent NASA/IPAC dust map gives $E(B-V)=0.34\pm0.01$ toward V723~Cas,
whose galactic coordinates are $(l, b)=(124\fdg9606,-8\fdg8068)$.
Therefore, we adopt $E(B-V)= 0.35$ and $d=3.85$~kpc in this paper.
These fitting results are summarized in Table \ref{physical_parameters}.

\section{Discussion}
\label{discussion_section}

\subsection{Brightness Confirmation of Model Light Curves}
\label{brightness_confirmation}
We examine whether or not the absolute brightness of our model
light curve is correct for classical novae with known distances.
\citet{har13} determined the distances of four novae,
V603~Aql, GK~Per, DQ~Her, and RR~Pic, with {\it HST} annual parallaxes.
We have already examined the case of RR~Pic in Section \ref{rr_pic}
and showed that the total brightness of our model light curve
reasonably reproduces the absolute brightness of RR~Pic
for the distance modulus of $(m-M)_{V,\rm RR~Pic}=8.7$ \citep{har13}.
In this subsection, we study the other three, i.e., GK~Per, V603~Aql, 
and DQ~Her.


\begin{figure}
\epsscale{1.15}
\plotone{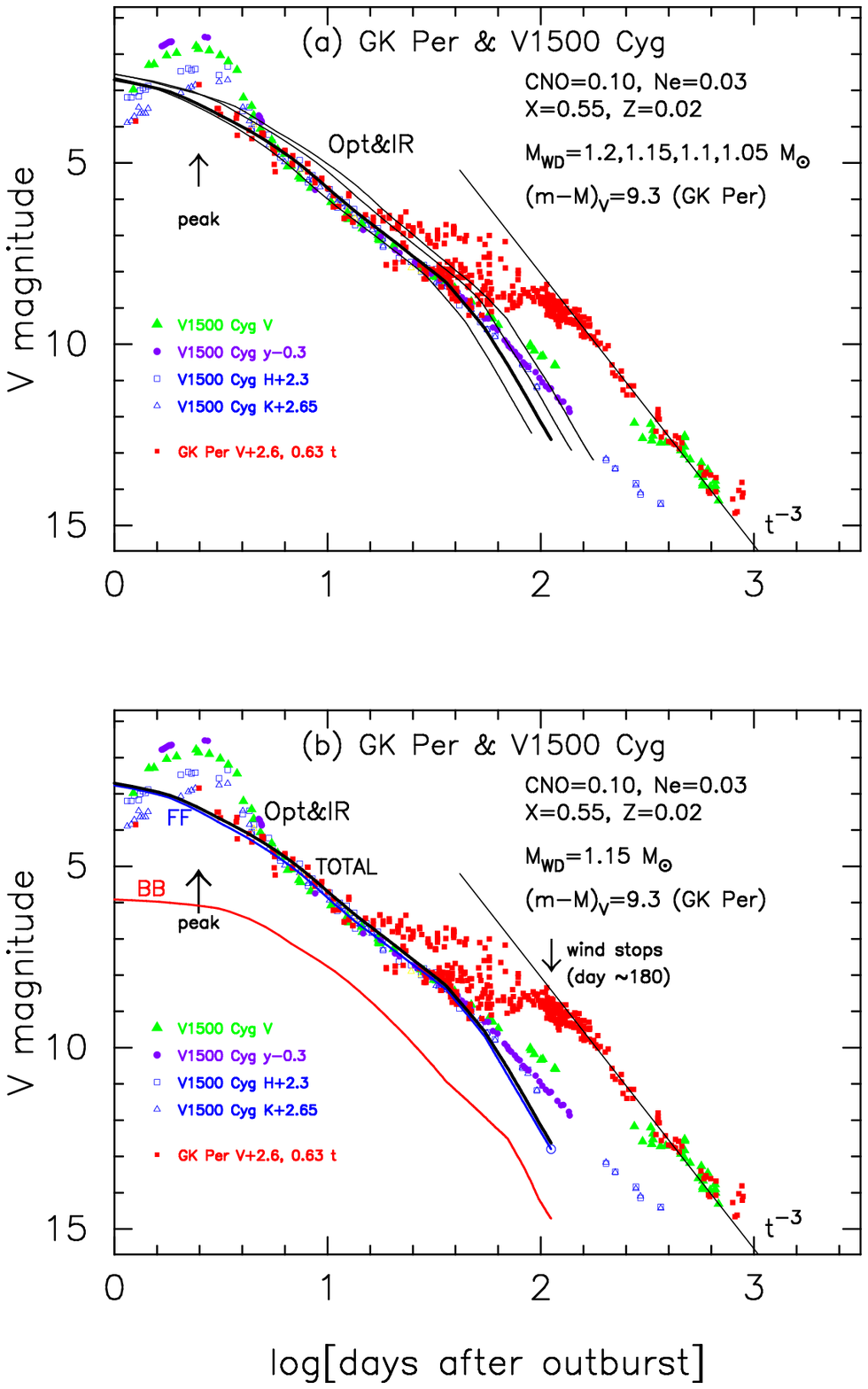}
\caption{
Light curves of GK~per in a logarithmic timescale.  The light curves
of V1500~Cyg are also added and overlapped to those of GK~Per. 
The light curves of GK~Per are shifted horizontally 
by $\Delta \log t = -0.20$ leftward
and vertically by $\Delta V=+2.6$ downward.
Visual observations (red filled squares) of GK~Per are taken from
\citet{chi01}, \citet{gor01},
\citet{ram01a, ram01b, ram01c, ram01d, ram01e, ram02, ram03},
\citet{sha01},
and \citet{wil01a, wil01b, wil01c, wil01d, wil02, wil19}.
For V1500~Cyg, the $y$ magnitudes are taken from \citet{loc76},
the $V$ magnitudes are from \citet{gal76}, \citet{ark76}, 
\citet{pfa76}, and \citet{tem79}, the NIR $H$
and $K$ magnitudes are taken from \citet{gal76}, \citet{kaw76}, 
and \citet{enn77}. 
(a) Assuming the chemical composition of
the envelope ``Ne nova 2,''  we plot four total $V$ flux light curves
for the WD mass models of 1.05, 1.1, 1.15, and $1.2~M_\sun$.
Among these four nova light curves, we found the
$1.15~M_\sun$ (thick black solid line) as a best fit model.  
(b) We plot three $V$ model light curves of photospheric (blackbody,
red solid line labeled ``BB''), free-free (blue solid line labeled
``FF''), and total flux (black solid line labeled ``TOTAL).
The photospheric emission does not much contribute to the total $V$ flux.
\label{gk_per_v1500_cyg_x55z02o10ne03_logscale_fig2}}
\end{figure}

\subsubsection{GK~Per 1901}
\label{discussion_gk_per}
The distances of GK~Per is obtained to be $d=477^{+28}_{-25}$~pc
by \citet{har13}.  The distance modulus in $V$ band is obtained to be
$(m-M)_{V,\rm GK~Per}=5\log 477^{+28}_{-25}/10 + 3.1\times0.3 = 9.3\pm0.1$,
where we adopt $E(B-V)=0.3$ \citep{wu89} after \citet{har13}.
The WD mass of GK~Per was estimated by \citet{mor02} to be 
$M_{\rm WD}=0.77^{+0.52}_{-0.24}~M_\sun$, being not accurately constrained.

For GK~Per, we assumed the chemical composition of ``Ne nova 2'' 
because no estimates are available in literature.
For this chemical composition, the absolute magnitudes
of free-free emission model light curves
were already determined in Table 3 of \citet{hac10k}.
Using the absolute magnitudes of free-free model light curves,
we calculated the total (free-free plus photospheric) 
$V$ flux light curves for the WD masses
of 1.05, 1.1, 1.15, and $1.2~M_\sun$.  We plot these four $V$ light curves
in Figure \ref{gk_per_v1500_cyg_x55z02o10ne03_logscale_fig2}(a), where  
we adopted $(m-M)_{V,\rm GK~Per}=9.3$.
Among the four WD masses, we obtained a best fit
for $M_{\rm WD}=1.15~M_\sun$ (a thick black solid line).
The other fluxes (blackbody and free-free fluxes) are
also plotted only for $M_{\rm WD}=1.15~M_\sun$ in Figure
\ref{gk_per_v1500_cyg_x55z02o10ne03_logscale_fig2}(b).  

In Figure \ref{gk_per_v1500_cyg_x55z02o10ne03_logscale_fig2},
we plot optical and NIR light curves of the very fast nova V1500~Cyg
as well as the fast nova GK~Per.  GK~Per shows a transition phase
in the middle part of the outburst.  We do not know how to fit
our model light curves with the observation in such an oscillatory 
light curve.  For this purpose, we overlap the light curve of V1500~Cyg
to that of GK~Per and select which part of oscillatory brightness
to be fitted.  In the figure, we shift the light curve of
GK~Per horizontally by $\Delta \log t = -0.20$ and vertically 
$\Delta V=+2.6$ mag to overlap it to the light curves of V1500~Cyg.
The two $V$ light curves reasonably overlap in the early phase and
in the very later phase.  In the middle part of the light curves,
the lower bound of oscillatory brightness of GK~Per reasonably overlaps
that of V1500~Cyg.  Therefore, we fit our model light curves to
the lower bound of oscillatory brightness during the transition phase
of GK~Per.  Note that our model light curve reasonably fits with
the early $V$ light curve but deviates from the visual observation
in the nebular phase.
This deviation in visual magnitudes is owing to strong emission lines
such as [\ion{O}{3}], which are not included in our model
\citep[see][for details]{hac06kb}.

To summarize, we are able to reproduce the absolute brightness
for the distance modulus of $(m-M)_{V,\rm GK~Per}=9.3$
and $M_{\rm WD}=1.15~M_\sun$.
Photospheric emission (red solid line labeled ``BB'') 
does not contribute to the total $V$ flux (black solid line
labeled ``TOTAL'') as shown in the Figure
\ref{gk_per_v1500_cyg_x55z02o10ne03_logscale_fig2}(b).  
\citet{hac06kb} showed that nova light curves follow a universal
decline law if free-free emission dominates the spectrum.
Figure \ref{gk_per_v1500_cyg_x55z02o10ne03_logscale_fig2}(b)  
confirms that the nova light curves follow the universal decline law.


\begin{figure}
\epsscale{1.15}
\plotone{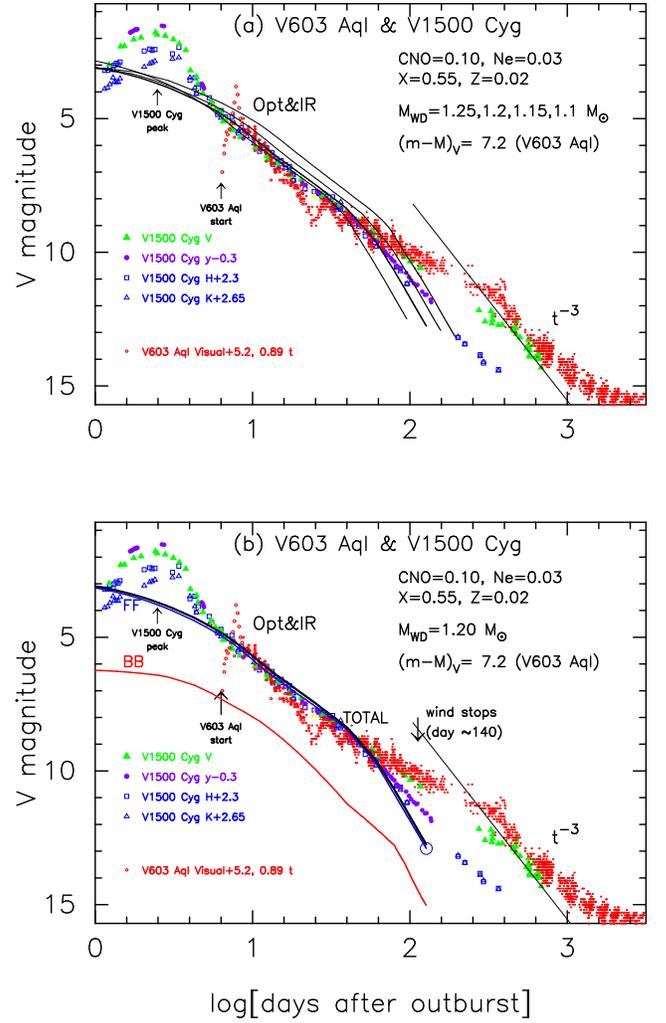}
\caption{
Same as Figure \ref{gk_per_v1500_cyg_x55z02o10ne03_logscale_fig2},
but for V603~Aql.  Visual data of V603~Aql are taken from \citet{cam23}
(red small open circles) and from AAVSO (red small dots).  
The light curves of V603~Aql are shifted horizontally by 
$\Delta \log t = -0.05$ leftward and vertically by $\Delta V=+2.4$ downward.
(a) Assuming the chemical composition of the envelope ``Ne nova 2,''
we plot four total $V$ flux light curves for the WD mass models of 1.1, 1.15,
1.2, and $1.25~M_\sun$.  Among these four nova light curves, we found the
$1.2~M_\sun$ (thick black solid line) as a best fit model.  
(b) We plot three $V$ model light curves of photospheric (blackbody,
red solid line labeled ``BB''), free-free (blue solid line labeled
``FF''), and total flux (black solid line labeled ``TOTAL) for the
best-fit $1.2~M_\sun$ WD.  The photospheric emission also 
does not much contribute to the total $V$ flux.
\label{v603_aql_v1500_cyg_x55z02o10ne03_logscale_2fig}}
\end{figure}

\subsubsection{V603~Aql 1918}
\label{discussion_v603_aql}
The distance of V603~Aql is obtained to be $d=249^{+9}_{-8}$~pc 
by \citet{har13}.  The distance modulus in $V$ band
is calculated to be $(m-M)_{V,\rm V603~Aql}=5\log 249^{+9}_{-8}/10 
+ 3.1\times0.07 = 7.2\pm0.07$, where we adopt
$E(B-V)=0.07$ \citep{gal74} after \citet{har13}.
The WD mass of V603~Aql was 
obtained by \citet{are00} to be $M_{\rm WD}=1.2\pm0.2~M_\sun$.

The chemical composition of V603~Aql is not available, so 
we assume ``Ne nova 2'' in this paper partly because 
we already estimated the absolute magnitudes of free-free
emission model light curves for ``Ne nova 2'' \citep{hac10k}.
We calculated the total (free-free plus photospheric)
$V$ flux light curves for the WD masses
of 1.1, 1.15, 1.2, and $1.25~M_\sun$.  We plot these four $V$ light
curves in Figure \ref{v603_aql_v1500_cyg_x55z02o10ne03_logscale_2fig}(a),
where we adopted $(m-M)_{V,\rm V603~Aql}=7.2$.  
Among the four WD masses, we obtained a best fit
for $M_{\rm WD}=1.2~M_\sun$ (a thick black solid line).
The other fluxes (blackbody and free-free fluxes) are
also plotted only for $M_{\rm WD}=1.2~M_\sun$ in Figure
\ref{v603_aql_v1500_cyg_x55z02o10ne03_logscale_2fig}(b).  
The brightness of our model light curves are consistent with
both the distance modulus of $(m-M)_V=7.2$ and
the WD mass of $M_{\rm WD}=1.2~M_\sun$.
This confirms that the absolute magnitudes of our model light curves
are reasonable.

In Figure \ref{v603_aql_v1500_cyg_x55z02o10ne03_logscale_2fig},
we add optical and NIR light curves of the very fast nova V1500~Cyg.
V603~Aql shows a transition phase in the middle part of the outburst
like GK~Per.  In order to know how to fit our model light curves
with the oscillatory light curve, we again overlap the light curve of
V1500~Cyg to that of V603~Aql and select which part of oscillatory brightness
to be fitted.  In the figure, we shift the light curve of
V603~Aql horizontally by $\Delta \log t = -0.05$ and vertically 
$\Delta V=+5.2$ mag to overlap it to the light curves of V1500~Cyg.
Moreover, we set the start of the light curves about 7 days later
(indicated by an arrow) than the start of V1500~Cyg. 
The two $V$ light curves reasonably overlap in the early phase and
in the very later phase.  In the middle part of the light curves,
the upper bound of oscillatory brightness of V603~Aql reasonably
overlaps that of V1500~Cyg.  Therefore, we fit our model light curves to
the upper bound of oscillatory brightness during the transition phase
of V603~Aql.  Again note that our model light curve reasonably fits with
the early $V$ light curve but deviates from the visual observation
in the later nebular phase.

To summarize, we are able to reproduce the absolute brightness
for the distance modulus of $(m-M)_{V, \rm V603~Aql}=7.2$
and $M_{\rm WD}=1.20~M_\sun$.  Photospheric emission does not
contribute to the total $V$ flux as shown in the Figure
\ref{v603_aql_v1500_cyg_x55z02o10ne03_logscale_2fig}(b).


\begin{figure}
\epsscale{1.15}
\plotone{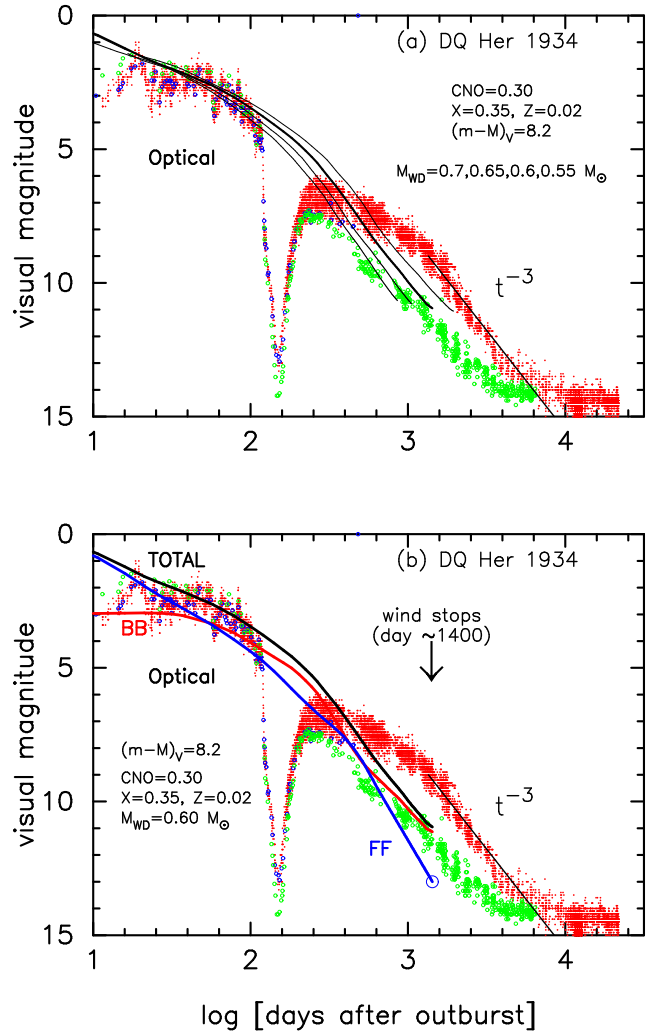}
\caption{
Same as Figure \ref{gk_per_v1500_cyg_x55z02o10ne03_logscale_fig2},
but for DQ~Her.  Visual magnitudes of DQ~Her are taken from AAVSO
(red small dots) and \citet{gap56} (blue small open circles),
and photographic magnitudes (green small open circles) are from \citet{gap56}.
(a) Assuming the chemical composition of the envelope ``CO nova 2,''
we plot four total $V$ flux light curves for the WD mass models of 0.55, 0.6,
0.65, and $0.7~M_\sun$. 
(b) We plot three $V$ model light curves of photospheric (``BB''),
free-free (``FF''), and total flux (``TOTAL) for the
$0.6~M_\sun$ WD.  The photospheric emission significantly
contributes to the total $V$ flux in contrast to the fast novae
GK~Per and V603~Aql.
\label{all_mass_dq_her_x35z02c10o20_m060_logscale_bb_ff_total_2fig}}
\end{figure}

\subsubsection{DQ~Her 1934}
\label{discussion_dq_her}
The trigonometric parallax distance of DQ~Her is $d=386^{+33}_{-29}$~pc
\citep{har13}.
Adopting $A_V=3.1\times E(B-V)=3.1\times 0.1=0.31$ \citep{ver87}, we obtain
the distance modulus in $V$ band as $(m-M)_V=A_V+5\log(d/{\rm10~pc})
=0.31 + 5 \log (386^{+33}_{-29}/ 10)= 8.24\pm0.18$.
Thus, we adopt $(m-M)_{V,\rm DQ~Her} = 8.2$.
The WD mass of DQ~Her was obtained by \citet{hor93}
to be $M_{\rm WD}=0.60\pm0.07~M_\sun$.
The chemical composition of ejecta was estimated by \citet{pet90} and
\citet{wil78} as listed in Table \ref{pw_vul_chemical_abundance}.
Here, we adopt the chemical composition of ``CO nova 2'' because
the averaged value is $X=0.31$ and close to that of ``CO nova 2.''

For the chemical composition of ``CO nova 2,''  the absolute magnitudes
of free-free emission model light curves were already determined 
in Table 2 of \citet{hac10k}.  So we calculated the total 
(free-free plus photospheric) $V$ flux light curves
for the WD masses of 0.55, 0.60, 0.65, and $0.70~M_\sun$
and plotted them in Figure
\ref{all_mass_dq_her_x35z02c10o20_m060_logscale_bb_ff_total_2fig}(a).  
It is remarkable that all the four $V$ light curves fit reasonably
to the observed visual magnitudes, at least, in the early phase
before the dust blackout started.
Therefore, we cannot select a best one among these four light curves.
However, this again confirms that the absolute brightness of
our model light curves are reasonable at least in the early decline
phase before the dust blackout.

In Figure
\ref{all_mass_dq_her_x35z02c10o20_m060_logscale_bb_ff_total_2fig}(b),
we adopt $M_{\rm WD}=0.60~M_\sun$ from the central value 
estimated by \citet{hor93} and
plot our total $V$ flux, free-free, and photospheric blackbody
light curves for the $0.6~M_\sun$ WD.
The photospheric emission (red solid line labeled ``BB'')
significantly contribute to the total $V$ flux (black solid line labeled 
``TOTAL'').   If we do not include the photospheric
emission, our free-free model $V$ flux (blue solid line labeled ``FF'')
does not fit to the observed one. 
This good agreement with the observed brightness suggests that
photospheric emission is necessary to reproduce the light curves
of slow novae like DQ~Her as well as free-free emission.

To summarize, our model light curves of total $V$ flux reasonably
reproduce the absolute brightnesses of optical light curves of novae
with known distances, at least, in the early phase
before the nebular phase or dust blackout starts.
For slower novae, photospheric emission dominates the spectrum in $V$ band.
For faster novae, on the other hand, free-free emission 
dominates the spectrum in $V$ band and therefore fast novae
follow the universal decline law.


\begin{figure*}
\epsscale{0.65}
\plotone{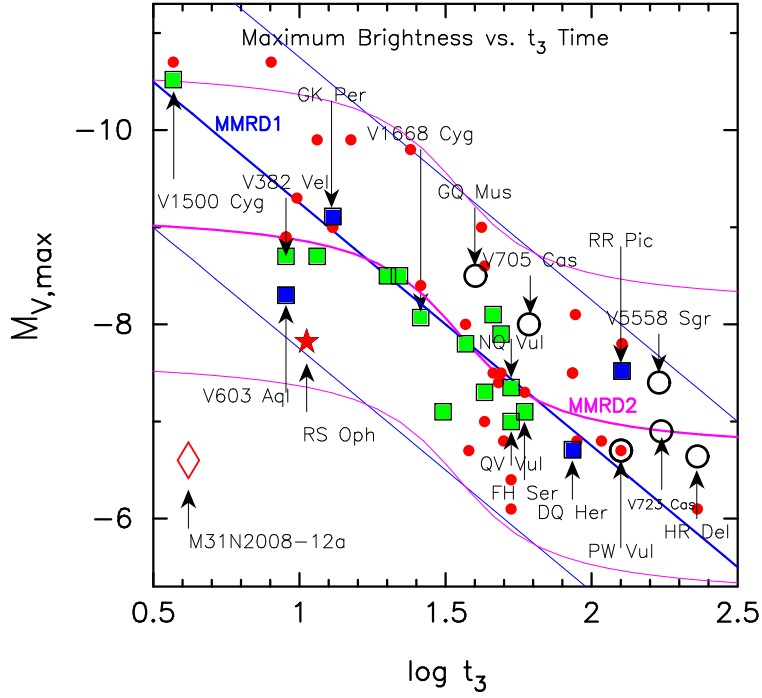}
\caption{
Maximum magnitude vs. rate of decline (MMRD) relation.
Large open black circles: six novae studied in the present work,
PW~Vul, V705~Cas, GQ~Mus, V5558~Sgr, HR~Del, and V723~Cas.
Red filled circles: individual novae taken from Table 5 of \citet{dow00}.
Blue filled squares: four novae calibrated with the annual parallax
method, taken from \citet{har13}.
Green filled squares: novae calibrated with the time-stretching method,
taken from \citet{hac14k}.
Blue solid line labeled ``MMRD1:'' Kaler-Schmidt law \citep{sch57}
and its $\pm1.5$~mag lines
(see Equation (\ref{kaler-schmidt-law})).
Magenta solid line labeled ``MMRD2:''
Della Valle \& Livio's law \citep{del95} and its $\pm1.5$~mag lines
(see Equation (\ref{della-valle-livio-law})).
Red filled star: the recurrent nova RS Oph as
an example for a very high mass accretion rate
and very short timescale $t_3$ \citep{hac14k}.
Red open diamond: the 1~yr recurrence period M31 nova, M31N2008-12a,
taken from \citet{tan14}, i.e., $M_{g, \rm max}=-6.6$ (maximum in $g$-band)
and $t_{3,g}\approx4.2$ days (measured in the $g$-band light curve). 
See text for more details.
\label{max_t3_point_B_scale_pw_vul_no3}}
\end{figure*}

\subsection{Do Slow Novae Follow the MMRD Relation?}
\label{discussion_mmrd}

In this subsection we discuss whether or not slow novae follow the MMRD
relation.   Theoretical free-free emission light curves of novae
clearly shows a trend that a more massive WD has a brighter maximum
magnitude (smaller $M_{V, {\rm max}}$) and a faster decline rate
(smaller $t_3$ time).  The relation between $t_3$ and 
$M_{V, {\rm max}}$ for novae is called
``Maximum Magnitude vs. Rate of Decline'' (MMRD) relation.

Figure \ref{max_t3_point_B_scale_pw_vul_no3} shows observed data
points of $(t_3,M_{V,\rm max})$ for many classical/recurrent novae.
A blue solid line flanking with $\pm1.5$~mag lines indicates
the relation of ``Kaler-Schmidt's law'' 
\citep[labeled ``MMRD1,'' see ][]{sch57},
which is Equation (\ref{kaler-schmidt-law}). 
A magenta solid line flanking with $\pm1.5$~mag lines indicates
the relation of ``della Valle-Livio's law'' 
\citep[labeled ``MMRD2,'' see ][]{del95},
which is Equation (\ref{della-valle-livio-law}). 
Red filled circles are novae taken from \citet{dow00}, the distances
of which were mainly derived from the nebular expansion parallax method.
We show the six novae studied in the present work
by large black open circles, i.e., PW~Vul, V705~Cas, GQ~Mus, V5558~Sgr,
HR~Del, and V723~Cas.  Green filled squares are novae taken from
\citet{hac10k}, the distance moduli of which are based
on the time-stretching method.  Blue filled squares indicate the four
novae, V603~Aql, DQ~Her, GK~Per, and RR~Pic, the distances of which were
determined by \citet{har13} with {\it HST} annual parallaxes.

\citet{hac06kb} found that nova light curves follow a universal
decline law when free-free emission dominates the continuum spectrum in 
optical and NIR regions.  Using this property, \citet{hac10k} found that,
if two nova light curves overlap each other after one of
the two is squeezed/stretched by a factor
of $f_s$ ($t'=t/f_s$) in the time direction, the brightnesses of the
two novae obey the relation of $m'_V = m_V - 2.5 \log f_s$,
which is the same as Equation (\ref{simple_final_scaling_flux}).
Based on this property, they derived a MMRD relation of
$M_{V, \rm max}= 2.5\log t_3 - 11.6$ for the chemical composition of
``CO nova 2,'' i.e., Equation (35) in \citet{hac10k}. 
We can again derive a similar trend, $M_{V, \rm max}= 2.5\log t_3 - 11.65$,
i.e., Equation (\ref{theoretical_MMRD_relation}), for another chemical
composition, ``CO nova 4,'' in Appendix \ref{absolute_magnitudes}.
Note that these two relations are in good agreement with Kaler-Schmidt's law.
The main trend of the MMRD relation is governed by the WD mass:
the more massive a WD is, the steeper the decline of a nova light curve is.
\citet{hac10k} further showed that the maximum brightness of a nova 
also depends on the initial envelope mass (ignition mass).
This initial envelope mass depends on the mass-accretion rate to the WD
\citep[see, e.g.,][for a recent result]{kat14shn}.
\citet{hac10k} concluded that the scatter of individual MMRD
points is due to various mass-accretion rates to the WD even for
the same WD mass.  The brighter the nova is,
the smaller the mass accretion rate to the WD is
\citep[see Figure 15 of][]{hac10k}.

\citet{har13} concluded that DQ~Her and GK~Per almost follow
the MMRD relation (MMRD1) but V603~Aql and RR~Pic do not
(see Figure \ref{max_t3_point_B_scale_pw_vul_no3}).
We examine the reason for about 1~mag faintness of V603~Aql 
compared with the blue solid line (MMRD1 relation).
\citet{hac14k} analyzed the light curves of V1500~Cyg and V603~Aql
and concluded that both of these novae harbor a $\sim1.2~M_\sun$ WD
for the envelope chemical composition of ``Ne nova 2.''
We estimated the initial envelope masses for these two novae,
that is, $M_{\rm env}=0.92\times10^{-5}~M_\sun$ for the $1.2~M_\sun$ WD
model of V1500~Cyg and $M_{\rm env}= 0.47\times10^{-5}~M_\sun$
for V603~Aql.  The initial envelope mass
corresponds to the envelope mass at optical maximum
(see Figure \ref{v603_aql_v1500_cyg_x55z02o10ne03_logscale_2fig}).
Thus, the difference in the ignition masses makes the 
apparent difference in the start time of the outbursts and
in the peak brightness as shown in Figure 
\ref{v603_aql_v1500_cyg_x55z02o10ne03_logscale_2fig}.
This is the main source for scatter of individual
MMRD points around the proposed MMRD relation.
Therefore, a smaller initial envelope mass is the main reason
that V603~Aql is fainter by $\sim1.2$ mag than the blue solid line
of MMRD1.
 
Now we first examine the case of PW~Vul among the seven novae
studied in the present work.
The MMRD point of PW~Vul locates slightly (0.2 mag) above 
Kaler-Schmidt's law (MMRD1), a blue thick solid line in Figure 
\ref{max_t3_point_B_scale_pw_vul_no3}.
Remember that our free-free emission model light curves 
usually follow the averaged MMRD relation of MMRD1
(Kaler \& Schmidt's law).  Thus, the agreement of PW~Vul with
the MMRD1 relation indicates that the initial envelope mass of PW~Vul
was a typical one for $\sim0.83~M_\sun$ WDs and also that
free-free emission dominates the spectrum in $V$ band
(see Figure \ref{all_mass_pw_vul_x55z02c10o10_vrijhk_composite}).

The next is V705~Cas.  This MMRD point is much ($\sim0.8$ mag) 
brighter than both of the MMRD1 and MMRD2 relations. 
The photospheric emission is not as much to make 
the $t_3$ time longer by a factor of 2.0 (because $2.5\Delta\log t_3=
2.5 \log 2.0\approx2.5\times0.3\approx0.8$), as clearly shown in Figure
\ref{all_mass_v705_cas_x55z02c10o10_composite}.  Therefore, we attribute
the difference in the maximum brightness to the difference in the initial
envelope mass.  To confirm this, we compared 
the rise time of UV~1455\AA\  flux between PW~Vul and V705~Cas
in Figures \ref{all_mass_pw_vul_x55z02c10o10_vrijhk_composite}
and \ref{all_mass_v705_cas_x55z02c10o10_composite}.
The UV~1455\AA\  flux had already risen at the optical maximum
in PW~Vul while it had not yet in V705~Cas.
This can be seen more easily in Figure
\ref{v705_cas_pw_vul_v_bv_ub_color_curve_logscale}.
The longer time before UV~1455\AA\  peak means that the envelope mass
is more massive in V705~Cas than in PW~Vul.  
The more massive initial envelope mass makes 
the brighter optical maximum of V705~Cas.  Thus, it locates
above the blue solid line of MMRD1 relation.   

The third object is GQ~Mus.  The MMRD point of this object is also
much ($\sim0.8$~mag) above both the MMRD1 (blue solid line)
and MMRD2 (magenta solid line).  GQ~Mus shows a bump of $\sim 1$ mag
brighter than the model light curve in the very early phase.
This feature is very similar to that of V1500~Cyg.
One could suppose that this bump is the origin of the deviation
from the blue solid line of MMRD1 relation, but V1500~Cyg does not show
such a large deviation (see Figure \ref{max_t3_point_B_scale_pw_vul_no3}).
Therefore, we suppose that photospheric emission
is the main source of the deviation.   We found, from Figure
\ref{all_mass_gq_mus_x35z02c10o20_x55z02c10o10_combine_bb}(a),
that the $t_3$ time of the total flux (black solid line)
is 2.3 times longer compared with the case of free-free only
(blue solid line).  This effect makes the $t_3$ time longer
by about $\Delta \log t_3 = \log 2.3 \approx 0.35$.
This corresponds to the increase in the brightness
by $2.5 \Delta \log t_3 = 2.5\times 0.35=0.9$ mag in the MMRD diagram,
being roughly consistent with the present position of GQ~Mus in
Figure \ref{max_t3_point_B_scale_pw_vul_no3}.
Thus, we conclude that GQ~Mus is above the MMRD relation because
of photospheric emission effect.

The remaining objects are static-to-wind transition novae,
RR~Pic, V5558~Sgr, HR~Del, and V723~Cas.
These MMRD points are also much (0.8--1.2 mag)
above the blue solid line of MMRD1 but HR~Del and V723~Cas are
consistent with the MMRD2.  From Figure 
\ref{all_mass_v723_cas_x55z02c10o10_composite_m51}
of V723~Cas, we see that the $t_3$ time estimated along our total $V$
flux model (black solid line) is $\sim2.2$ times longer than 
the case of free-free only (blue solid line).  
Note again that our free-free emission model light curves 
usually follow the averaged MMRD relation of MMRD1
(Kaler \& Schmidt's law). 
The photospheric emission effect raises the brightness by
$2.5 \Delta \log t_3 = 2.5 \log 2.2 \approx 0.8$~mag compared with
the MMRD1 (blue solid line), being roughly consistent with the position
of V723~Cas in Figure \ref{max_t3_point_B_scale_pw_vul_no3}.  Similarly
for HR~Del, we obtain, from Figure 
\ref{all_mass_hr_del_x55z02c10o10_composite_2fig_no2}(a),
a factor of $\sim2.2$ and a raise of 
$2.5 \Delta \log t_3 = 2.5 \log 2.2 \approx 0.8$~mag
owing to the photospheric emission effect.
This is also consistent with the position of HR~Del in
Figure \ref{max_t3_point_B_scale_pw_vul_no3}.
V5558~Sgr and RR~Pic showed much brighter ($\sim 1.2$ mag)
optical maxima than the MMRD1 (blue solid line).
These two novae showed prominent amplitudes of oscillations
during the multiple-peak.  We think that these large peaks
are related to more massive envelopes compared with those
of HR~Del and V723~Cas.  In fact, the amplitude of multiple-peak is
decreasing in V5558~Sgr, suggesting reduction of the envelope mass due to
mass loss.  The same explanation is possible in RR~Pic,
whose MMRD point is also 1.2 mag brighter than the MMRD1 relation (blue
solid line).

It is interesting to see the position of the recurrent nova RS Oph
(red filled star) and the 1 yr recurrence period M31 nova, M31N2008-12a
(red open diamond) in Figure \ref{max_t3_point_B_scale_pw_vul_no3}.
RS~Oph locates 1.2 mag below the MMRD1.
This faintness corresponds to a much smaller envelope mass 
at optical maximum, suggesting a massive WD and very high
mass accretion rate.  This situation is very consistent with
the total picture of recurrent novae; a very massive WD close to the
Chandrasekhar mass and a high mass accretion rate
to the WD \citep[e.g.,][]{hac01kb}.  In this figure, we adopt
$M_{V, {\rm max}} = -7.8$ and $t_3 = 10.5$ days with
the distance of $d= 1.4$ kpc \citep{hac06b, bar06, hac14k}, 
absorption of $A_V = 3.1 E(B-V) = 3.1 \times 0.65 = 2.0$
\citep{hac14k}, $m_{V, {\rm max}}= 5.0$ \citep{ros87},
and $t_3 = 10.5$ days from optical light curve fitting with
our free-free model light curves \citep{hac01kb, hac06b, hac07kl}.
The 1 yr recurrence period M31 nova, M31N2008-12a is depicted by
a red open diamond.  It is very faint, i.e., $M_{g, \rm max}=-6.6$ 
(maximum in $g$-band) and $t_{3,g}\approx4.2$ days 
(measured in the $g$-band light curve), taken from \citet{tan14}. 
The 1 yr recurrence period is close to the shortest recurrence period
of novae, suggesting a very massive
WD close to the Chandrasekhar mass and a very high accretion rate
\citep[see, e.g.,][]{tan14, kat14shn}, thus a very small envelope mass.
These support our conclusion that the peak brightness of a nova depends on
the initial envelope mass as well as the WD mass itself.

To summarize,
the primary parameter of the MMRD relation is the WD mass and the
secondary parameter is the initial envelope mass.  Variations in the initial
envelope mass is the origin of scatter from the averaged MMRD relation,
MMRD1.  More massive envelopes correspond to the region
above the MMRD1 line and less massive envelopes correspond to
the region below the MMRD1 line.
Photospheric emission is the third factor of the MMRD relation
but becomes more important in slow novae,
because it makes $t_3$ time longer in low mass WDs.




\begin{deluxetable*}{lllllllllllllll}
\tabletypesize{\scriptsize}
\tablecaption{Light Curves of CO Novae\tablenotemark{a}
\label{light_curves_of_novae_co4}}
\tablewidth{0pt}
\tablehead{
\colhead{$m_{\rm ff}$} &
\colhead{0.55$M_\sun$} &
\colhead{0.6$M_\sun$} &
\colhead{0.65$M_\sun$} &
\colhead{0.7$M_\sun$} &
\colhead{0.75$M_\sun$} &
\colhead{0.8$M_\sun$} &
\colhead{0.85$M_\sun$} &
\colhead{0.9$M_\sun$} &
\colhead{0.95$M_\sun$} &
\colhead{1.0$M_\sun$} &
\colhead{1.05$M_\sun$} &
\colhead{1.1$M_\sun$} &
\colhead{1.15$M_\sun$} &
\colhead{1.2$M_\sun$} \\
\colhead{(mag)} &
\colhead{(day)} &
\colhead{(day)} &
\colhead{(day)} &
\colhead{(day)} &
\colhead{(day)} &
\colhead{(day)} &
\colhead{(day)} &
\colhead{(day)} &
\colhead{(day)} &
\colhead{(day)} &
\colhead{(day)} &
\colhead{(day)} &
\colhead{(day)} &
\colhead{(day)} 
}
\startdata
  3.000     & 0.0 & 0.0 & 0.0 & 0.0 & 0.0 & 0.0 & 0.0 & 0.0 & 0.0 & 0.0 & 0.0 & 0.0 & 0.0 & 0.0 \\

  3.250     &  3.429     &  2.630     &  2.591     &  2.210     &  2.090     &  1.400     &  1.153     &  1.060     & 0.960     & 0.859     & 0.761     & 0.689     & 0.621     & 0.566     \\
  3.500     &  9.489     &  7.150     &  5.251     &  4.480     &  4.200     &  2.810     &  2.399     &  2.130     &  1.920     &  1.735     &  1.505     &  1.372     &  1.244     &  1.125     \\
  3.750     &  16.73     &  11.92     &  10.28     &  6.990     &  6.360     &  4.500     &  3.686     &  3.220     &  2.890     &  2.605     &  2.263     &  2.033     &  1.845     &  1.685     \\
  4.000     &  25.84     &  18.62     &  15.93     &  10.22     &  8.600     &  6.560     &  5.106     &  4.370     &  3.890     &  3.485     &  3.035     &  2.706     &  2.449     &  2.243     \\
  4.250     &  35.20     &  26.12     &  21.82     &  14.76     &  11.43     &  8.680     &  6.586     &  5.580     &  5.010     &  4.375     &  3.822     &  3.392     &  3.071     &  2.811     \\
  4.500     &  44.97     &  33.92     &  27.89     &  20.55     &  15.39     &  10.84     &  8.296     &  6.830     &  6.200     &  5.345     &  4.625     &  4.102     &  3.732     &  3.433     \\
  4.750     &  56.92     &  41.91     &  34.15     &  26.12     &  19.57     &  13.21     &  10.14     &  8.120     &  7.440     &  6.545     &  5.574     &  4.831     &  4.411     &  4.082     \\
  5.000     &  73.18     &  53.71     &  41.54     &  32.00     &  23.94     &  16.31     &  12.77     &  10.04     &  8.960     &  7.785     &  6.754     &  5.783     &  5.216     &  4.770     \\
  5.250     &  93.75     &  67.24     &  51.78     &  38.17     &  28.92     &  19.55     &  15.39     &  12.16     &  10.73     &  9.085     &  7.984     &  6.923     &  6.235     &  5.656     \\
  5.500     &  117.3     &  82.44     &  62.96     &  44.79     &  34.27     &  23.01     &  17.83     &  14.32     &  12.62     &  10.44     &  8.994     &  7.933     &  7.155     &  6.536     \\
  5.750     &  143.2     &  100.6     &  76.05     &  54.33     &  41.01     &  27.27     &  20.39     &  16.29     &  14.36     &  11.90     &  10.06     &  8.753     &  7.828     &  7.115     \\
  6.000     &  169.1     &  119.9     &  90.21     &  64.79     &  48.34     &  31.96     &  23.29     &  18.37     &  16.22     &  13.43     &  11.21     &  9.623     &  8.538     &  7.732     \\
  6.250     &  196.2     &  138.8     &  105.5     &  76.07     &  56.51     &  37.00     &  27.03     &  20.66     &  18.27     &  15.09     &  12.53     &  10.61     &  9.349     &  8.392     \\
  6.500     &  225.9     &  159.4     &  122.1     &  88.03     &  65.54     &  42.74     &  31.12     &  23.62     &  20.77     &  16.85     &  13.94     &  11.71     &  10.28     &  9.142     \\
  6.750     &  259.4     &  181.6     &  138.6     &  100.2     &  75.15     &  49.14     &  35.57     &  26.89     &  23.51     &  18.93     &  15.51     &  12.90     &  11.30     &  9.962     \\
  7.000     &  297.1     &  205.6     &  156.3     &  113.5     &  84.87     &  56.14     &  40.49     &  30.41     &  26.51     &  21.22     &  17.29     &  14.26     &  12.36     &  10.80     \\
  7.250     &  340.5     &  232.5     &  175.9     &  127.7     &  95.18     &  63.63     &  45.86     &  34.25     &  29.98     &  23.67     &  19.22     &  15.75     &  13.54     &  11.71     \\
  7.500     &  393.1     &  264.4     &  199.4     &  143.1     &  106.3     &  71.08     &  51.74     &  38.45     &  33.37     &  26.21     &  21.15     &  17.30     &  14.76     &  12.70     \\
  7.750     &  455.2     &  300.8     &  225.9     &  160.7     &  120.4     &  79.16     &  57.70     &  42.99     &  36.91     &  29.00     &  23.22     &  18.82     &  15.97     &  13.63     \\
  8.000     &  519.0     &  345.8     &  255.8     &  182.1     &  135.9     &  88.73     &  64.06     &  47.67     &  40.84     &  32.03     &  25.48     &  20.47     &  17.28     &  14.64     \\
  8.250     &  592.2     &  394.2     &  290.9     &  206.4     &  153.7     &  99.40     &  71.42     &  52.84     &  45.17     &  35.39     &  27.97     &  22.27     &  18.72     &  15.75     \\
  8.500     &  656.8     &  444.5     &  331.9     &  235.5     &  175.4     &  112.0     &  80.12     &  58.82     &  50.36     &  39.11     &  30.89     &  24.42     &  20.40     &  17.03     \\
  8.750     &  730.1     &  499.8     &  370.0     &  267.3     &  199.2     &  127.4     &  89.99     &  65.73     &  56.68     &  43.55     &  34.11     &  26.93     &  22.36     &  18.63     \\
  9.000     &  804.7     &  551.1     &  403.0     &  299.8     &  224.8     &  144.9     &  102.5     &  74.22     &  64.08     &  48.99     &  38.34     &  29.98     &  24.81     &  20.53     \\
  9.250     &  866.8     &  600.1     &  440.8     &  331.6     &  253.0     &  163.6     &  116.7     &  84.49     &  73.31     &  55.38     &  43.03     &  33.63     &  27.83     &  22.92     \\
  9.500     &  935.9     &  647.3     &  484.0     &  367.3     &  279.2     &  185.1     &  131.6     &  96.54     &  83.61     &  63.32     &  49.08     &  38.00     &  31.37     &  25.68     \\
  9.750     &  1000.     &  700.4     &  533.6     &  399.8     &  301.6     &  208.1     &  148.5     &  108.9     &  93.88     &  72.14     &  55.91     &  43.18     &  35.47     &  28.95     \\
  10.00     &  1059.     &  750.2     &  570.9     &  429.1     &  322.4     &  227.8     &  167.2     &  122.6     &  105.7     &  81.11     &  62.98     &  48.79     &  39.93     &  32.54     \\
  10.25     &  1123.     &  795.2     &  605.3     &  460.9     &  342.1     &  246.8     &  179.6     &  137.8     &  117.2     &  91.19     &  70.78     &  54.70     &  44.67     &  36.30     \\
  10.50     &  1190.     &  842.8     &  641.7     &  488.9     &  363.1     &  267.1     &  193.5     &  150.2     &  125.0     &  101.7     &  79.43     &  61.21     &  49.89     &  40.43     \\
  10.75     &  1261.     &  893.3     &  680.3     &  518.7     &  385.2     &  283.6     &  208.8     &  162.4     &  133.2     &  111.3     &  86.87     &  67.75     &  54.85     &  44.76     \\
  11.00     &  1336.     &  946.7     &  721.2     &  550.2     &  408.7     &  301.0     &  225.9     &  175.6     &  141.9     &  118.4     &  94.26     &  73.77     &  59.08     &  48.84     \\
  11.25     &  1415.     &  1003.     &  764.5     &  583.5     &  433.6     &  319.5     &  244.7     &  187.1     &  151.2     &  126.0     &  100.1     &  79.34     &  63.62     &  52.65     \\
  11.50     &  1500.     &  1063.     &  810.4     &  618.9     &  459.9     &  339.0     &  259.8     &  199.1     &  160.9     &  134.0     &  106.4     &  85.27     &  68.48     &  56.40     \\
  11.75     &  1589.     &  1127.     &  859.0     &  656.3     &  487.8     &  359.7     &  275.6     &  211.7     &  171.3     &  142.4     &  113.0     &  90.78     &  73.68     &  60.10     \\
  12.00     &  1684.     &  1194.     &  910.5     &  696.0     &  517.4     &  381.7     &  292.4     &  225.1     &  182.3     &  151.4     &  120.0     &  96.66     &  79.07     &  63.80     \\
  12.25     &  1784.     &  1265.     &  965.0     &  738.0     &  548.7     &  405.0     &  310.1     &  239.3     &  193.9     &  160.9     &  127.4     &  102.6     &  83.82     &  67.73     \\
  12.50     &  1890.     &  1341.     &  1022.     &  782.5     &  581.9     &  429.6     &  329.0     &  254.3     &  206.2     &  170.9     &  135.3     &  108.9     &  88.86     &  71.88     \\
  12.75     &  2002.     &  1421.     &  1084.     &  829.7     &  617.0     &  455.7     &  348.9     &  270.2     &  219.3     &  181.6     &  143.6     &  115.5     &  94.19     &  76.28     \\
  13.00     &  2122.     &  1505.     &  1148.     &  879.6     &  654.3     &  483.3     &  370.0     &  287.0     &  233.1     &  192.9     &  152.5     &  122.6     &  99.88     &  80.94     \\
  13.25     &  2248.     &  1595.     &  1217.     &  932.5     &  693.7     &  512.5     &  392.4     &  304.9     &  247.7     &  204.8     &  161.8     &  130.0     &  105.8     &  85.88     \\
  13.50     &  2381.     &  1690.     &  1290.     &  988.0     &  735.4     &  543.5     &  416.0     &  323.8     &  263.2     &  217.5     &  171.7     &  137.9     &  112.2     &  91.11     \\
  13.75     &  2523.     &  1791.     &  1367.     &  1048.     &  779.7     &  576.4     &  441.1     &  343.8     &  279.6     &  230.9     &  182.2     &  146.3     &  118.9     &  96.64     \\
  14.00     &  2673.     &  1898.     &  1448.     &  1111.     &  826.5     &  611.2     &  467.7     &  365.0     &  297.0     &  245.1     &  193.3     &  155.2     &  126.0     &  102.5     \\
  14.25     &  2832.     &  2010.     &  1535.     &  1177.     &  876.2     &  648.0     &  495.9     &  387.5     &  315.4     &  260.1     &  205.1     &  164.6     &  133.5     &  108.7     \\
  14.50     &  3000.     &  2130.     &  1626.     &  1248.     &  928.8     &  687.0     &  525.7     &  411.2     &  334.9     &  276.1     &  217.5     &  174.5     &  141.5     &  115.3     \\
  14.75     &  3178.     &  2257.     &  1723.     &  1323.     &  984.4     &  728.4     &  557.3     &  436.4     &  355.6     &  292.9     &  230.7     &  185.1     &  150.0     &  122.3     \\
  15.00     &  3367.     &  2391.     &  1826.     &  1402.     &  1043.     &  772.2     &  590.8     &  463.1     &  377.5     &  310.8     &  244.7     &  196.3     &  158.9     &  129.7     \\

\hline
X-ray\tablenotemark{b}
 & 8210  & 6150 & 4700  & 3650 & 2640   & 1900   & 1370   & 980   & 730   & 540   & 370   & 250  & 169 & 112 \\
\hline
$\log f_{\rm s}$\tablenotemark{c}
 & 0.60  & 0.47 & 0.39  & 0.29 & 0.17   & 0.06   & $-0.05$   & $-0.15$   & $-0.24$   & $-0.33$   & $-0.43$   & $-0.55$  & $-0.67$ & $-0.77$ \\
\hline
$M_{\rm w}$\tablenotemark{d}
 & 5.5 & 5.1  & 4.6 & 4.2   & 3.8   & 3.3   & 2.9   & 2.5   & 2.2   & 1.8   & 1.5  & 1.2 & 0.9 & 0.7 
\enddata
\tablenotetext{a}{chemical composition of
the envelope is assumed
to be that of ``CO nova 4'' in Table \ref{chemical_composition}.}
\tablenotetext{b}{duration of supersoft X-ray phase in units of days.}
\tablenotetext{c}{stretching factor against the $0.83~M_\sun$ model
which is the best fit light curve for the PW Vul
 UV 1455\AA\  observation
in Figure \ref{all_mass_pw_vul_x55z02c10o10_vrijhk_universal_no2}.}
\tablenotetext{d}{absolute magnitudes at the bottom point in Figure
\ref{all_mass_pw_vul_x55z02c10o10_vrijhk_absolute_mag} by assuming
$(m-M)_V = 13.0$  (PW Vul).}
\end{deluxetable*}

\section{Conclusions}
\label{conclusions}
     There have been suggested several scaling laws for classical nova
light curves \citep[e.g.,][for a summary]{hac08kc}.  
\citet{hac06kb} proposed that classical nova light curves follow
a universal shape when continuum flux is dominated by free-free emission.
Using this property, \citet{hac10k} theoretically explained the main trend
of the MMRD relations.  These results were confirmed only for fast novae.
In this paper, we examined seven novae of slow evolution, in which
photospheric emission could considerably contribute to the continuum
spectra in $V$ band rather than free-free emission.
We obtain the following main results:

\noindent
{\bf 1.} Based on various observational estimates in literature,
we estimated physical parameters of the slow nova PW~Vul.
We adopted the distance modulus of $(m-M)_V=13.0$ in $V$ band,
extinction of $E(B-V)=0.55$, and distance of $d=1.8$~kpc for PW~Vul.  

\noindent
{\bf 2.} We divide approximately a nova spectrum into two components, one
is photospheric and the other is optically thick free-free emission.
During the optically thick wind phase of the slow nova PW~Vul,
free-free emission dominates the continuum spectrum in NIR bands
while photospheric emission contributes, to some extent,
to the continuum spectrum in $V$ band. 

\noindent
{\bf 3.} We calculated the total $V$ model light curves
(the sum of free-free plus photospheric emission) 
of classical novae for the chemical composition of $X= 0.55$, $Y=0.23$,
$Z=0.02$, and $X_{\rm CNO}=0.20$ (``CO nova 4''), which is close to
that of PW~Vul.  By simultaneous fitting of the total $V$ light curve model
and the blackbody UV~1455\AA\  light curve model, we determine the WD mass
of PW~Vul to be $\sim 0.83~M_\sun$.

\noindent
{\bf 4.} Using the distance modulus of $(m-M)_V=13.0$ for PW~Vul 
and properties of the universal decline law,
we determined the absolute magnitudes of free-free emission light curves
for various WD masses with the envelope chemical composition of ``CO nova 4.''
Based on the universal decline law, we also derived the MMRD relations
for ``CO nova 4.''  This theoretical MMRD relations 
are consistent with the ever proposed empirical formulae (Appendix
\ref{absolute_magnitudes}).

\noindent
{\bf 5.}  We also analyzed the moderately-fast nova V705~Cas and
estimated the WD mass to be $\sim 0.78~M_\sun$.  Even for this 
WD mass, we found that free-free emission still dominates the continuum
spectrum in $V$ and NIR bands.  We obtained the distance modulus of
$(m-M)_V=13.4$ and the color excess of $E(B-V)=0.45$,
which are consistent with those obtained from other observations.

\noindent
{\bf 6.}  We reanalyzed the fast nova GQ~Mus.  Fitting our model 
light curves with optical $V$, UV~1455\AA, and supersoft X-ray 
light curve observations, we confirmed that the WD mass is 
$\sim 0.65~M_\sun$ for an assumed chemical composition of $X= 0.35$, 
$Y=0.33$, $Z=0.02$, and $X_{\rm CNO}=0.30$ (``CO nova 2'').
For this low WD mass, we found that photospheric emission is more
important and dominates the continuum spectrum in $V$ band.
We consistently obtained the distance modulus of $(m-M)_V=15.7$,
color excess of $E(B-V)=0.45$, and distance of $d=7.3$~kpc.

\noindent
{\bf 7.}  We further analyzed four very slow novae, RR~Pic, V5558~Sgr,
HR~Del, and V723~Cas, and estimated their WD masses as low as 
$\sim0.5$--$0.55~M_\sun$.  
We also consistently obtained the distance moduli, color excesses,
and distances of V5558~Sgr, HR~Del, and V723~Cas
as $(m-M)_V=13.9$, 10.4, and 14.0, $E(B-V)=0.7$, 0.15, and 0.35, 
and $d=2.2$, 0.97, and 3.9~kpc, respectively.  
We found that, in optical $V$ band, photospheric emission is more important 
than free-free emission in these four novae.

\noindent
{\bf 8.}  We confirmed that our total $V$ flux model light curves
reasonably reproduce the absolute brightnesses of four novae with
known distances, i.e., RR~Pic, GK~Per, V603~Aql, and DQ~Her.
We found that free-free emission dominates the spectra in $V$ band
for the fast novae GK~Per and V603~Aql but photospheric emission
significantly contributes to the total $V$ flux for the slow novae
RR~Pic and DQ~Her.

\noindent
{\bf 9.}  The four very slow novae, RR~Pic, V5558~Sgr, HR~Del, and V723~Cas
lie about 0.8--1.3~mag above Kaler-Schmidt's MMRD relation.
In these novae, photospheric emission dominates the continuum spectra
in $V$ band and makes their $t_3$ times much longer ($\sim 2$ times)
than that of free-free emission only.  Because the model light curves
of free-free emission follow Kaler-Schmidt's MMRD relation,
the total (free-free plus photospheric) flux light curves 
raise the MMRD brightness by $2.5\log 2\approx 0.8$~mag. 
This is the reason that the MMRD points of these novae
are about 1~mag brighter than Kaler-Schmidt's MMRD relation.

\acknowledgments
     We are grateful to Angelo Cassatella for fruitful discussion
and critical reading of the manuscript.
We also thank 
the American Association of Variable Star Observers (AAVSO)
and Variable Star Observers League of Japan (VSOLJ)
for the archival data of PW~Vul, V705~Cas, GQ~Mus, 
RR~Pic, V5558~Sgr, HR~Del, and V723~Cas.
This research has been supported in part by the Grant-in-Aid for
Scientific Research (22540254, 24540227) 
of the Japan Society for the Promotion of Science.



\appendix


\begin{figure*}
\epsscale{0.75}
\plotone{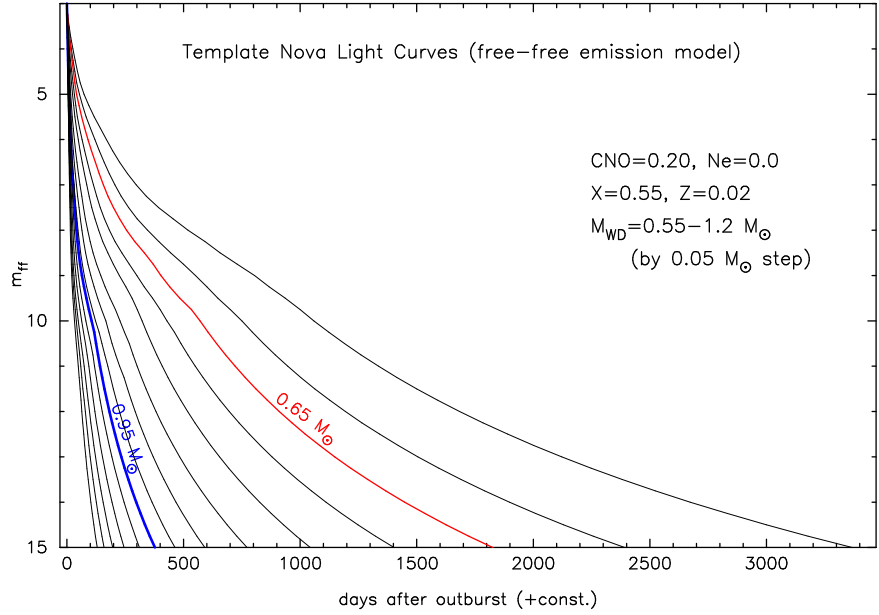}
\caption{
Magnitudes of our free-free emission model light curves 
for WD masses of $0.55 - 1.2~M_\sun$ in $0.05~M_\sun$ steps,
numerical data of which are tabulated
in Table \ref{light_curves_of_novae_co4}.
We adopt the chemical composition ``CO nova 4'' in Table
\ref{chemical_composition} for these WD envelopes.
The decay timescale depends on the WD mass.
Two light curves are highlighted by a red thick solid line ($0.65 ~M_\sun$)
and a blue thick solid line ($0.95 ~M_\sun$).
\label{all_mass_light_curve_model_x55z02c10o10}}
\end{figure*}


\begin{figure*}
\epsscale{0.95}
\plotone{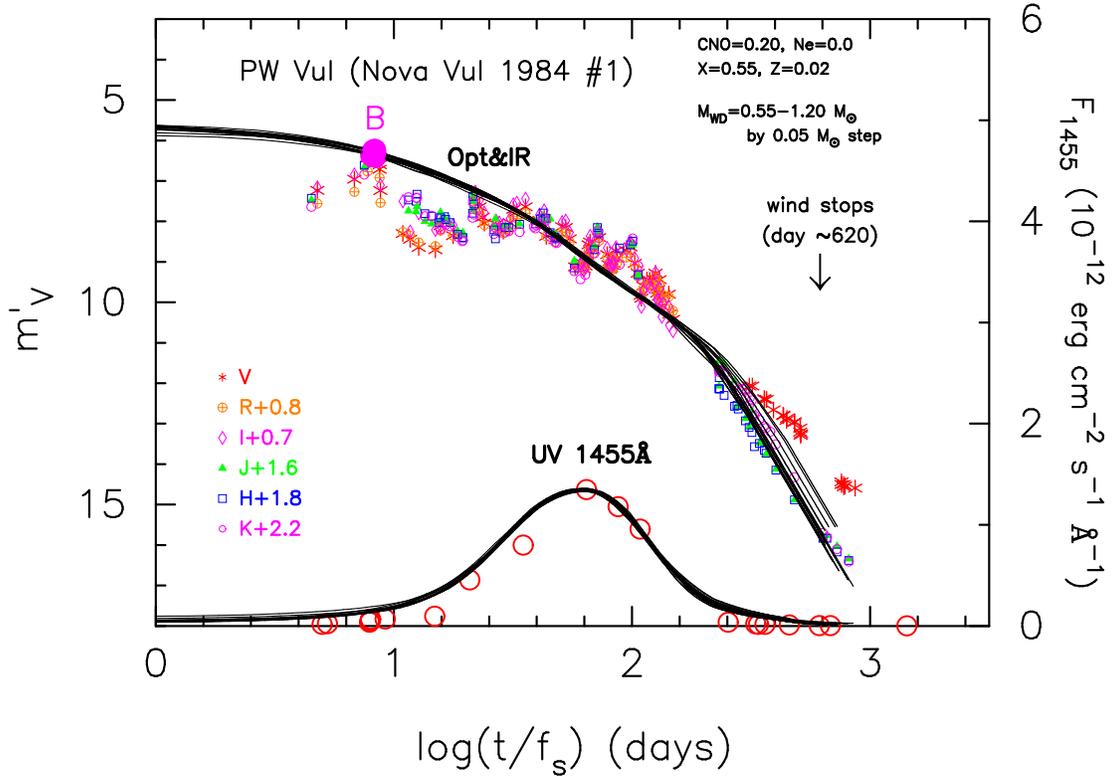}
\caption{
Same as Figure \ref{all_mass_pw_vul_x55z02c10o10_vrijhk_calib_no2},
but all the model light curves of free-free emission and UV~1455\AA\ 
blackbody emission for various WD masses,
which are rescaled to overlap each other.
Each timescaling factor of $f_{\rm s}$
is tabulated in Table \ref{light_curves_of_novae_co4}.
The right edge of each free-free emission model light curve
corresponds to the epoch when the optically thick winds stop.
The end epoch of winds, day $\sim620$ for the $0.83~M_\sun$ WD, is denoted
by an arrow.  Point B corresponds to the peak of $V$ magnitude of 
the PW~Vul outburst.  Here, we assume the start of the day ($t=0$)
as JD 2445910.0 for the observational points of PW~Vul.  
\label{all_mass_pw_vul_x55z02c10o10_vrijhk_universal_no2}}
\end{figure*}


\begin{figure*}
\epsscale{0.95}
\plotone{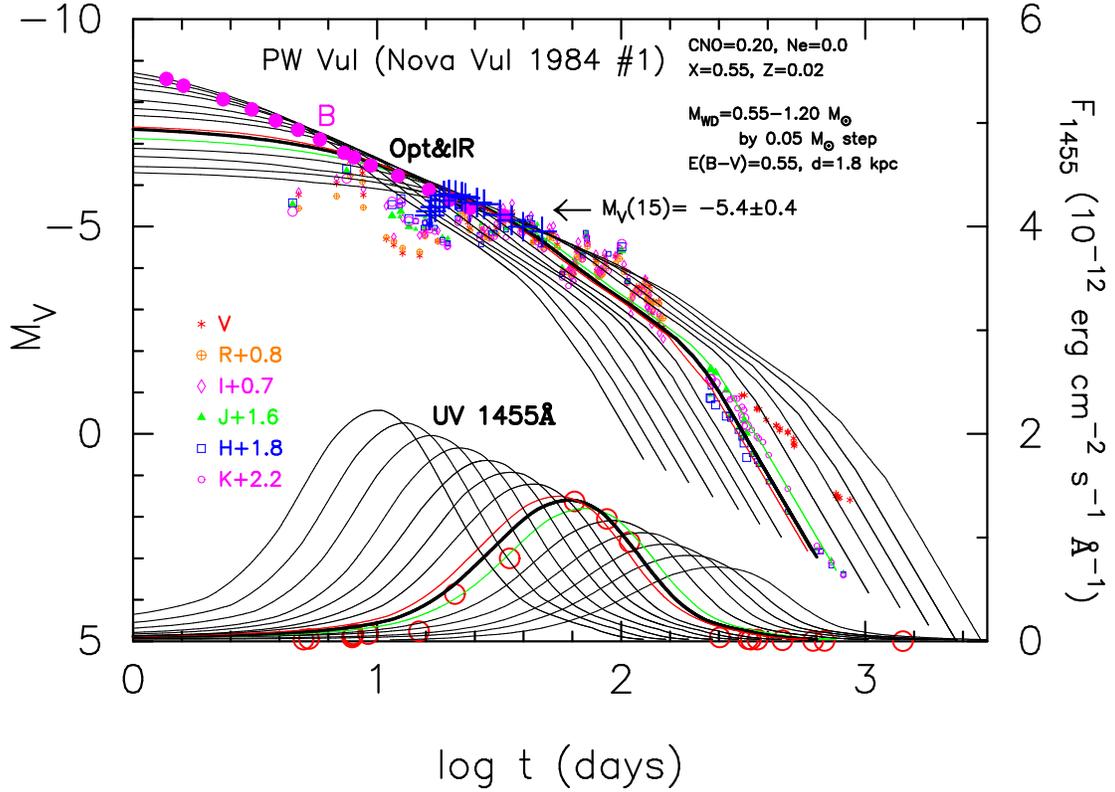}
\caption{
Same as Figure \ref{all_mass_pw_vul_x55z02c10o10_vrijhk_universal_no2},
but for absolute magnitudes and real timescales.
We have calibrated the free-free
model light curves by a distance modulus of $(m-M)_V = 13.0$ for
PW~Vul and restored the absolute magnitude of each free-free emission
light curve (labeled ``OPT\&IR'') by Equations (\ref{real_timescale_flux})
and $(m-M)_{V, \rm PW~Vul}=13.0$.   The position at point B in Figure
\ref{all_mass_pw_vul_x55z02c10o10_vrijhk_universal_no2}
is indicated by a magenta filled circle on each light curve.
We also show the magnitude,
$M_V(15)$, 15 days after the optical maximum, by a blue cross on each
light curve.  We obtain an average value of $M_V(15) = -5.4 \pm 0.4$
among $0.7$--$1.05 ~M_\sun$ WDs.  UV~1455\AA\  model light curves
are also restored in the real timescale and flux at the distance of
10~kpc without absorption.  Black thick solid lines denote 
those for the $0.83~M_\sun$ WD model,
red thin solid ones do for the $0.85~M_\sun$, and
green thin solid ones do for the $0.8~M_\sun$ WD.
\label{all_mass_pw_vul_x55z02c10o10_vrijhk_absolute_mag}}
\end{figure*}


\begin{figure}
\epsscale{0.95}
\plotone{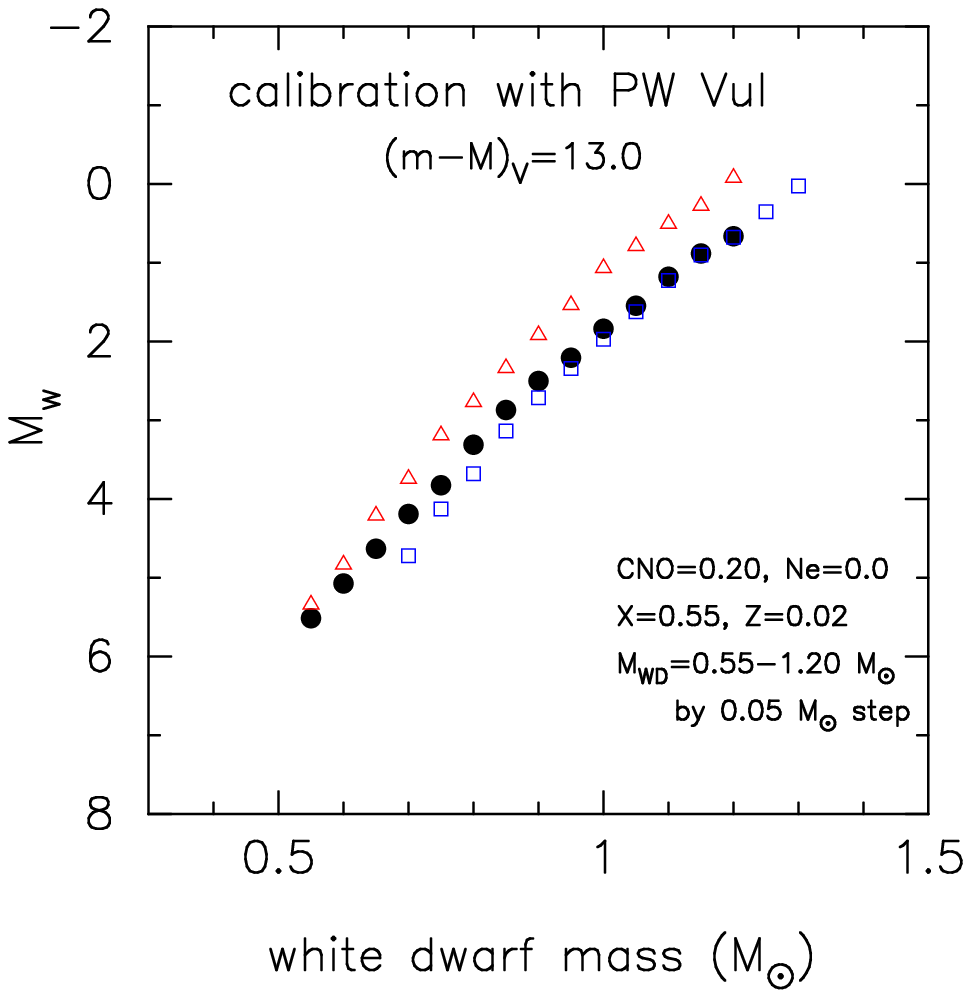}
\caption{
The absolute magnitude at the end point of each free-free emission
model light curve, $M_{\rm w}$, is plotted against the WD mass for 
the chemical composition of ``CO nova 4'' (filled circles).
We also added two other chemical composition models
of ``CO nova 2'' (open triangles) and ``Ne nova 2'' (open squares)
taken from \citet{hac10k}.  See text for more details.
\label{bottom_ff_line_x55z02c10o10_absolute_mag}}
\end{figure}

\section{Absolute Magnitudes of Free-Free Model Light Curves}
\label{absolute_magnitudes}

We have already calibrated the absolute magnitude of the $0.83~M_\sun$
free-free model light curve for ``CO nova 4''
in Section \ref{calibration_co4_pw_vul},
which is defined by $M_{\rm w}=3.0$ at the end point
of the free-free emission model light curve.
In this appendix, we calibrate all free-free 
model light curves for various WD masses using $M_{\rm w}=3.0$
of the $0.83~M_\sun$ WD model, i.e., the PW~Vul data.
In other words, we will determine the absolute magnitudes, $M_{\rm w}$,
for all WD mass light curves in Figure 
\ref{all_mass_light_curve_model_x55z02c10o10}.

\subsection{Model light curves of free-free emission}
\label{free_free_model_light_curve}

In terms of free-free emission, \citet{hac06kb} obtained model light
curves of novae in $0.05 ~M_\sun$ steps for masses in the range 
$M_{\rm WD}= 0.55$ -- $1.2~M_\sun$ with the
chemical composition ``CO nova 4.''  The flux is calculated from
\begin{equation}
F_\nu^{\{M_{\rm WD}\}} (t) = C \left[ {{\dot M_{\rm wind}^2}
\over {v_{\rm ph}^2 R_{\rm ph}}} \right]^{\{M_{\rm WD}\}}_{(t)},
\label{free-free_calculation_original}
\end{equation}
where $\dot M_{\rm wind}$ is the wind
mass-loss rate, $v_{\rm ph}$ is the wind velocity at the photosphere,
$R_{\rm ph}$ is the photospheric radius of each wind solution, and
$C$ is the proportionality constant \citep[see Equation (9) of][]{hac06kb}.
Note that the flux $F_\nu$ is independent of the frequency $\nu$ 
in the case of optically-thin free-free emission.  The details of 
calculations are presented in \citet{hac06kb, hac10k}.
Then the magnitude of model light curves are calculated as
\begin{equation}
m_{\rm ff} = -2.5 ~ \log ~ \left[ {{\dot M_{\rm wind}^2}
\over {v_{\rm ph}^2 R_{\rm ph}}} \right]^{\{M_{\rm WD}\}}_{(t)}
+ ~ G^{\{M_{\rm WD}\}}.
\label{template-wind-free-free-emission}
\end{equation}
The numerical data entering in Equation 
(\ref{template-wind-free-free-emission})
are tabulated in Table \ref{light_curves_of_novae_co4}.
Subscript $(t)$ denotes the time dependence, while superscript 
$\{M_{\rm WD}\}$ indicates a model parameter.  
The last row (15th mag) of each column  in Table 
\ref{light_curves_of_novae_co4} corresponds to the end of the wind phase
in each light curve sequence.
In other words, we define the constant $G^{\{M_{\rm WD}\}}$ in Equation
(\ref{template-wind-free-free-emission}) such that
the last (lowest) point of each light curve (the end
of an optically thick wind phase) is 15th mag.  This helps to shorten
the table.  The magnitudes of free-free emission $m_{\rm ff}$
are plotted in Figure \ref{all_mass_light_curve_model_x55z02c10o10}.

\subsection{Time-normalized light curves}
The free-free emission model light curves in Figure 
\ref{all_mass_light_curve_model_x55z02c10o10}
have a very similar shape.  For the chemical compositions
of ``CO nova 2'' and ``Ne nova 2,''  \citet{hac10k} showed
that model light curves corresponding to different masses have
a homologous behaviour, in the sense that they overlap each other
if properly squeezed or stretched along time.
Here we show that the same property applies to the case of 
chemical composition ``CO nova 4.''
Figure \ref{all_mass_pw_vul_x55z02c10o10_vrijhk_universal_no2} demonstrates
that the free-free emission model light curves in Figure 
\ref{all_mass_light_curve_model_x55z02c10o10} overlap each other
if they are properly squeezed/stretched along time.

The time-scaling factor, $f_{\rm s}$, of each model was determined
by increasing or decreasing $f_{\rm s}$ until the model UV~1455\AA\ 
light curve shape matches the observational points.  We also normalize
the peak flux of our model UV~1455\AA\  light curve
to match the observational peak.
For example, the evolution of the $0.8~M_\odot$ model evolves
1.15 times slower than the PW~Vul observation, so $f_{\rm s}
\approx 1.15$.  The $0.85 ~M_\odot$ model evolves 1.1 times faster than
that, so $f_{\rm s} \approx 0.9$.  The scaling factor $\log f_s$ thus
obtained are tabulated in Table \ref{light_curves_of_novae_co4}.

If we squeeze the timescale of our model light curves
with $t'=t/f_s$  as shown in
Figure \ref{all_mass_pw_vul_x55z02c10o10_vrijhk_universal_no2},
these light curves are written as
\begin{equation}
{m'}_V^{\{M_{\rm WD}\}} (t')= 
-2.5 ~ \log ~ \left[ {{\dot M_{\rm wind}^2}
\over {v_{\rm ph}^2 R_{\rm ph}}} \right]^{\{M_{\rm WD}\}}_{(t'f_s)}
+K_V,
\label{relative_magnitude_two_free-free_convert}
\end{equation}
where $K_V$ is a constant common for all WD masses.
Because they all overlap each other (i.e., the universal decline law),
{\it we regard that all these are the same phenomena.}  
Then, it indicates that   
\begin{equation}
\left[ {{\dot M_{\rm wind}^2}
\over {v_{\rm ph}^2 R_{\rm ph}}} \right]^{\{M_{\rm WD}\}}_{(t'f_s)}
=
\left[ {{\dot M_{\rm wind}^2}
\over {v_{\rm ph}^2 R_{\rm ph}}} \right]^{\{0.83~M_\sun\}}_{(t'f_s)}
=
\left[ {{\dot M_{\rm wind}^2}
\over {v_{\rm ph}^2 R_{\rm ph}}} \right]^{\{0.83~M_\sun\}}_{(t')},
\label{relative_magnitude_two_free-free_equal}
\end{equation}
for all $M_{\rm WD}$.  Note that $f_s=1$ for the $0.83~M_\sun$ WD.  

In general, if we squeeze the timescale of a physical phenomenon
by a factor of $f_{\rm s}$ (i.e., $t' = t/f_{\rm s}$),  we covert
the frequency to $\nu ' = f_{\rm s} \nu$ and the flux of free-free
emission to $F'_{\nu '} = f_{\rm s} F_\nu$
because
\begin{equation}
{{d } \over {d t'}} = f_{\rm s} {{d} \over {d {t}} }.
\label{time-derivation-flux}
\end{equation}
Substituting $F'_{\nu'} = F'_\nu$ (independent of the frequency
in optically-thin free-free emission) into $F'_{\nu '} = f_{\rm s} F_\nu$
and integrating $F'_\nu = f_{\rm s} F_\nu$ with the $V$-filter response
function,
we have the following relation, i.e.,
\begin{eqnarray}
m'_V (t/f_s) =  m_V (t) - 2.5 \log f_{\rm s}.
\label{simple_final_scaling_flux}
\end{eqnarray}
Substituting Equation (\ref{relative_magnitude_two_free-free_equal})
into (\ref{relative_magnitude_two_free-free_convert}), and
then Equation (\ref{relative_magnitude_two_free-free_convert}) into
(\ref{simple_final_scaling_flux}), we obtain the apparent 
$V$ magnitudes of
\begin{equation}
m_V^{\{M_{\rm WD}\}} (t) = 2.5 \log f_{\rm s} 
-2.5 ~ \log ~ \left[ {{\dot M_{\rm wind}^2}
\over {v_{\rm ph}^2 R_{\rm ph}}} \right]^{\{0.83~M_\sun\}}_{(t/f_s)}
+ K_V,
\label{simple_scaling_flux}
\end{equation}
where note that $f_s$ is the time-scaling factor for the WD with mass
of $M_{\rm WD}$, not for the $0.83 M_\sun$ WD.


\begin{figure}
\epsscale{1.15}
\plotone{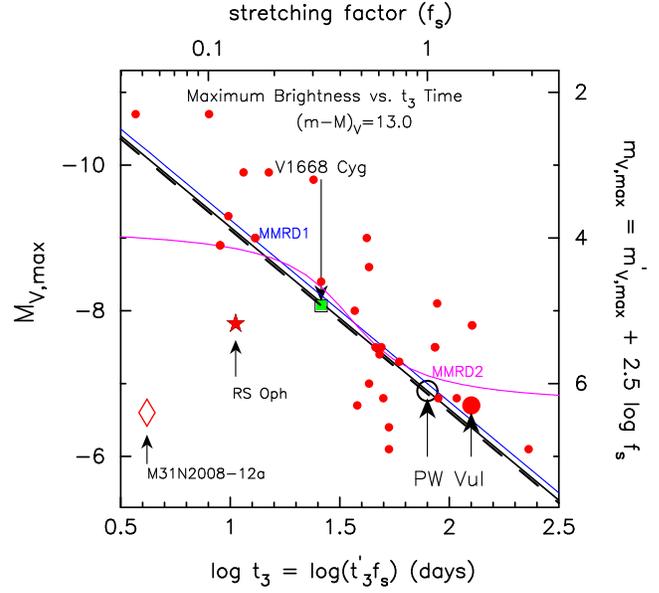}
\caption{
Maximum magnitude vs. rate of decline (MMRD) relation.
Black solid line, i.e., Equation (\ref{theoretical_MMRD_relation}):
our model MMRD relation calibrated with PW~Vul (large black open circle).
Black dashed line: another model MMRD relation calibrated with V1668~Cyg
\citep[green filled square,][]{hac10k}, 
i.e., $M_{V,{\rm max}} = 2.5 \log t_3 -11.6$.
Blue solid line labeled ``MMRD1,'' i.e., Equation (\ref{kaler-schmidt-law}):
Kaler-Schmidt law \citep{sch57}.  Magenta solid line labeled ``MMRD2,''
i.e., Equation (\ref{della-valle-livio-law}): della Valle \& Livio's law
\citep{del95}.   Red filled star: the recurrent nova RS Oph as
an example for a very high mass accretion rate
and very short timescale $t_3$ (very small $f_{\rm s}$) \citep{hac10k}.
Red open diamond: the 1~yr recurrence period M31 nova, M31N2008-12a,
taken from \citet{tan14}, i.e., $M_{g, \rm max}=-6.6$ (maximum in $g$-band)
and $t_{3,g}\approx4.2$ days (measured in the $g$-band light curve). 
Red filled circles: individual novae taken from Table 5 of \citet{dow00}.
Large red filed circle: PW~Vul taken from Table 5 of \citet{dow00}.
Large open black circle: PW~Vul estimated from our model light curve
of $0.83~M_\sun$ WD.  See text for more details.
\label{max_t3_point_B_scale_pw_vul_no2}}
\end{figure}

\subsection{Absolute magnitudes of nova light curves}

The corresponding absolute magnitudes of the light curves can be 
readily obtained from Equation (\ref{simple_scaling_flux}) and
from the distance modulus of PW~Vul, i.e.,
\begin{eqnarray}
M_V^{\{M_{\rm WD}\}} (t) &=& m_V^{\{M_{\rm WD}\}} (t) 
- (m-M)_{V, \rm PW~Vul} \cr\cr
&=& 2.5 \log f_{\rm s} 
-2.5 ~ \log ~ \left[ {{\dot M_{\rm wind}^2}
\over {v_{\rm ph}^2 R_{\rm ph}}} \right]^{\{0.83~M_\sun\}}_{(t/f_s)} \cr\cr
& & + K_V - (m-M)_{V, \rm PW~Vul} \cr\cr
&=& 2.5 \log f_{\rm s} + {m'}_V^{\{M_{\rm WD}\}} (t/f_s)
- (m-M)_{V, \rm PW~Vul},
\label{real_timescale_flux}
\end{eqnarray}
where $(m-M)_{V, \rm PW~Vul}=13.0$ is the distance modulus of PW~Vul 
harboring a $0.83~M_\sun$ WD and
we use Equation (\ref{relative_magnitude_two_free-free_equal}), i.e.,
\begin{equation}
\left[ {{\dot M_{\rm wind}^2}
\over {v_{\rm ph}^2 R_{\rm ph}}} \right]^{\{M_{\rm WD}\}}_{(t)}
=
\left[ {{\dot M_{\rm wind}^2}
\over {v_{\rm ph}^2 R_{\rm ph}}} \right]^{\{0.83~M_\sun\}}_{(t/f_s)},
\label{relative_magnitude_two_free-free_equal_no2}
\end{equation}
and Equation (\ref{relative_magnitude_two_free-free_convert}), i.e.,
\begin{equation}
{m'}_V^{\{M_{\rm WD}\}} (t/f_s)= 
-2.5 ~ \log ~ \left[ {{\dot M_{\rm wind}^2}
\over {v_{\rm ph}^2 R_{\rm ph}}} \right]^{\{M_{\rm WD}\}}_{(t)}
+K_V,
\label{relative_magnitude_two_free-free_convert_no2}
\end{equation}
to derive the last line of the above equation.  The last line 
in Equation (\ref{real_timescale_flux})
simply means that the model light curve ${m'}_V^{\{M_{\rm WD}\}}(t)$
in Figure \ref{all_mass_pw_vul_x55z02c10o10_vrijhk_universal_no2}
is shifted horizontally by $\log f_s$ and vertically by
$2.5 \log f_s - (m-M)_{V, \rm PW~Vul}$ to retrieve the absolute
magnitude and real timescale.  These retrieved absolute magnitudes are
plotted in Figure
\ref{all_mass_pw_vul_x55z02c10o10_vrijhk_absolute_mag}
on the real timescale.
We also tabulate the absolute magnitude, $M_{\rm w}$,
at the end point of winds in Table \ref{light_curves_of_novae_co4}
and plot them in Figure \ref{bottom_ff_line_x55z02c10o10_absolute_mag}.
These values of ``CO nova 4'' are in between those for ``CO nova 2''
and ``Ne nova 2.''  This confirms that our calibration of absolute
magnitude is reasonable. 
Then, we retrieve the absolute magnitudes
of all model light curves in Table \ref{light_curves_of_novae_co4} as
\begin{equation}
M_V = m_{\rm ff} - (m_{\rm w} - M_{\rm w})
    = m_{\rm ff} - (15.0 - M_{\rm w}).
\end{equation}

It should be noted that $K_V$ is a constant common to all WD masses 
\citep[see][]{hac10k}.  
Using Equations (\ref{template-wind-free-free-emission}) and 
(\ref{relative_magnitude_two_free-free_convert}), we obtain
$m_{\rm ff} - m'_V=15.0 - 16.0 = -1.0=G^{\{0.83 ~M_\sun \} }-K_V$ 
at the bottom of the light curve (end of winds), where
we directly read $m'_V=m_{\rm w}=16.0$ for the $0.83~M_\sun$ model
from Figure \ref{all_mass_pw_vul_x55z02c10o10_vrijhk_calib_no2}.
We recall that $G^{\{M_{\rm WD}\}}$ was defined to satisfy $m_{\rm ff}=15$ at
the end point of optically thick winds in Equation 
(\ref{template-wind-free-free-emission}).  We determine
the value of $K_V$ from $K_V = G^{\{0.83 ~M_\sun \} } + 1.0$.

It is interesting to verify the consistency of our theoretical light
curves with the empirical finding that the absolute magnitude
15 days after optical maximum, $M_V(15)$, is almost constant for novae.
The value was first proposed by \citet{bus55} with 
$M_V(15)= -5.2 \pm 0.1$, followed by \citet{coh85} 
with $M_V(15)= -5.60 \pm 0.43$,
\citet{van87} with $M_V(15)= -5.23 \pm 0.39$,
\citet{cap89} with $M_V(15)= -5.69 \pm 0.42$,
and \citet{dow00} with $M_V(15)= -6.05 \pm 0.44$.
The decline rates of our model light curves depend slightly on
the chemical composition. 
We have already obtained $M_V(15)= -5.95 \pm 0.25$
for $0.55$--$1.2 ~M_\odot$ WDs with ``CO nova 2'' 
and $M_V(15)= -5.6 \pm 0.3$ for $0.70$--$1.3 ~M_\sun$ WDs
with ``Ne nova 2'' \citep{hac10k}.  This value is
$M_V(15)= -5.4 \pm 0.4$ for 0.7--$1.05~M_\sun$ WDs
with ``CO nova 4'' as shown in Figure
\ref{all_mass_pw_vul_x55z02c10o10_vrijhk_absolute_mag}.
In this figure, we plot, by magenta filled circles, the $V$ maxima
of each model light curve corresponding to the $V$ maximum of PW~Vul
in Figure \ref{all_mass_pw_vul_x55z02c10o10_vrijhk_universal_no2}
and, by blue crosses, the absolute magnitudes of each model light curve
15 days after $V$ maximum, i.e., 15 days from each magenta filled circle.
The obtained values are roughly consistent 
with the above empirical relations.

\subsection{MMRD relation}
\label{subsec_mmrd}
The clear trend appearing from Figure 
\ref{all_mass_pw_vul_x55z02c10o10_vrijhk_absolute_mag}
is that a more massive WD is systematically brighter at maximum
(smaller $M_{V, {\rm max}}$; see magenta filled circles near the mark B)
and has a faster decline rate (smaller $t_2$ or $t_3$ time).
The relation between $t_3$ (or $t_2$)
and $M_{V, {\rm max}}$ for a nova is usually called 
``Maximum Magnitude vs. Rate of Decline'' (MMRD) relation.
Here $t_3$ ($t_2$) time is defined by 3-mag (2-mag) decay time 
from its maximum in units of day.  Now we derive a theoretical MMRD
relation for the ``CO nova 4'' chemical composition.
Apparent maximum brightness $m_{V, {\rm max}}$
of each WD mass model light curve is expressed as
$m_{V,{\rm max}} = m'_{V,{\rm max}} + 2.5 \log f_{\rm s}$ when
the $t_3$ time is squeezed as $t_3 = f_{\rm s} t'_3$.
Eliminating $f_{\rm s}$ from these two relations, we have
$m_{V,{\rm max}} = 2.5 \log t_3 + m'_{V,{\rm max}} - 2.5 \log t'_3$.
We obtained $t'_3 = 80$ days and $m'_{V,{\rm max}} = 6.1$, where
we measured $t'_3$ and $m'_{V,{\rm max}}$
along our model light curves in Figure 
\ref{all_mass_pw_vul_x55z02c10o10_vrijhk_universal_no2}.  
Then, we obtain our MMRD relation as
\begin{eqnarray}
M_{V,{\rm max}} & = & m_{V,{\rm max}}  - (m-M)_V \cr
& = & 2.5 \log t_3 + m'_{V,{\rm max}} - 2.5 \log t'_3 - (m-M)_V \cr
& = & 2.5 \log t_3 -11.65,
\label{theoretical_MMRD_relation}
\end{eqnarray}
where we use $(m-M)_V=13.0$ for PW~Vul.
Figure \ref{max_t3_point_B_scale_pw_vul_no2} shows this theoretical
MMRD relation.  This figure also shows the MMRD relation calibrated
with V1668~Cyg \citep[black dashed line taken from][]{hac10k}.
We also indicate the time-scaling factor
$f_{\rm s}$ against the PW~Vul light curves
in the upper axis of the same figure.

For comparison, two empirical MMRD relations are plotted in the same figure,
i.e., Kaler-Schmidt's law \citep[blue solid line labeled 
``MMRD1'':][]{sch57}, i.e.,
\begin{equation}
M_{V, {\rm max}} = -11.75 + 2.5 \log t_3,
\label{kaler-schmidt-law}
\end{equation}
and della Valle \& Livio's law 
\citep[magenta solid line labeled ``MMRD2'':][]{del95}, i.e.,
\begin{equation}
M_{V, {\rm max}} = -7.92 -0.81 \arctan \left(
{{1.32-\log t_2} \over {0.23}} \right),
\label{della-valle-livio-law}
\end{equation}
where we use a relation of $t_2  \approx 0.6 \times t_3$ 
for the optical light curves that follow the universal decline law
\citep{hac06kb}.

Figure \ref{max_t3_point_B_scale_pw_vul_no2} also shows 
observational points for individual novae (red filled circles),
taken from Table 5 of \cite{dow00}.  Note that we remeasured the $t_3$
time of PW~Vul along our model light curve, which resulted 
in $t_3= 80$~days and $M_{V, \rm max}= -6.9$ as shown by a large
black open circle.  This data point is leftside to 
Downes \& Duerbeck's (2000) estimate ($t_3=126$~day and 
$M_{V, \rm max}=-6.7$) denoted by a large red filled circle.

The scatter of the observed data with respect to the MMRD formulae in
Figure \ref{max_t3_point_B_scale_pw_vul_no2} is generally large,
and larger than the observational errors.  
This strongly suggests that the scatter is due to the presence
of a second parameter.  \citet{hac10k} pointed out that the main parameter
governing the MMRD relation is the WD mass (represented by the time-scaling
factor $f_{\rm s}$ in Figure \ref{max_t3_point_B_scale_pw_vul_no2}),
the second parameter being the initial envelope mass 
(or the mass accretion rate of the WD).  In turn, the initial 
envelope mass (ignition mass) depends on the mass accretion rate
to the WD \citep[see, e.g., Figure 3 of][]{kat14shn} such that
the lower the mass accretion rate is, the larger the envelope mass is.
This simply means that, for the same WD mass, novae are brighter/fainter
for lower/higher mass accretion rates.  \citet{hac10k} clearly showed
the dependence of maximum brightness on the initial envelope
mass (see their Figures 8 and 15).  We further show fainter examples of 
maximum brightness, RS~Oph and M31N2008-12a, both of which
are recurrent novae with $\sim20$ and $\sim$1 yr recurrence periods,
respectively.  We conclude that this second parameter,
the initial envelope mass, can reasonably explain the scatter
of individual novae from the empirical MMRD relations so far proposed
\citep[see Figure 15 of ][]{hac10k}.

\section{Time-stretching Method}
\label{time-stretching_method_appendix}

The distance modulus of PW~Vul can be estimated, in a very different way,
from a resemblance between PW~Vul and other optically well-observed novae.
\citet{hac06kb} found that nova light curves follow a universal
decline law when free-free emission dominates the spectrum in 
optical and NIR regions.  Using this property, \citet{hac10k} found that,
if two nova light curves overlap each other after one of
the two is squeezed/stretched by a factor
of $f_s$ ($t'=t/f_s$) in the time direction, the brightnesses of the
two novae obey the relation of 
\begin{equation}
m'_V = m_V - 2.5 \log f_s,
\label{time-stretching_method}
\end{equation}
which is the same as Equation (\ref{simple_final_scaling_flux}).
Using this property with calibrated nova light curves, we can estimate
the absolute magnitude of a target nova.  Figure
\ref{pw_vul_v1668_cyg_v1500_cyg_v1974_cyg_v_bv_ub_color_curve_logscale_no3}
shows time-normalized light curves of PW~Vul, V1668~Cyg, 
V1974~Cyg, and V533~Her against that of V1500~Cyg, 
similar to Figure 41 of \citet{hac14k}, but we used the reanalyzed data.
The $V$ light curve, $B-V$ and $U-B$ color curves of PW~Vul are well
squeezed to match the other ones.  Note that the dereddened 
$(B-V)_0$ and $(U-B)_0$ of each nova also follow 
a general course in the color-color diagram
\citep[i.e., overlap each other; see][for the general
course of $UBV$ color evolution of novae]{hac14k}.
These five novae obey the relations of
\begin{eqnarray}
(m-M)_{V, \rm V1500~Cyg} &=& 12.3\cr
&=&(m-M)_{V,\rm PW~Vul} + \Delta V - 2.5 \log 0.182 \cr
&\approx& 13.0 - 2.6 + 1.85 = 12.25\cr
&=&(m-M)_{V,\rm V1668~Cyg} + \Delta V - 2.5 \log 0.44 \cr
&\approx& 14.25 - 2.9 + 0.90 = 12.25\cr
&=& (m-M)_{V,\rm V1974~Cyg} + \Delta V - 2.5 \log 0.42 \cr
&\approx& 12.2 - 0.8 + 0.95 = 12.35\cr
&=& (m-M)_{V,\rm V533~Her} + \Delta V - 2.5 \log 0.54 \cr
&\approx& 10.8 + 0.8 + 0.67 = 12.27,
\label{time-stretching_brightness_pw_vul}
\end{eqnarray}
where $\Delta V$ is the difference of apparent brightness obtained in Figure 
\ref{pw_vul_v1668_cyg_v1500_cyg_v1974_cyg_v_bv_ub_color_curve_logscale_no3}
by which the $V$ light curve of each nova is shifted up or down against
that of V1500~Cyg.  The time-scaling factors are also obtained in this
figure as $f_s=0.182$ for PW~Vul,
$f_s=0.44$ for V1668~Cyg, $f_s=0.42$ for V1974~Cyg, and
$f_s=0.54$ for V533~Her, each against that of V1500~Cyg.
We obtained these stretching factors by shifting horizontally 
each light curve to overlap them.  The apparent distance moduli of 
V1500~Cyg, V1668~Cyg, and V1974~Cyg were calibrated as
$(m-M)_{V,\rm V1500~Cyg} = 12.3$, $(m-M)_{V,\rm V1668~Cyg} = 14.25$,
and $(m-M)_{V,\rm V1974~Cyg} = 12.2$ in \citet{hac14k}.
These three are all consistent with each other.
The distance modulus of V533~Her was also calculated
to be $(m-M)_{V,\rm V533~Her} = 10.8$.
\citet{hac14k} obtained the distance-modulus of PW~Vul based on this
time-stretching method as $(m-M)_V=13.0\pm0.1$,
which is consistent with Equation (\ref{distance_modulus_eq_pw_vul_no1}).
This is a strong support to our adopted values of $E(B-V)=0.55$ 
and $d=1.8$~kpc.


\begin{thebibliography}{}

%




\bibitem[Aller (1984)]{all84}
Aller, L. H. 1984, Physics of Thermal Gaseous Nebulae (Dortlechit:
D. Reidel).


\bibitem[Andreae \& Drechsel (1990)]{and90}
Andreae, J., \& Drechsel, H. 1990, Physics of Classical Novae,
eds. A. Cassatella and R. Viotti (Berin: Springer-Verlag), 204

\bibitem[Andreae et al. (1991)]{and91}
Andreae, J., Drechsel, H., Snijders, M. A. J., \& Cassatella, A. 
1991, \aap, 244, 111

\bibitem[Andre\"a et al. (1994)]{and94}
Andre\"a, J., Drechsel, H., \& Starrfield, S. 1994, \aap, 291, 869






\bibitem[Arenas et al. (2000)]{are00}
Arenas, J., Catal\'an, M. S., Augusteijn, T., \& Retter, A. 2000, \mnras,
311, 135


\bibitem[Arkhipova \& Zaitseva (1976)]{ark76}
Arkhipova, V. P., \& Zaitseva, G. V. 1976, SvAL, 2, 35

\bibitem[Arkhipova et al. (2000)]{ark00}
Arkhipova, V. P., Burlak, M. A., \& Esipov, V. F. 2000, 
Astronomy Letters, 26, 372










\bibitem[Barnes \& Evans (1970)]{bar70}
Barnes, T. G., \& Evans, N. R. 1970, \pasp, 82, 889

\bibitem[Barry et al. (2006)]{bar06}
Barry, R. K., Mukai, K., Sokoloski, J. L., et al.
2006, RS Ophiuchi (2006) and the Recurrent Nova Phenomenon, 
ASP Conference Series, 401, 52























\bibitem[Budding (1983)]{bud83}
Budding, E. 1983, \iaucirc, 3853, 2




\bibitem[Buscombe \& de Vaucouleurs (1955)]{bus55}
Buscombe, W., \& de Vaucouleurs, G. 1955, The Observatory, 75, 170



\bibitem[Campbell \& Shapley (1923)]{cam23}
Campbell, L., \& Shapley, H. 1923, 
Annals of the Astronomical Observatory of Harvard College, 81, 113


\bibitem[Candy et al. (1967)]{can67}
Candy, M. P., Alcock, G. E. D., \& Zissell, R. E. 1967, \iaucirc, 2022, 1


\bibitem[Capaccioli et al. (1989)]{cap89}
Capaccioli, M., della Valle, M., Rosino, L., D'Onofrio, M. 1989, \aj, 97, 1622


\bibitem[Cassatella et al. (2002)]{cas02}
Cassatella, A., Altamore, A., \& Gonz\'alez-Riestra, R. 2002, \aap, 384, 1023







\bibitem[Child (1901)]{chi01}
Child, L. 1901, \mnras, 61, 483

\bibitem[Chincarini (1964)]{chi64}
Chincarini, G. 1964, \pasp, 76, 289





\bibitem[Chochol \& Pribulla (1997)]{cho97b}
Chochol, D., \& Pribulla, T. 1997, CoSka, 27, 53


\bibitem[Chochol \& Pribulla (1998)]{cho98}
Chochol, D., \& Pribulla, T. 1998, CoSka, 28, 121





\bibitem[Cohen (1985)]{coh85}
Cohen, J. G. 1985, \apj,  292, 90


\bibitem[Cohen \& Rosenthal (1983)]{coh83}
Cohen, J. G., \& Rosenthal, A. J. 1983, \apj, 268, 689





%





\bibitem[Dawson (1926)]{daw26}
Dawson, B. H. 1926, \aj, 36, 148


\bibitem[de Freitas Pacheco \& Codina (1985)]{pac85}
de Freitas Pacheco, J. A., \& Codina, S. J. 1985, \mnras, 214, 481

\bibitem[della Valle (1991)]{del91}
della Valle, M. 1991, \aap, 252, L9


\bibitem[della Valle \& Livio (1995)]{del95}
della Valle, M., \& Livio, M. 1995, \apj, 452, 704 






\bibitem[Diaz \& Steiner (1989)]{dia89}
Diaz, M. P., \& Steiner, J. E. 1989, \apj, 339, L41

\bibitem[Diaz \& Steiner (1994)]{dia94}
Diaz, M. P., \& Steiner, J. E. 1994, \apj, 425,252






\bibitem[Downes \& Duerbeck (2000)]{dow00}
Downes, R. A., \& Duerbeck, H. W. 2000, \aj, 120, 2007



\bibitem[Drechsel et al. (1977)]{dre77}
Drechsel, H., Rahe, J., Duerbeck, H. W., Kohoutek, L., \& Seitter, W. C.
1977, \aaps, 30, 323


\bibitem[Duerbeck (1981)]{due81}
Duerbeck, H. W. 1981, \pasp, 93, 165


\bibitem[Duerbeck et al. (1984)]{due84}
Duerbeck, H. W., Geffert, M., Nelles, B., Dummler, R., \& Nolte, M. 1984,
IBVS, 2641, 1









\bibitem[Ennis et al. (1977)]{enn77}
Ennis, D., Becklin, E. E., Beckwith, S., et al.
1977, \apj, 214, 478



\bibitem[Evans et al. (1990)]{eva90}
Evans, A., Callus, C. M., Whitlock, P. A., \& Laney, D. 
1990, \mnras, 246, 527 

\bibitem[Evans et al. (2003)]{eva03}
Evans, A., Gehrz, R. D., Geballe, T. R., et al. 2003, \aj, 126, 1981











\bibitem[Friedjung (1992)]{fri92}
Friedjung, M. 1992, \aap, 262, 487

\bibitem[Friedjung et al. (2010)]{fri10}
Friedjung, M., Dennefeld, M., \& Voloshina, I. 2010, \aap, 521, A84







\bibitem[Gallagher \& Holm (1974)]{gal74}
Gallagher, J. S., \& Holm, A. V. 1974, ApJL, 189, L123 



\bibitem[Gallagher \& Ney (1976)]{gal76}
Gallagher, J. S., \& Ney, E. P. 1976, ApJL, 204, L35


\bibitem[Gaposchkin (1956)]{gap56}
Gaposchkin, S. 1956, \aj, 61, 36






\bibitem[Gehrz et al. (1988)]{geh88}
Gehrz, R. D., Harrison, T. E., Ney, E. P., et al. 1988, \apj, 329, 894



\bibitem[Gehrz et al. (1998)]{geh98}
Gehrz, R. D., Truran, J. W., Williams, R. E., \& Starrfield, S. 
1998, \pasp, 110, 3




\bibitem[Goranskij et al. (2000)]{gor00}
Goranskij, V. P., Shugarov, S. Y., Katysheva, N. A., et al. 
2000, IBVS, 4852

\bibitem[Goranskij et al. (2007)]{gor07}
Goranskij, V. P., Katysheva, N. A., Kusakin, A. V., et al. 
2007, Astrophysical Bulletin, 62, 125


\bibitem[Gore (1901)]{gor01}
Gore, J. E. 1901, \mnras, 62, 156

\bibitem[Gonz\'alez-Riestra et al. (1996)]{gon96}
Gonz\'alez-Riestra, R., Shore, S. N., Starrfield, S., \& Krautter, J.
1996, \iaucirc, 6295, 1




\bibitem[Grevesse \& Anders (1989)]{gre89}
Grevesse, N., \& Anders, E. 1989, Cosmic Abundances of Matter,
ed. C. J. Waddington (New York: AIP), 1

\bibitem[Grygar (1969)]{gry69}
Grygar, J. 1969, IBVS, 371, 1  



\bibitem[Hachisu \& Kato (2001)]{hac01kb}
Hachisu, I., \& Kato, M. 2001, \apj, 558, 323




\bibitem[Hachisu \& Kato (2004)]{hac04k}
Hachisu, I., \& Kato, M. 2004, ApJL, 612, L57



\bibitem[Hachisu \& Kato (2006)]{hac06kb}
Hachisu, I., \& Kato, M. 2006, \apjs, 167, 59

\bibitem[Hachisu \& Kato (2007)]{hac07k}
Hachisu, I., \& Kato, M. 2007, \apj, 662, 552


\bibitem[Hachisu \& Kato (2010)]{hac10k}
Hachisu, I., \& Kato, M. 2010, \apj, 709, 680


\bibitem[Hachisu \& Kato (2014)]{hac14k}
Hachisu, I., \& Kato, M. 2014, ApJ, 785, 97

\bibitem[Hachisu et al. (2008)Hachisu, Kato, \& Cassatella]{hac08kc}
Hachisu, I., Kato, M., \& Cassatella, A. 2008, \apj, 687, 1236




\bibitem[Hachisu et al. (2006)]{hac06b}
Hachisu, I., Kato, M., Kiyota, S., et al. 2006, ApJL, 651, L141

\bibitem[Hachisu et al. (2007)]{hac07kl}
Hachisu, I., Kato, M., \& Luna, G. J. M. 2007, \apj, 659, L153




%


\bibitem[Harrison et al. (2013)]{har13}
Harrison, T. E., Bornak, J., McArthur, B. E., \& Benedict, G. F.
2013, \apj, 767, 7 

\bibitem[Harman \& O'Brien (2003)]{har03}
Harman, D. J., \& O'Brien, T. J. 2003, \mnras, 344, 1219



\bibitem[Hassall et al. (1990)]{has90}
Hassall, B. J. M., Snijders, M. A. J., Harris, A. W., et al. 1990, in
Physics of Classical Novae, ed. A. Cassatella \& R. Viotti (Berin: Springer),
202

\bibitem[Hauschildt et al. (1994)]{hau94}
Hauschildt, P. H., Starrfield, S., Shore, S. N., et al.
1994, \aj, 108, 1008

\bibitem[Hauschildt et al. (1995)]{hau95}
Hauschildt, P. H., Starrfield, S., Shore, S. N., Allard, F., \&
Baron, E. 1995, \apj, 447, 829

\bibitem[Hauschildt et al. (1997)]{hau97}
Hauschildt, P. H., Shore, S. N., Schwarz, G. J., et al., \apj, 490, 803













\bibitem[Horne et al. (1993)]{hor93}
Horne, K., Welsh, W., \& Wade, R. A. 1993, \apj, 410, 357


\bibitem[Hric et al. (1998)]{hri98}
Hric, L., Petr\'ik, K., Urban, Z., \& Han\v{z}l, D. 1998, \aaps, 133, 211


\bibitem[Hutchings (1970)]{hut70}
Hutchings, J. B. 1970,
Publications of the Dominion Astrophysical Observatory, 13, 347




\bibitem[Iijima (2006)]{iij06}
Iijima, T. 2006, \aap, 451, 563


\bibitem[Iijima (2007a)]{iij07a}
IIjima, T. 2007a, CBET, 934, 1

\bibitem[Iijima (2007b)]{iij07b}
Iijima, T. 2007b, CBET, 1006, 1


\bibitem[Iijima et al. (1998)]{iij98}
Iijima, T., Rosino, L., \& della Valle, M. 1998, \aap, 338, 1006








\bibitem[Kamath \& Ashok (1999)]{kam99}
Kamath, U. S., \& Ashok, N. M. 1999, \aaps, 136, 107








\bibitem[Kato \& Hachisu (1994)]{kat94h}
Kato, M., \& Hachisu, I., 1994, \apj, 437, 802




\bibitem[Kato \& Hachisu (2009)]{kat09h}
Kato, M., \& Hachisu, I., 2009, \apj, 699, 1293

\bibitem[Kato \& Hachisu (2011)]{kat11h}
Kato, M., \& Hachisu, I., 2011, \apj, 743, 157


\bibitem[Kato et al. (2009)]{kat09hc}
Kato, M., Hachisu, I., \& Cassatella, A. 2009, \apj, 704, 1676

\bibitem[Kato et al. (2011)]{kat11hcg}
Kato, M., Hachisu, I., Cassatella, A., \& Gonz\'alez-Riestra, R.
2011, \apj, 727, 72 

\bibitem[Kato et al. (2012)]{kat12mh}
Kato, M., Miko{\l}ajewska, J., \& Hachisu, I. 2012, \apj, 750, 5 

\bibitem[Kato et al. (2014)]{kat14shn}
Kato, M., Saio, H., Hachisu, I., \& Nomoto, K. 2014, \apj, 793, 136




\bibitem[Kawara et al. (1976)]{kaw76}
Kawara, K., Maihara, T., Noguchi, K., et al.
1976, \pasj, 28, 163






\bibitem[Kiss \& Sarneczky (2007)]{kis07}
Kiss, L. \& Sarneczky, K. 2007, CBET, 1039, 1











\bibitem[Kohoutek (1981)]{koh81}
Kohoutek, L. 1981, \mnras, 196, 87P



\bibitem[Kolotilov et al. (1995)]{kol95}
Kolotilov, E. A., Munari, U., \& Yudin, B. F. 1995, \mnras, 275, 185




\bibitem[Kosai et al. (1984)]{kos84}
Kosai, H., Takana, W., Watanabe, T., et al.
1984, \iaucirc, 3963, 2





%

\bibitem[Krautter et al. (1984)]{kra84}
Krautter, J., Beuermann, K., Leitherer, C., et al.
1984, \aap, 137, 307





\bibitem[K\"urster \& Barwig (1988)]{kur88}
K\"urster, M., \& Barwig, H. 1988, \aap, 199, 201

%

%















\bibitem[Lockwood \& Millis (1976)]{loc76}
Lockwood, G. W., \& Millis, R. L. 1976, \pasp, 88, 235



\bibitem[Lyke \& Campbell (2009)]{lyk09}
Lyke, J. E., \& Campbell, R. D. 2009, \aj, 138, 1090











\bibitem[Malakpur (1975)]{mal75}
Malakpur, I. 1975, \apss, 38, 403


\bibitem[Mannery (1970)]{man70}
Mannery, E. J. 1970, \pasp, 82, 626 

\bibitem[Marshall et al. (2006)]{mar06}
Marshall, D. J., Robin, A. C., Reyl\'e, C., Schultheis, M., \& Picaud, S.
2006, \aap, 453, 635













\bibitem[Miroshnichenko (1988)]{mir88}
Miroshnichenko, A. S. 1988, SvA, 32, 298

\bibitem[Mollerus (1969)]{mol69}
Mollerus, B. 1969, \aap, 3, 376

\bibitem[Morales-Rueda et al. (2002)]{mor02}
Morales-Rueda, L., Still, M. D., Roche, P., Wood, J. H., Lockley, J. J. 
2002, \mnras, 329, 597

\bibitem[Morisset \& P\'equignot (1996)]{mor96}
Morisset, C., \& P\'equignot, D. 1996, \aap, 312, 135






\bibitem[Munari et al. (1996)]{mun96}
Munari, U., Goranskij, V. P., Popova, A. A., et al. 1996, \aap, 315, 166



\bibitem[Munari et al. (2007)]{mun07c}
Munari, U., Orio, M., Valentini, M., et al. 2007, CBET, 1010, 1







\bibitem[Munari et al. (1994b)]{mun94b}
Munari, U., Yudin, B. F., Kolotilov, E. A., et al. 1994b, \aap, 284, L9


\bibitem[Nha (1967)]{nha67}
Nha, Il-Seong 1967, IBVS, 238, 1

\bibitem[Naito et al. (2007)]{nai07}
Naito, H., Matsuda, K., \& Yamaoka, H. 2007, CBET, 934, 2





\bibitem[Nakano et al. (1993)]{nak93}
Nakano, S., Kanatsu, K., Kawanishi, K., et al. 1993, \iaucirc, 5902, 1 








\bibitem[Nakano et al. (2007)]{nak07a}
Nakano, S., Sakurai, Y., Itagaki, K., \& Koff, R. 2007, \iaucirc, 8832, 1









\bibitem[Ness et al. (2008)]{nes08}
Ness, J.-U., Schwarz, G., Starrfield, S., et al.
2008, \aj, 135, 1328


\bibitem[Nishimaki et al. (2008)]{nis08}
Nishimaki, Y., Yamamuro, T., Motohara, K., Miyata, T., \& Tanaka, M.
2008, \pasj, 60, 191


\bibitem[Nomoto (1982)]{nom82}
Nomoto, K. 1982, \apj, 253, 798





\bibitem[O'Connell (1968)]{oco68}
O'Connell, D. J. K. 1968, IBVS, 313, 1

\bibitem[\"Oegelman et al. (1987)]{oge87}
\"Oegelman, H., Krautter, J., \& Beuermann, K. 1987, \aap, 177, 110







\bibitem[Onderli\v{c}ka \& Vete\v{s}n\'{i}k (1968)]{ond68}
Onderli\v{c}ka, B., \& Vete\v{s}n\'{i}k, M. 1968, BAICz, 19, 99



\bibitem[Orio et al. (2001)]{ori01}
Orio, M., Covington, J., \& \"Ogelman, H. 2001, \aap, 373, 542










\bibitem[Payne-Gaposchkin (1957)]{pay57}
Payne-Gaposchkin, C. 1957, The Galactic Novae (Amsterdam: North-Holland)



\bibitem[P\'equignot et al. (1993)]{peq93}
P\'equignot, D., Petitjean, P., Boisson, C., \& Krautter, J. 1993, \aap,
271, 219




\bibitem[Petitjean et al. (1990)]{pet90}
Petitjean, P., Boisson, C., \& Pequignot, D. 1990, \aap, 240, 433

\bibitem[Pfau (1976)]{pfa76}
Pfau, W. 1976, IBVS, 1106, 1


\bibitem[Poggiani (2008)]{pog08}
Poggiani, R. 2008, NewA, 13, 557


\bibitem[Poggiani (2010)]{pog10}
Poggiani, R. 2010, NewA, 15, 657

\bibitem[Poggiani (2012)]{pog12}
Poggiani, R. 2012, Memorie della Societa Astronomica Italiana, 83, 753








\bibitem[Prialnik \& Kovetz (1995)]{pri95}
Prialnik, D., \& Kovetz, A. 1995, \apj, 445, 789


\bibitem[Rafanelli \& Rosino (1978)]{raf78}
Rafanelli, P., \& Rosino, L. 1978, \aaps, 31, 337





\bibitem[Rambaut (1901a)]{ram01a}
Rambaut, A. A. 1901a, \mnras, 61, 348

\bibitem[Rambaut (1901b)]{ram01b}
Rambaut, A. A. 1901b, \mnras, 61, 390

\bibitem[Rambaut (1901c)]{ram01c}
Rambaut, A. A. 1901c, \mnras, 61, 467

\bibitem[Rambaut (1901d)]{ram01d}
Rambaut, A. A. 1901d, \mnras, 61, 544

\bibitem[Rambaut (1901e)]{ram01e}
Rambaut, A. A. 1901e, \mnras, 62, 78

\bibitem[Rambaut (1902)]{ram02}
Rambaut, A. A. 1902, \mnras, 62, 586

\bibitem[Rambaut (1903)]{ram03}
Rambaut, A. A. 1903, \mnras, 63, 509


\bibitem[Rieke \& Lebofsky (1985)]{rie85}
Rieke, G. H., \& Lebofsky, M. J. 1985, \apj, 288, 618






\bibitem[Robb \& Scarfe (1995)]{rob95}
Robb, R. M., \& Scarfe, C. D. 1995, \mnras, 273, 347

\bibitem[Robinson \& Ashbrook (1968)]{rob68}
Robinson, L. J., \& Ashbrook, J. 1968, IBVS, 252, 1








\bibitem[Rosino \& Iijima (1987)]{ros87}
Rosino, L., \& Iijima, T. 1987, \apss, 130, 157










\bibitem[Rudy et al. (2007b)]{rud07b}
Rudy, R. J., Lynch, D. K., Russell, R. W., \& Woodward, C. E. 2007b,
\iaucirc, 8884, 2


\bibitem[Rudy et al. (2002)]{rud02}
Rudy, R. J., Venturini, C. C., Lynch, D. K., Mazuk, S., \& Puetter, R. C.
2002, \apj, 573, 794










\bibitem[Saizar et al. (1991)]{sai91}
Saizar, P., Starrfield, S., Ferland, G. J., et al. 1991, \apj, 367, 310




\bibitem[Sanyal (1974)]{san74}
Sanyal, A. 1974, ApJS, 28, 115








\bibitem[Schlafly \& Finkbeiner (2011)]{schl11}
Schlafly, E. F., \& Finkbeiner, D. P. 2011, \apj, 737, 103 





\bibitem[Schmidt et al. (1987)]{sch87a}
Schmidt, G. D., Smith, P., \& Elston, R. 1987, \iaucirc, 4415, 2

\bibitem[Schmidt \& Stockman (1987)]{sch87b}
Schmidt, G. D., \& Stockman, H. S. 1987, \iaucirc, 4458, 1

\bibitem[Schmidt (1957)]{sch57}
Schmidt, Th. 1957, ZA, 41, 182


\bibitem[Schwarz et al. (1997)]{sch97}
Schwarz, G. J., Starrfield, S., Shore, S. N., \& Hauschildt, P. H. 
1997, \mnras, 290, 75


\bibitem[Schwarz et al. (2011)]{sch11}
Schwarz, G. J., Ness, J.-U., Osborne, J. P.,  et al. 2011, \apjs, 197, 31



\bibitem[Seaton (1979)]{sea79}
Seaton, M. J. 1979, \mnras, 187, 73P







\bibitem[Shanley et al. (1995)]{sha95}
Shanley, L., \"Ogelman, H., Gallagher, J. S., Orio, M., \& Krautter, J.
1995, ApJL, 438, L95


\bibitem[Sharp (1901)]{sha01}
Sharp, M. C. 1901, \mnras, 61, 398

\bibitem[Shen et al. (1964)]{she64}
Shen, L.-Z., et al. 1964, AcASn, 12, 83





\bibitem[Shore et al. (1994)]{sho94}
Shore, S. N., Starrfield, S., Gonzalez-Riestrat, R., Hauschildt, P. H., \&
Sonneborn, G. 1994, Natur, 369, 539








\bibitem[Slavin et al. (1994)]{sla94}
Slavin, A. J., O'Brien, T. J., \& Dunlop, J. S. 1994, \mnras, 266, L55

\bibitem[Slavin et al. (1995)]{sla95}
Slavin, A. J., O'Brien, T. J., \& Dunlop, J. S. 1995, \mnras, 276, 353







\bibitem[Solf (1983)]{sol83}
Solf, J. 1983, \apj, 273, 647




\bibitem[Spencer Jones (1925)]{spe25}
Spencer Jones, H. 1925, The Observatory, 48, 261

\bibitem[Spencer Jones (1931)]{spe31}
Spencer Jones, H. 1931, Annals of the Cape Observatory, 10, 9.1






\bibitem[Stokes (1967)]{sto67}
Stokes, A. J. 1967, IBVS, 224, 1


\bibitem[Strope et al. (2010)]{str10}
Strope, R., Schaefer, B. E., \& Henden, A. A. 2010, \aj, 140, 34



\bibitem[Tanaka et al. (2011)]{tan11}
Tanaka, J., Nogami, D., Fujii, M., et al.
2011, \pasj, 63, 911

\bibitem[Tang et al. (2014)]{tan14}
Tang, S., Bildsten, L., Wolf, W. M., et al. 2014, \apj, 786, 61 







\bibitem[Tempesti (1979)]{tem79}
Tempesti, P. 1979, AN, 300, 51


\bibitem[Terzan (1968)]{ter68}
Terzan, A. 1968, JO, 51, 329 








\bibitem[Tylenda (1978)]{tyl78}
Tylenda, R. 1978, Acta Astron., 28, 333



\bibitem[van den Bergh \& Younger (1987)]{van87}
van den Bergh, S., \& Younger, P. F. 1987, \aaps, 70, 125


\bibitem[van Genderen (1963)]{gen63}
van Genderen, A. M. 1963, BAN, 17, 293







\bibitem[Verbunt (1987)]{ver87}
Verbunt, F. 1987, \aaps, 71, 339














\bibitem[Warner (1995)]{war95}
Warner, B. 1995, Cataclysmic Variable Stars, (Cambridge: 
Cambridge Univ. Press)






\bibitem[Whitelock et al. (1984)]{whi84}
Whitelock, P. A., Carter, B. S., Feast, M. W., et al.
1984, \mnras, 211, 421


%

%

%


\bibitem[Williams (1901a)]{wil01a}
Williams, A. S. 1901a, \mnras, 61, 337

\bibitem[Williams (1901b)]{wil01b}
Williams, A. S. 1901b, \mnras, 61, 396

\bibitem[Williams (1901c)]{wil01c}
Williams, A. S. 1901b, \mnras, 61, 480

\bibitem[Williams (1901d)]{wil01d}
Williams, A. S. 1901d, \mnras, 61, 550

\bibitem[Williams (1902)]{wil02}
Williams, A. S. 1902, \mnras, 62, 589

\bibitem[Williams (1919)]{wil19}
Williams, A. S. 1919, \mnras, 79, 362


\bibitem[Williams \& Gallagher (1979)]{wil79}
Williams, R. E., \& Gallagher, J. S. 1979, \apj, 228, 482

\bibitem[Williams et al. (1991)]{wil91}
Williams, R. E., Hamuy, M., Phillips, M. M., et al.
1991, \apj, 376, 721




\bibitem[Williams et al. (1996)]{wil96}
Williams, P. M., Longmore, A. J., \& Geballe, T. R. 1996, \mnras, 279, 804


\bibitem[Williams et al. (1978)]{wil78}
Williams, R. E., Woolf, N. J., Hege, E. K., Moore, R. L., \&
Kopriva, D. A. 1978, \apj, 224, 171








\bibitem[Wright \& Barlow (1975)]{wri75}
Wright, A. E., \& Barlow, M. J. 1975, \mnras, 70, 41

\bibitem[Wu et al. (1989)]{wu89}
Wu, C.-C., Holm, A. V., Panek, R. J., et al.
1989, \apj, 339, 443



\bibitem[Yamaoka et al. (2007)]{yam07b}
Yamaoka, H., Haseda, K., \& Nakamura, Y. 2007, \iaucirc, 8832, 2




%
\end{thebibliography}
\end{document}